\pdfoutput=1
\RequirePackage{color}
\documentclass{JINST}
\usepackage{xspace}
\usepackage{multirow}

\def\aprime{\ensuremath{A^{\prime}}\xspace}
\def\atoee{\ensuremath{\aprime\!\to\! e^+e^-}\xspace}

\def\htomm{\ensuremath{H\to\!\mu^+\mu^-}\xspace}

\title{A novel approach to the bias-variance problem in bump hunting}

\author{Mike Williams\\
Laboratory for Nuclear Science, Massachusetts Institute of Technology, Cambridge, MA 02139}

\abstract{
This study explores various data-driven methods for performing background-model selection, and for assigning uncertainty on the signal-strength estimator that arises due to the choice of background model.
The performance of these methods is evaluated in the context of several realistic example problems.
Furthermore, a novel strategy is proposed that greatly simplifies the process of performing a bump hunt when little is assumed to be known about the background.
This new approach is shown to greatly reduce the potential bias in the signal-strength estimator, without degrading the sensitivity by increasing the variance, and to produce confidence intervals with valid coverage properties.
}

\begin{document}

\section{Introduction}

%A critical aspect of {\em bump hunting}, {\em i.e.}\ searching for unknown particles by scanning an invariant mass spectrum,
%is choosing a background model, especially when the signal-to-background ratio $S/B$ is small.
Typically in a bump hunt, {\em i.e.}\ a search for unknown particles using a scan of an invariant mass spectrum, model selection involves choosing what type of series best describes the background and at what order it should be truncated.
If the background-only probability density function (PDF) has sufficient complexity to fully describe the data in the absence of a signal contribution, then the signal-strength estimator $\hat{S}$ will likely be unbiased.
One way to achieve this is to add complexity  to the background model; however, increasing the background-model complexity increases the variance on $\hat{S}$, which degrades the sensitivity of the search.
Conversely, the use of an overly simplistic background model will often produce biased $\hat{S}$ values, along with invalid confidence intervals (CIs) and $p$-values.
How does one know how much complexity is required?

A common approach is to simulate ensembles of data samples using plausible background models, then fit each sample to various alternative model PDFs to determine how much complexity is required to achieve an unbiased $\hat{S}$.
This approach works well for cases where the underlying PDFs are well known.
The potential bias on $\hat{S}$ can be estimated using {\em spurious} signal yields observed in fits to simulated data and accounted for as a systematic uncertainty~\cite{ref:atlas}.
Another approach is to use a data-driven method to choose how much complexity is required in the background model, {\em e.g.}, at what order to truncate a series of background terms.
It is worth stressing that while many well-known data-driven methods exist to perform this task, none are guaranteed to provide unbiased $\hat{S}$ values and, by default, most do not account for uncertainty due to model selection when producing CIs and $p$-values.
Therefore, these data-driven methods should first be applied to simulated data samples to demonstrate that they correctly select sufficiently complex background models when confronted with simulated data that is expected to be similar to the experimental data.
Both of these approaches quickly become infeasible when scanning over a large mass range---especially at masses $\mathcal{O}(\Lambda_{\rm QCD})$, where QCD resonances and other non-monotonic features are expected to appear (somewhere) in the spectrum.
Furthermore, both approaches really only demonstrate unbiased $\hat{S}$ values under the small set of background models simulated, which often do not include any peaking structures.

Ideally, a single data-driven method could be applied across the entire mass spectrum, which does not require performing large-scale simulation studies at each mass.
The method should produce unbiased $\hat{S}$ values, valid CIs and $p$-values, with minimal input required from the analyst.
Furthermore, it is desirable that the method is robust against the presence of small peaking-background structures that are sufficiently wider than the signal.
Qualitatively, a small structure is one that may go unnoticed by the analyst prior to performing the bump hunt,
%{\em e.g.}\ due to the blinding approach and/or because the contribution from the peaking structure is not statistically significant in any mass bin,
and sufficiently wide means dissimilar enough to the signal PDF to avoid inducing a huge variance on $\hat{S}$.
A quantitative assessment of these concepts is given below.

This article explores various data-driven methods for performing background-model selection, and for assigning uncertainty on  $\hat{S}$ that arises due to the choice of background model.
The performance of these methods is evaluated in the context of several realistic example problems.
Furthermore, a novel strategy is proposed that greatly simplifies the process of performing a bump hunt when little is assumed to be known about the background.
A key aspect of this approach, which seems to have gone unnoticed to date, is that the addition of terms to the background model that are orthogonal to the signal PDF on the fit domain do not---on their own---affect the variance of $\hat{S}$.
This new approach is shown to greatly reduce the potential bias in the signal-strength estimator without degrading the sensitivity and to produce CIs with valid coverage properties in all examples considered in this study.
The article is structured as follows: various principles and methods are presented in the context of a toy-model example in Sec.~\ref{sec:method};
two real-world examples are studied in Sec.~\ref{sec:real}; and a summary with detailed discussion is provided in Sec.~\ref{sec:sum}.

\section{Principles \& Methodology}
\label{sec:method}

%This section presents the general principles and methodology of a bump hunt.
The primary quantities of interest in a bump hunt at each test-mass value are $\hat{S}$ and its CI, from which an upper limit can be derived, and the {\em local} $p$-value, which gives the significance of the signal ignoring the trials factor.
If a mass scan is performed, the {\em global} $p$-value---which gives the significance of the largest excess including the trials factor---can be obtained either using Monte Carlo (this is a commonly used approach, {\em e.g.}, see Ref~\cite{ref:me} for a discussion on the procedure) or by employing asymptotic formulae~\cite{ref:LEE,ref:cowan}.
Since converting local $p$-values into global ones is well covered in the literature, this step will not be discussed in detail here.

\subsection{NULL Hypothesis \& Wide Model}

%The first crucial concept in a bump hunt is that of the background-only (NULL) hypothesis.
The NULL hypothesis, against which the signal-plus-background hypothesis is tested, is generally implicitly taken to be {\em the absence of any unknown particles in the mass spectrum}; however, in reality, the NULL is {\em the lack of any features in the data that cannot be explained by the background-only PDF}.
In the absence of a signal contribution, it is vital that the background model is able to describe the data well enough to provide an unbiased $\hat{S}$ estimate and a valid CI and $p$-value.

\begin{figure}[t]
  \centering
  \includegraphics[width=0.32\textwidth]{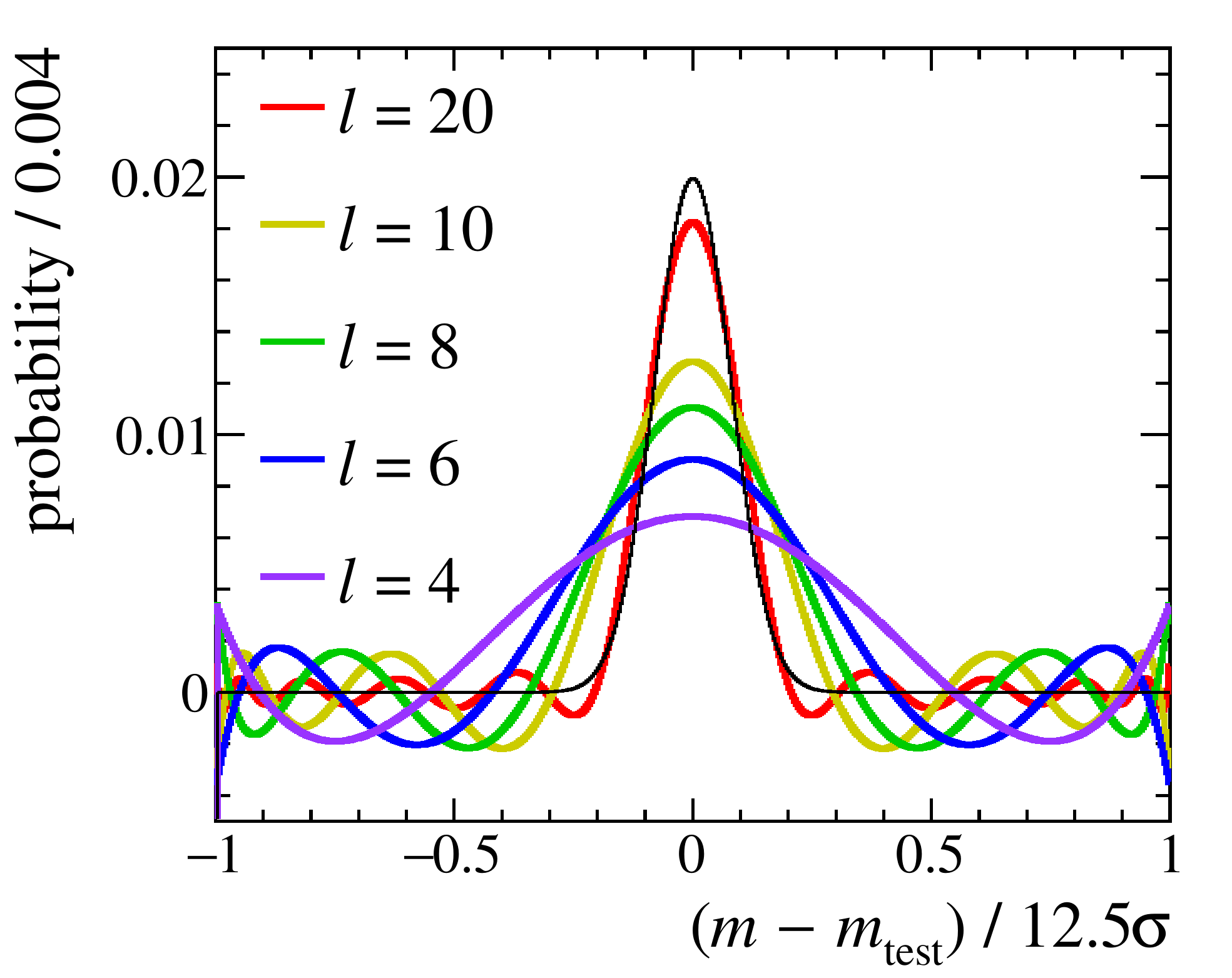}
  \includegraphics[width=0.32\textwidth]{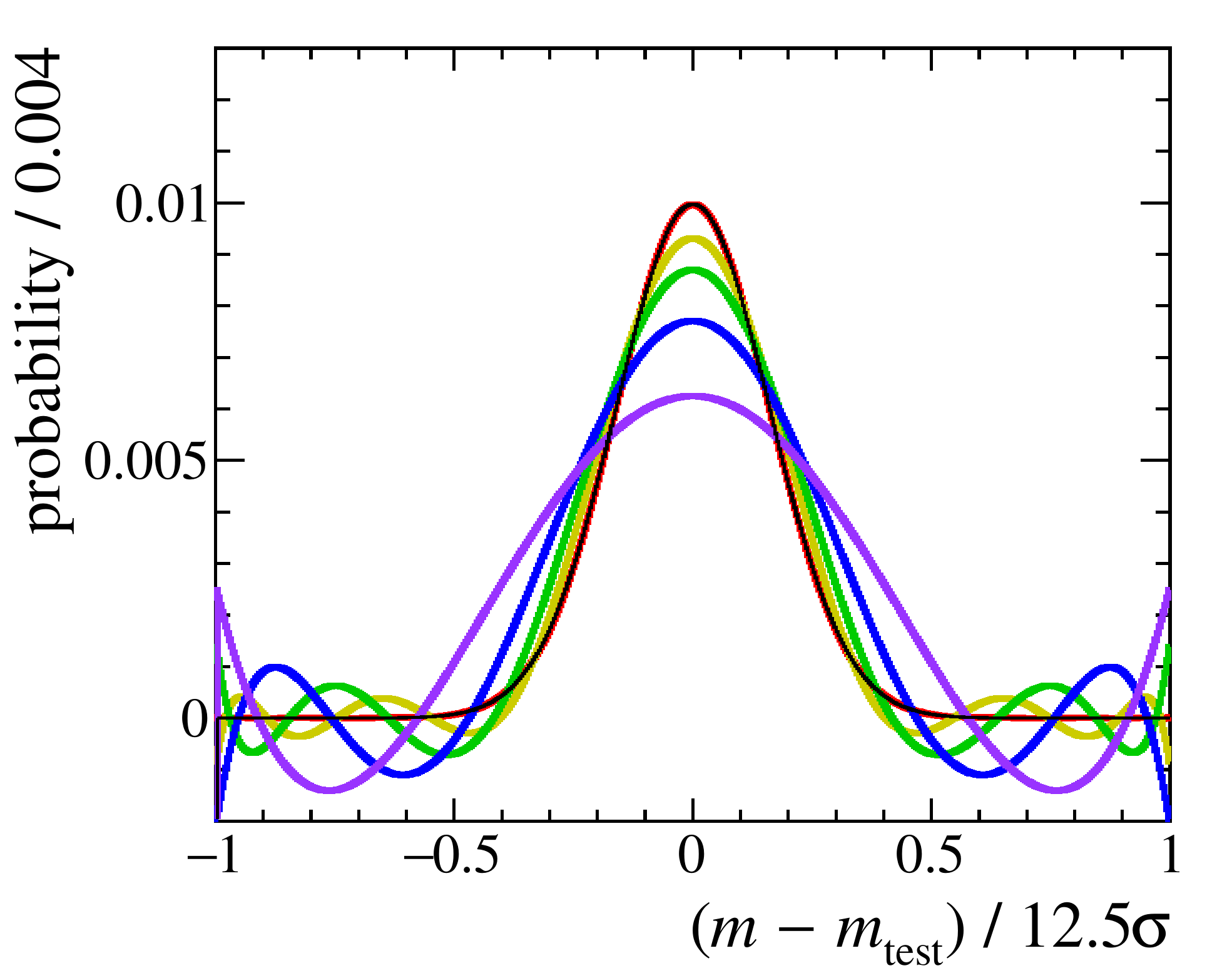}
  \includegraphics[width=0.32\textwidth]{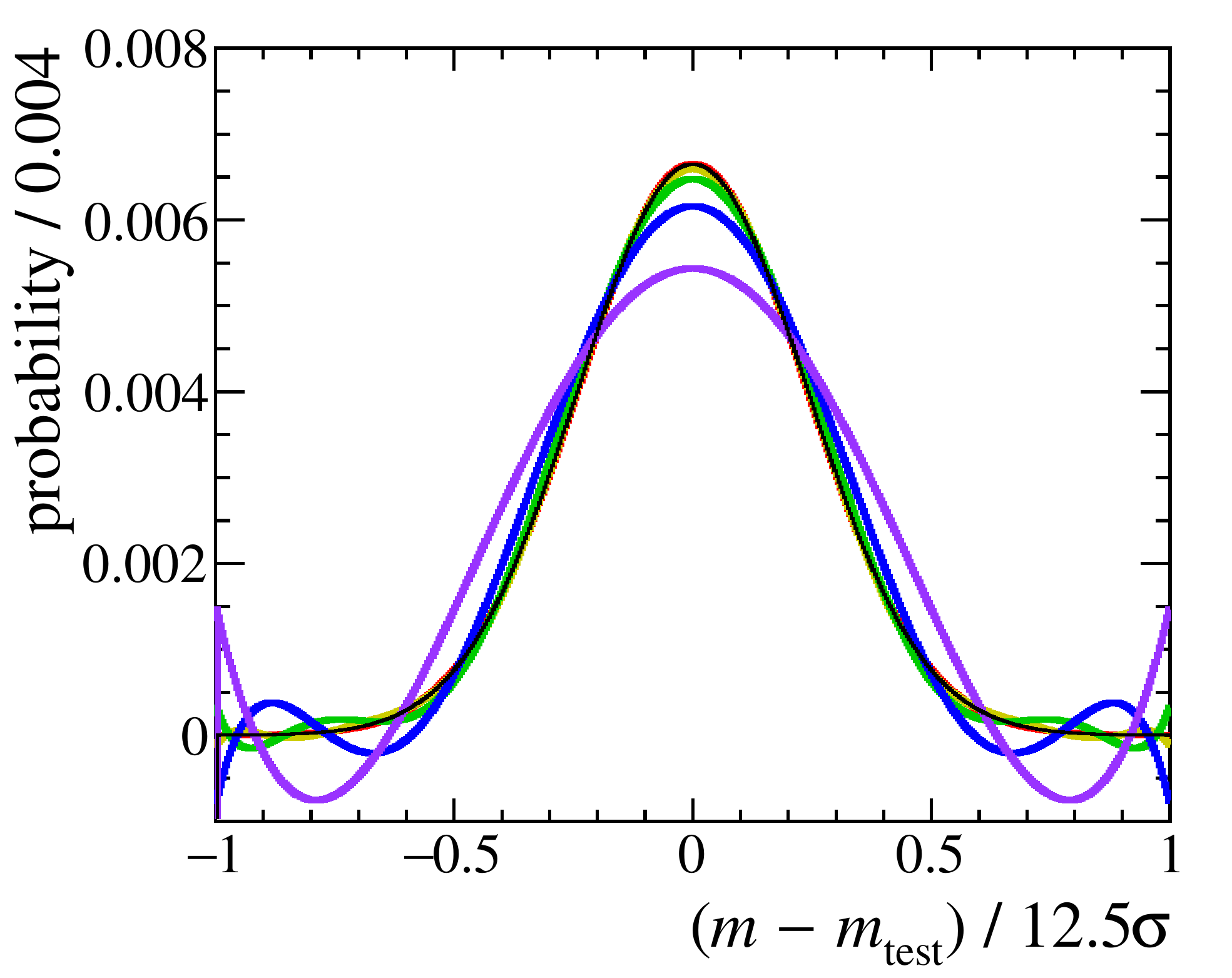}
\caption{Gaussian distributions with widths (left) $\sigma$, (middle) $2\sigma$, and (right) $3\sigma$ projected onto truncated series of Legendre polynomials (the legend labels denote the largest $\ell$ included in each truncated set).}
  \label{fig:gaus_ortho}
\end{figure}

In this study, the local fit region is always taken to have a length of $25\sigma$ (25 times the mass resolution) and be centered on the test-mass value, which includes a $\pm2.5\sigma$ region potentially containing the signal and a $10\sigma$-wide region on either side that is virtually signal free. In principle, the use of a larger fit region can be beneficial if information is known about the form of the background PDF on the larger region; however, if such information is not available, then the use of a larger fit region trades variance for background-model uncertainty (none of the results presented in this article qualitatively depend on the size of the fit region).
Each local fit region is transformed such that it spans the interval $[-1,1]$ with the test mass at zero.
All fits performed in this study maximize the binned likelihood, and all fit component values are computed as the integral of the function across each bin; therefore, when transforming the fit range from $[m-12.5\sigma,m+12.5\sigma]$ to $[-1,1]$ and {\em vice versa}, a one-to-one mapping of the bins is used making such transformations trivial to implement.

Figure~\ref{fig:gaus_ortho} shows a Gaussian signal PDF, along with peaking-background structures taken here to be Gaussian distributions with widths of two and three times that of the signal, each projected onto a truncated series of Legendre polynomials. The key observations to make here are as follows:
\begin{itemize}
  \item A Gaussian signal is an even function on this transformed fit interval; its odd moments are all zero. Real-world signals are typically predominantly even functions with minimal odd moments. Adding a large number of odd modes to the background-only PDF will have minimal impact on the variance of $\hat{S}$ due to this orthogonality, whereas additional even modes will increase the variance (for reasons discussed in the following bullet).
  The extent to which this statement holds for a non-even signal function is easily assessed by comparing its odd and even moments. If the odd moments are roughly an order of magnitude smaller than the even ones,  the impact on the variance of odd modes in the background model will be small.\footnote{This is also the case if the signal mass is not known precisely, {\em e.g.}\ if the uncertainty on the peak position is $\approx \sigma/5$, the odd moments are an order of magnitude smaller than the even ones. While this article focuses on discovery, the same approach advocated here can also be employed to perform mass measurements.}
  \item The signal can be described well using all even modes with $\ell \leq 20$, modulo some {\em ring out} that occurs at a level about an order of magnitude smaller than the peak height. Roughly speaking, the ring out must be comparable to the Poisson uncertainty in each mass bin to disfavor building up a signal-like structure in this manner; therefore, including even modes with $\ell \gtrsim 10$ forbids discovery of signals with $S \lesssim 10\sqrt{B}$. Furthermore, the extent to which a signal-like structure can be built up by the available even modes will be directly reflected in the variance of $\hat{S}$. These considerations provide a natural upper limit on $\ell$ to consider. {\em I.e.}\ if modes with $\ell > 10$ are required to describe the background-only data, the sensitivity will be driven by model uncertainty. In such a case, a dedicated strategy should be developed for this test mass (see comment on resonances in the next bullet).
  \item The large-distance (non-peaking) structure of the background PDF will generally be a mixture of even and odd modes with minimal contribution from large $\ell$ modes. If this is not the case, {\em e.g.}, due to the tail of a large resonance contribution just outside of the fit region, then some additional dedicated long-distance terms must be added to the PDF (see Sec.~\ref{sec:hmm}).
  \item A signal-like shape (peaking background) with a width $\approx 3\sigma$ is well described by a PDF consisting of all even modes $\ell \leq 10$ (wider structures can be described using fewer even modes). Such descriptions, however, are not perfect, and large peaking structures must have dedicated components in the PDF. {\em N.b.}, the presence of such a structure, which is predominantly made up of even modes (just like the signal), will likely result in a biased $\hat{S}$ near the centroid value of the peaking structure unless both the size and shape of the peaking background are highly constrained, though valid CIs and $p$-values will still be possible to obtain.
  \item Structures with widths $\lesssim 3\sigma$ will not be able to be accommodated by the background-only PDF unless they are insignificant (assuming we choose a maximum $\ell$ of 10); therefore, if such a structure exists and is not assigned a dedicated PDF component, the CI and $p$-value will likely be invalid---the test will fail.
\end{itemize}
Based on these observations, the largest allowed $\ell$, {\em i.e.}\ the largest mode where the series could be truncated, can be chosen based on the narrowest possible peaking background structure that might occur at the test mass.
Provided that the test regions are defined in terms of the mass resolution, then the relationship between $\ell_{\rm max}$ and the width of a peaking background that can be handled properly is straightforward.
For example, choosing $\ell_{\rm max} = 10$ corresponds to a NULL hypothesis of {\em the lack of any significant peaking structure with a width $\lesssim 3\sigma$}.
This approach has the desirable feature that the choice of $\ell_{\rm max}$ is largely driven by the potential peaking-background shapes that may occur, which must be considered in detail anyway, rather than on a large-scale simulation study at each mass.
In the event that no peaking-background structures are possible, the choice of $\ell_{\rm max}$ is driven by the variation of the background-only PDF that is not accounted for by any problem-specific PDF components. Choosing $\ell_{\rm max}=6$ or 8 will adequately describe most possible background PDFs while introducing minimal extra variance on $\hat{S}$.

The set of all modes with $\ell \leq \ell_{\rm max}$ and any additional problem-specific PDF components defines the largest possible background model (all background models considered will be subsets of this one), which is referred to as the {\em wide} model in the literature and this name is adopted throughout this study.
Since adding odd modes to the background model does not directly increase the variance---but does reduce the potential for bias on $\hat{S}$---it is sensible to include all odd modes up to (possibly even beyond) $\ell_{\rm max}$ in the background-only model.
Even modes, however, do affect the variance on $\hat{S}$; therefore, adding all even modes up to $\ell_{\rm max}$ will reduce the sensitivity of the search.
%As I will show below, it is important to include enough even modes to avoid $\hat{S}$ bias, but not so many that the sensitivity is greatly reduced.
Because of this, it is desirable to employ a data-driven model-selection procedure to remove {\em unnecessary} even modes from the background-only model.
Such a procedure will introduce its own contribution to the uncertainty, which must be accounted for to construct valid CIs and $p$-values.

\subsection{Toy-Model Example}
\label{sec:toy}

\begin{figure}[t]
  \centering
  \includegraphics[width=0.32\textwidth]{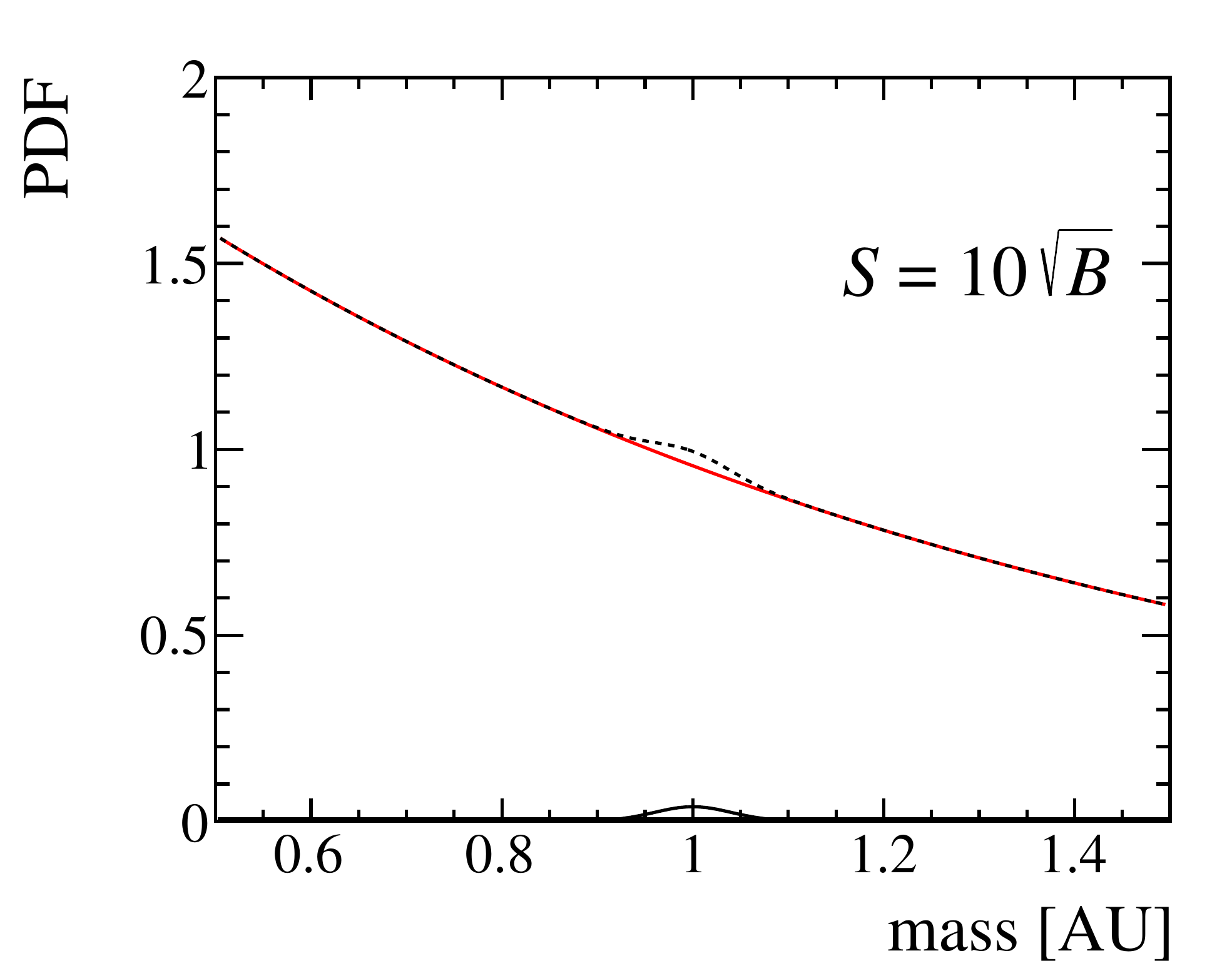}
  \includegraphics[width=0.32\textwidth]{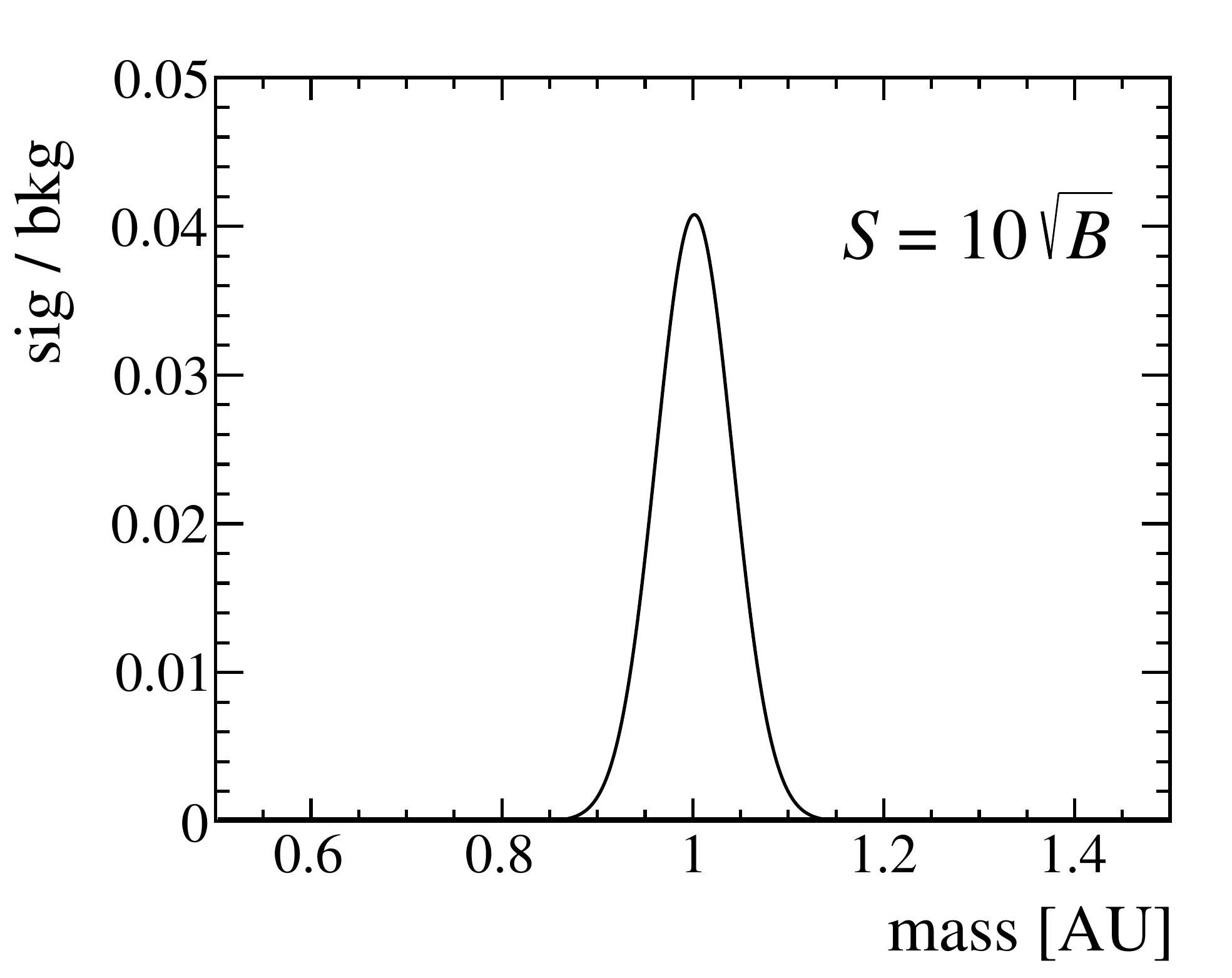}
  \includegraphics[width=0.32\textwidth]{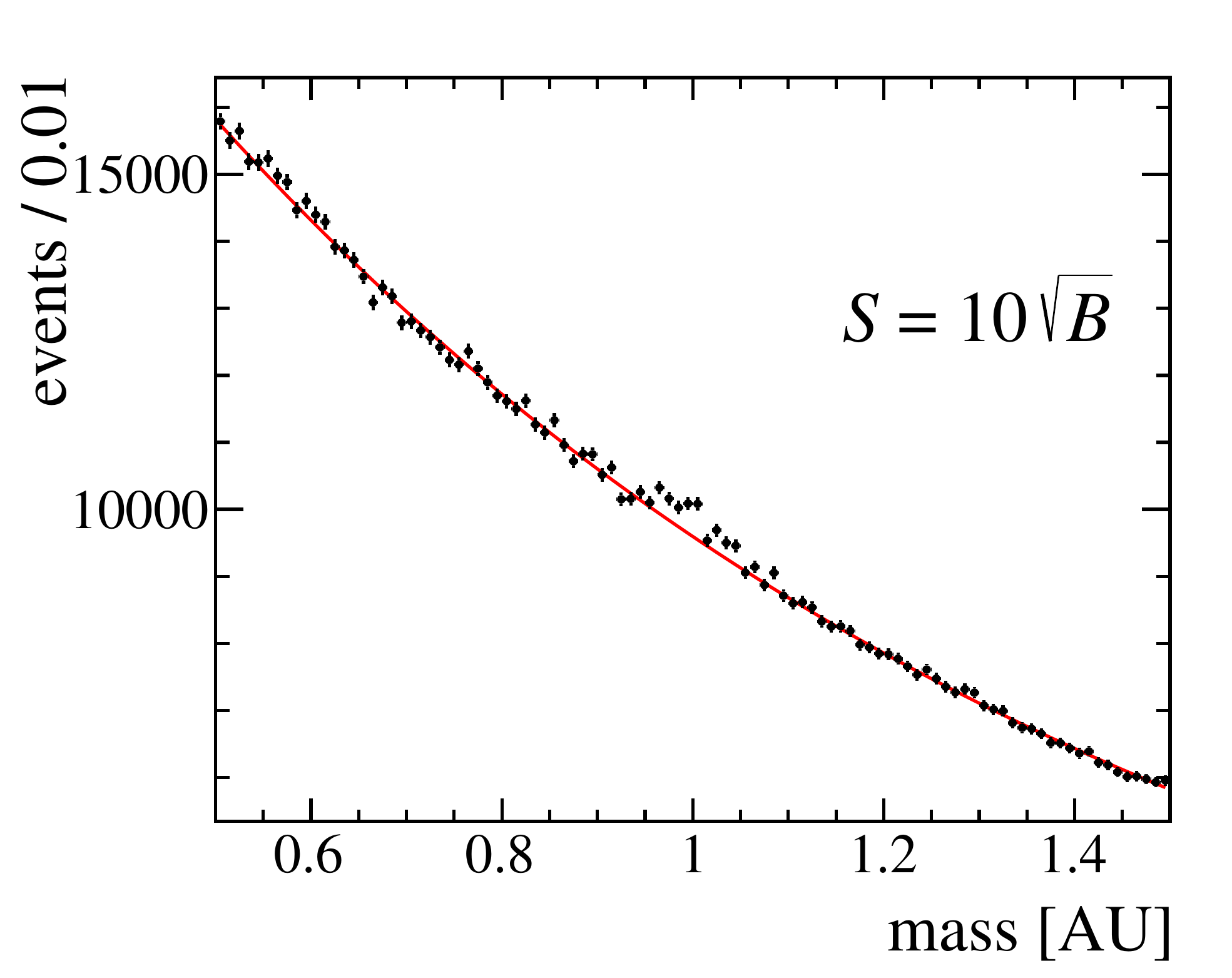}
  \includegraphics[width=0.32\textwidth]{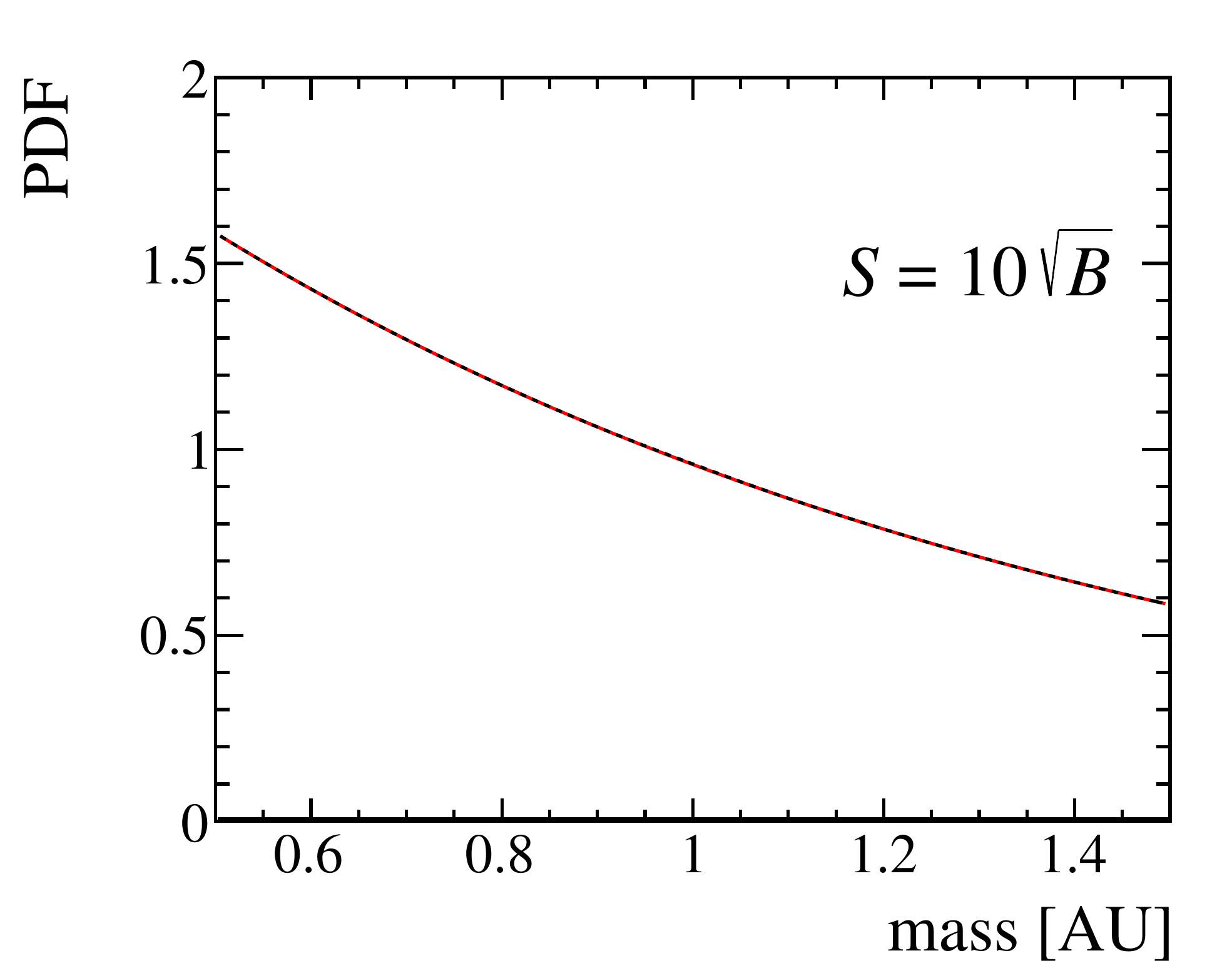}
  \includegraphics[width=0.32\textwidth]{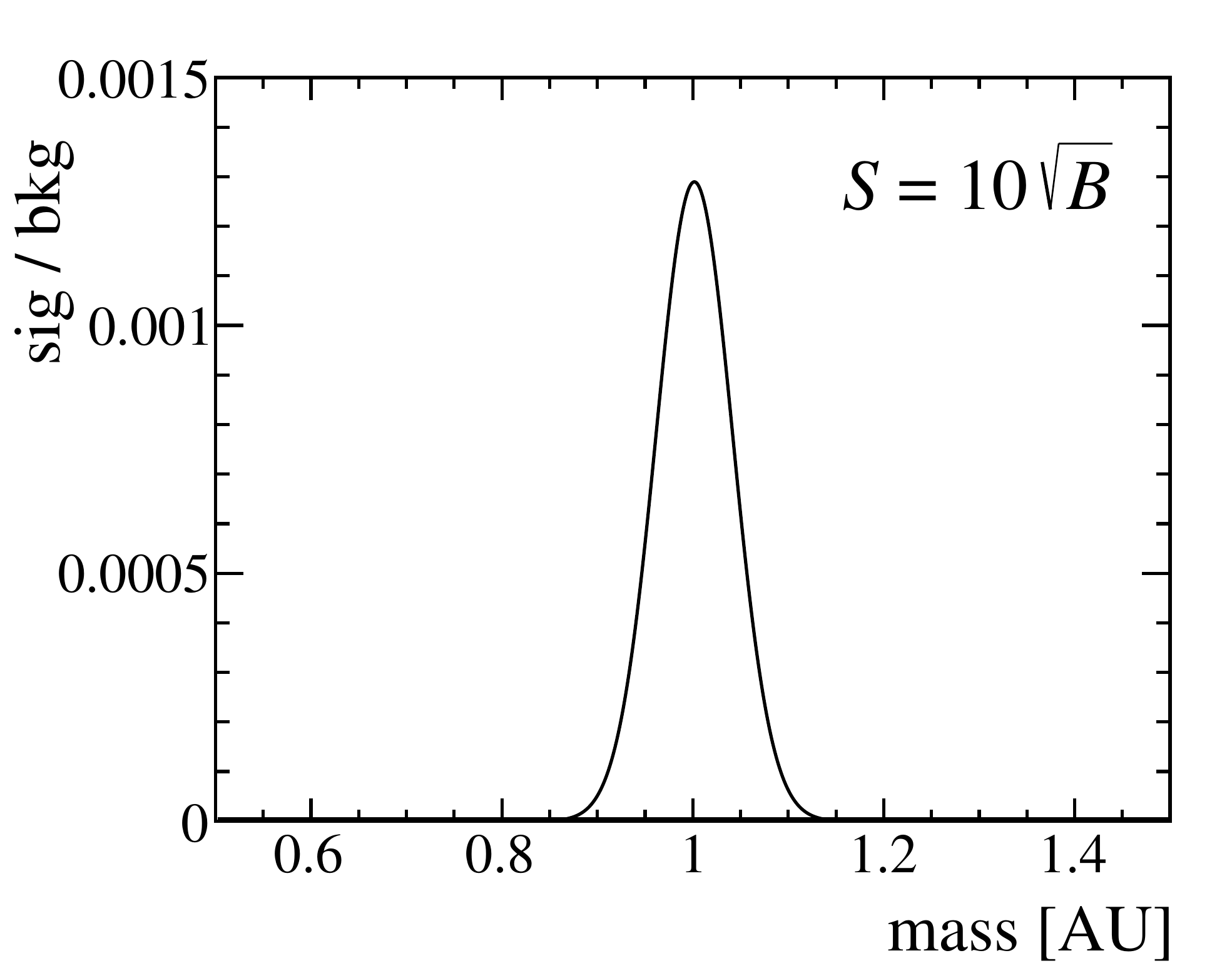}
  \includegraphics[width=0.32\textwidth]{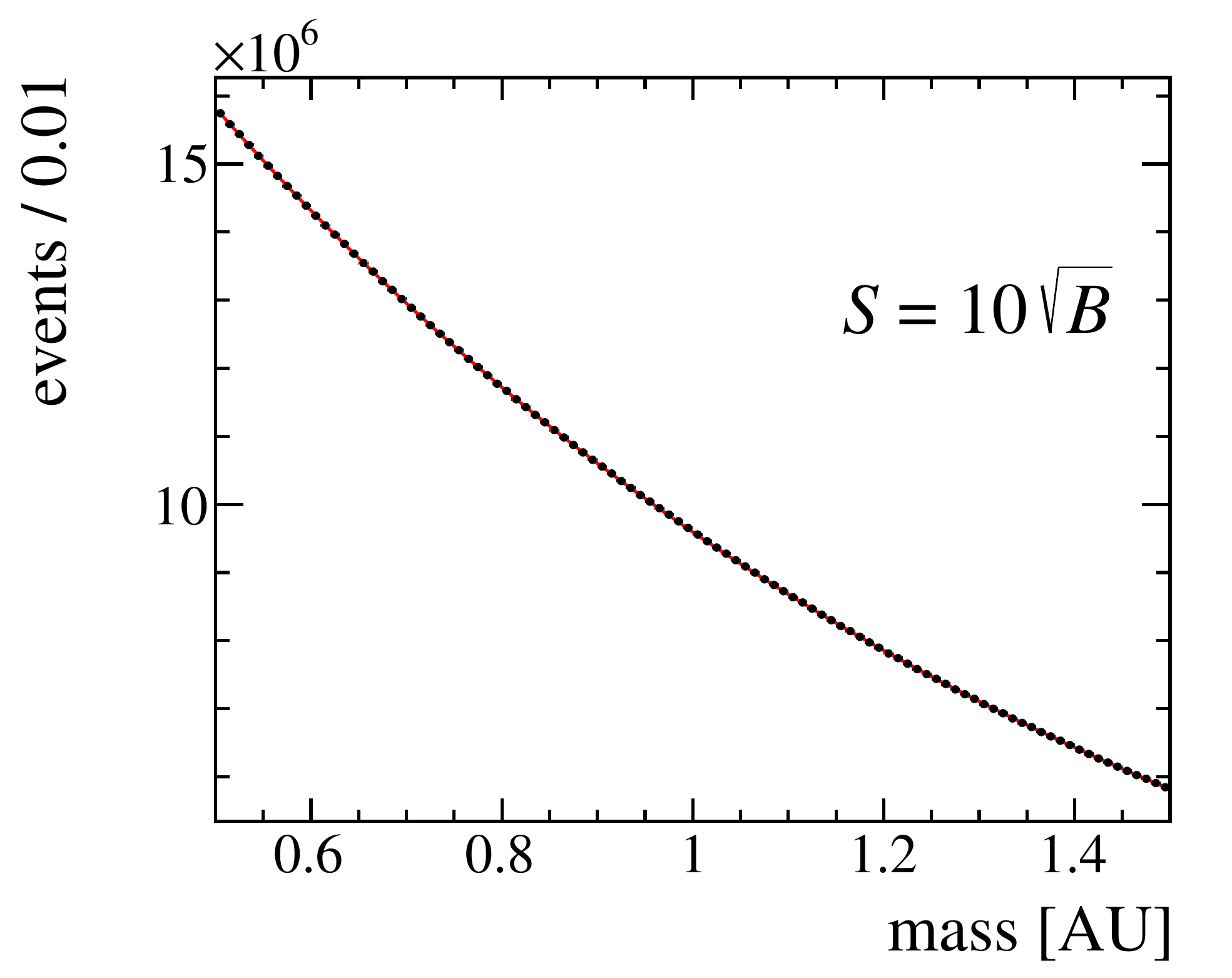}
\caption{Toy-model example for $S=10\sqrt{B}$, where $B$ is the background yield in a $\pm2\sigma$ window around the signal mass (this is the largest $S$ value considered) for (top) $\langle {\rm bkg} \rangle = 10^6$ and (bottom )$10^9$.
The left panels show the PDFs, including the (red) exponential background, (solid black) Gaussian signal, and (dashed black) total.
The middle panels present the signal-to-background ratios for $S=10\sqrt{B}$.
The right panels display example data sets sampled from the PDFs, with the background PDFs shown to help illuminate the signal contribution (though the signal is too small to be seen in the $\langle {\rm bkg} \rangle = 10^9$ data sample).
The signal has a mass of 1.00 and a width of 0.04 (in arbitrary units), making the full mass range considered a $25\sigma$ window.
}
  \label{fig:exp_pdf}
\end{figure}

A toy-model example problem is used throughout this section that consists of a Gaussian signal and an exponential background.
The expected background yield is taken to be $\langle {\rm bkg} \rangle = 10^6$ or $10^9$ and the expected signal yield is $S = s \sqrt{B}$, where $B$ is the background yield in a $\pm2\sigma$ window around the signal mass and values of $s=0,1,2,5$ and 10 are considered.
An ensemble of 1000 data sets is generated using this PDF for each expected background and signal yield, a total of $2\times5=10$ ensembles.
The yield in each mass bin is sampled from a Poisson distribution, an example data set is shown in Fig.~\ref{fig:exp_pdf}.
In this toy-model example, only one mass value is considered (no scan in mass is performed).
Nothing will be assumed to be known about the background-only PDF, except that it could contain a small peaking-background structure with a width $\gtrsim 3\sigma$; therefore, $\ell_{\rm max} = 10$ is chosen as the largest $\ell$ value that will be considered for the background model.
Despite its simplicity, this example suffices to illustrate the key principles of bump hunting.

\subsection{Model Selection \& Uncertainty}

The literature on data-driven model selection is vast (see, {\em e.g.}, Ref.~\cite{ref:overview}).
The general strategy is to reward goodness of fit while punishing model complexity.
The most commonly used methods are based on the penalized likelihood
\begin{equation}
  \Lambda = -2\log{\hat{\mathcal{L}}} + c \cdot n({\rm par}),
\end{equation}
where $\hat{\mathcal{L}}$ is the maximum likelihood value, $n({\rm par})$ is the number of unknown parameters in the model, and $c$ is a constant where:
\begin{itemize}
  \item $c=0$ corresponds to adding no penalty term to the likelihood;
  \item $c=2$ is the Akaike Information Criterion (AIC)~\cite{ref:aic}, which (in the asymptotic limit) estimates the Kullback-Leibler (KL) divergence between the model and the truth, up to an unknown constant ({\em i.e.}\ the model with the minimum AIC value is the one that minimizes the KL divergence in the asymptotic limit);
  \item and $c=\log{[n({\rm bins})]}$ (or $\log{[n({\rm candidates})]}$ for an unbinned fit) is the Bayesian Information Criterion (BIC)~\cite{ref:bic}, which is derived using a Bayesian approach and under certain conditions asymptotically selects the true model (if the true model is in the set of models considered).
\end{itemize}
Other penalty terms also exist in the literature (see, {\em e.g.}, Ref.~\cite{ref:hjort} and references therein).
%The fact that BIC asymptotically selects the true model may seem appealing, but does one ever really expect the true background model to be under consideration?
Since $\log{[n({\rm bins})]} > 2$ for $n({\rm bins}) > 7$, the BIC penalty term is always larger than that of AIC in a bump hunt; therefore, the model selected by BIC will be at most as complex as the one selected by AIC (often, it will be less complex).
This results in larger potential biases when using BIC, though also possibly smaller variance.
In this work, I will focus on AIC-based strategies, leaving discussion of alternative methods, including performing a data-driven optimization of $c$, until Sec.~\ref{sec:other}.
{\em N.b.}, the value $c=2$ is optimal under the AIC assumptions for any model space, {\em i.e.}\ using $c=2$ is not specific to the choice here of using Legendre polynomials.

Within the context of the toy-model problem presented in the previous subsection, the following approaches for determining $\hat{S}$, the CI, and local $p$-value are considered (all based on binned maximum-likelihood fits):
\begin{description}
  \item[(true)] $c=0$ using the true background PDF, representing the best-possible performance (this is assumed to be unachievable in reality, but is shown to provide a benchmark);
  \item[(wide)] $c=0$ using a background PDF that includes all Legendre modes with $\ell \leq 10$, this is the largest model considered (recall that this is referred to as the wide model);
  \item[(step)] a {\em stepwise-in} approach with $c=0$, where the background model is initially chosen as only the $\ell = 0$ mode, and higher-order modes are added successively but only if $\Delta \Lambda > 1$;\footnote{This stepwise-in approach is certainly {\em ad hoc}, but since it has been commonly used in physics analyses I choose to study its performance here rather than any of the better-motivated stepwise-based methods. For example, in non-discovery analyses, it is not uncommon for the background model to be chosen by fitting with a low-order function and then deciding manually whether higher-order terms are required based on the change in the $\chi^2$ value. Such an approach is equivalent to the stepwise-in one implemented here and is often mistaken for being the $F$ test.}
  \item[(aic)] the standard AIC approach, where background models with $\ell_{\max}$ from 0 to 10 are considered and the one that minimizes $\Lambda$ with $c=2$ is selected;
  \item[(aic-o)] an AIC-based approach, where each background model contains every odd mode with $\ell \leq 9$, and all possible subsets of even modes with $\ell \leq 10$ are considered ($2^6 = 64$ total models), with the background model that minimizes $\Lambda$ with $c=2$ selected;
  \item[(avg)] and a {\em frequentist model averaging} approach with $c=2$, where all models considered in the previous bullet are averaged over to obtain $\hat{S}$ (no model is selected).
\end{description}
The frequentist model averaging approach~\cite{ref:buckland} is natural to consider in a bump hunt, since one has no interest in the background model beyond its impact on $\hat{S}$, the CI, and local $p$-value. The variation considered here defines a weight for each of the 64 models $w_m \propto {\rm exp}(-\Lambda_m/2)$, where $m$ denotes the model and $\Lambda$ is defined using $c=2$ (this results in the exponent being half the AIC value for each model); the weights are then normalized such that they sum to unity.
The signal strength is the weighted average of the $\hat{S}$ values obtained from each of the 64 models $\hat{S}_{\rm avg} = \sum w_m \hat{S}_m$.

Traditionally, once a model is selected (however that is done), the CI for $\hat{S}$ and the $p$-value are determined using the profile likelihood---and any uncertainty due to the choice of model is ignored, though possibly investigated during separate systematic studies using an {\em ad hoc} approach.
Ignoring the model-selection uncertainty is, of course, valid when using the true background PDF, and assumed to be valid by construction when using the wide model.
Recall that the wide model is defined such that it has sufficient complexity to accommodate any structure that could be manifest in the data in the absence of a signal; therefore, by construction using the wide model results in an unbiased $\hat{S}$, a valid CI and $p$-value, and no model-selection uncertainty is required (provided that the assumptions made about the background when constructing the wide model are valid) since the model uncertainty is incorporated into the variance.
Otherwise, model-selection uncertainty can be sizable, even when using a well-known data-driven approach like AIC, and must be accounted for to obtain valid CIs and $p$-values.

There are a number of prescriptions for handling model-selection uncertainty in the literature, though most require calculating complicated problem-specific functions (see, {\em e.g.}, Ref.~\cite{ref:fic}).
Ref.~\cite{ref:iic} proposed a simple $\Lambda$-based approach that works as follows:
$\Lambda$ is calculated for all models with a non-zero choice for $c$ ($c=2$ is used in this study);
then the CI is obtained from a profile of $\Lambda$, including the penalty term, where the model index $m$ is treated as a discrete nuisance parameter.
This results in the $\Lambda$ value for each signal strength being taken as the minimum $\Lambda$ value from all models (including the model-dependent penalty term); {\em i.e.}\ $\Lambda$ is minimized at each signal-strength value in the same way it is for a continuous nuisance parameter.
The CIs are then defined as usual, {\em e.g.}, a 68.3\% interval is defined as the region where $\Lambda - \Lambda(\hat{S}) \equiv \Delta\Lambda < 1$.
Finally, in the frequentist model averaging approach, a symmetric CI is assigned about $\hat{S}_{\rm avg}$ defined using a variance of $\sigma^2_{\rm avg} = \sum w_m \left[(\hat{S}_m - \hat{S}_{\rm avg})^2 + \sigma_m^2 \right]$, where the first term provides an estimate of the bias for each model and the second is the per-model variance.
%\footnote{Physicists often employ a crude type of model averaging. For example, consider a fit where two background models are considered with the same number of parameters, and the likelihoods obtained using both background models are approximately equal. The use of the mean value of the two fit results for $\hat{S}$ and half the difference as a systematic uncertainty gives approximately the same result as the model-averaging approach.}

\begin{figure}[t]
  \centering
  \includegraphics[width=0.49\textwidth]{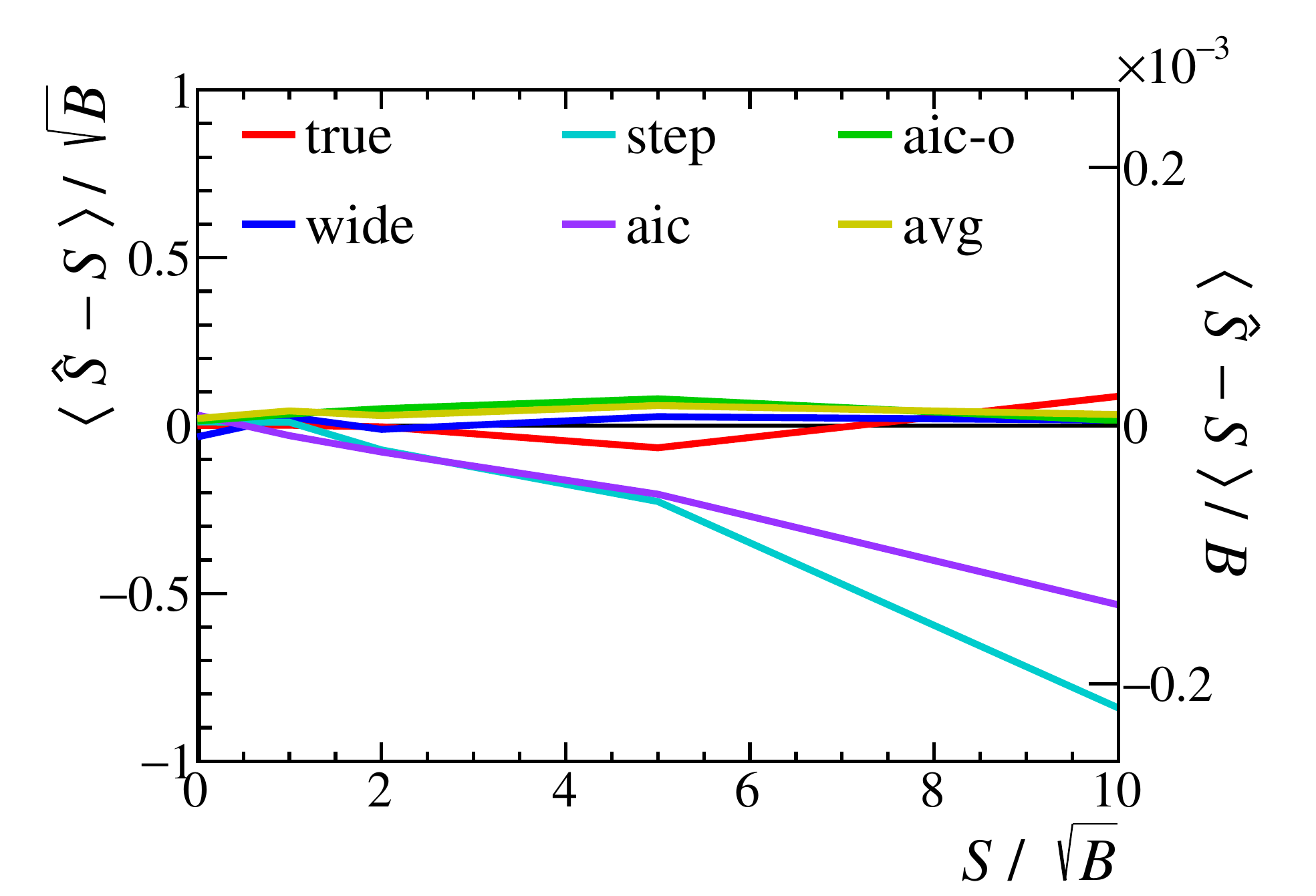}
  \includegraphics[width=0.49\textwidth]{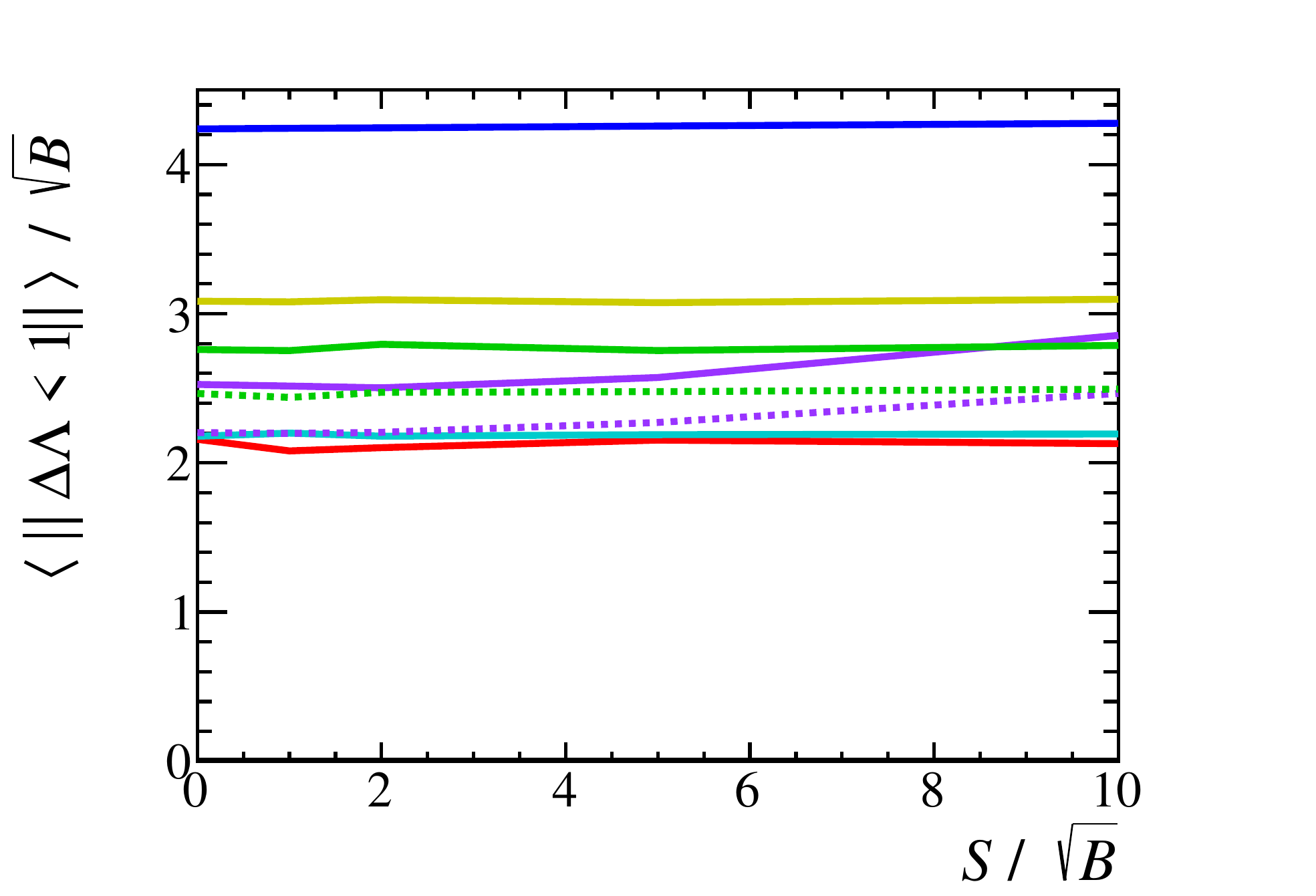}
  \includegraphics[width=0.49\textwidth]{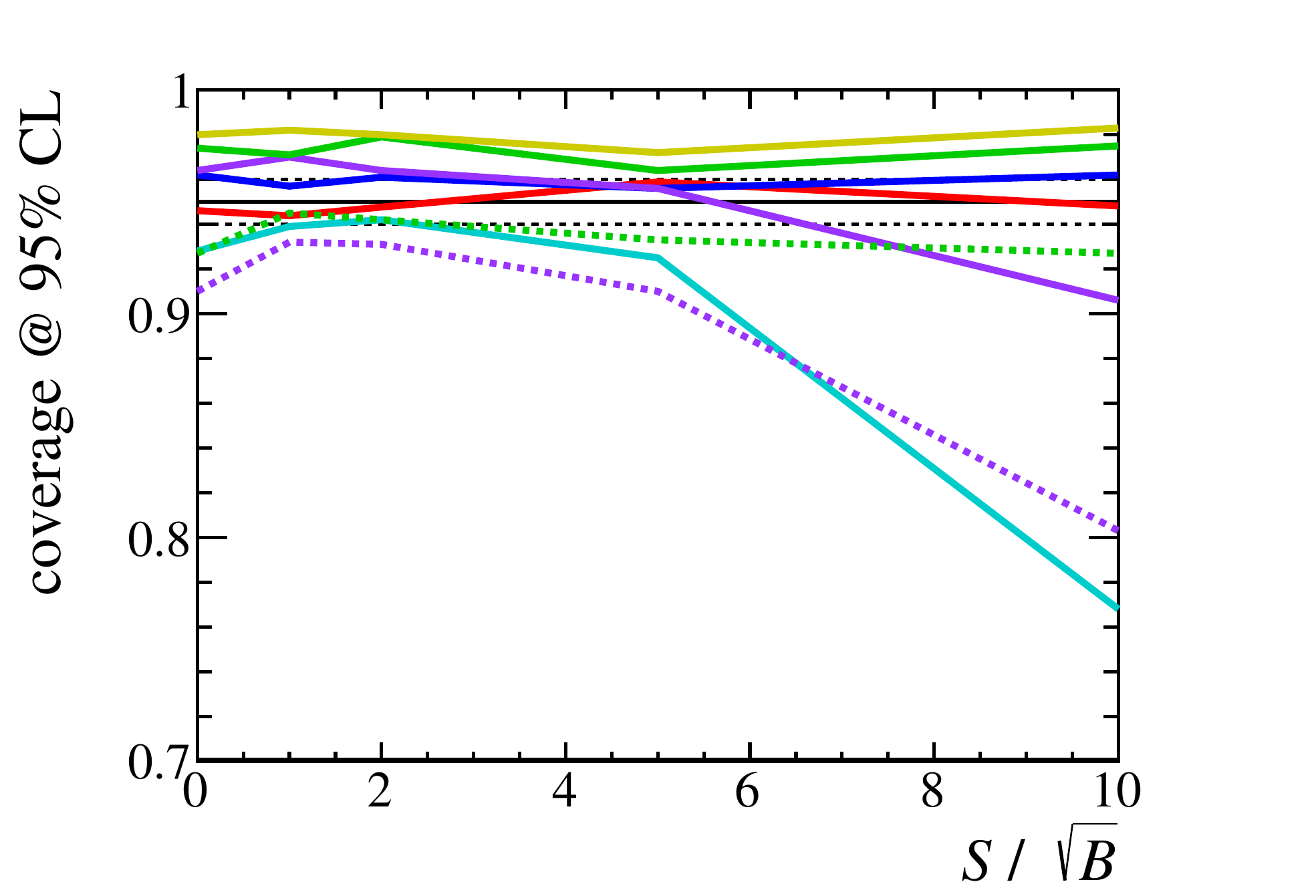}
  \includegraphics[width=0.49\textwidth]{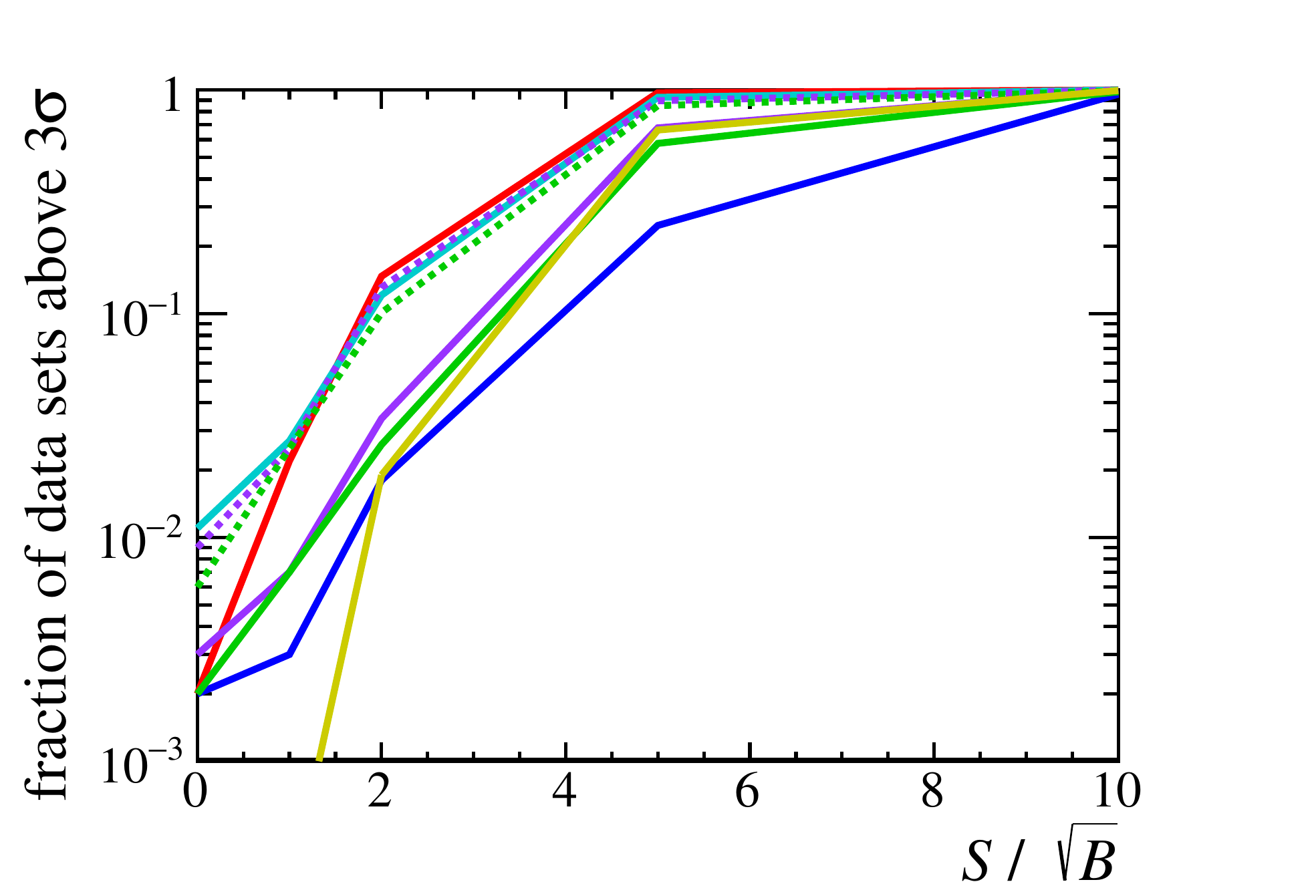}
\caption{
Results for the toy-model example with $\langle {\rm bkg} \rangle = 10^6$ versus the true signal $S$ relative to $\sqrt{B}$, where $B$ is the background yield in a $\pm2\sigma$ window around the signal mass
(see Table~\ref{tab:labels} for legend label definitions).
(top~left) The mean bias in signal estimator $\hat{S}$ relative to (left axis) $\sqrt{B}$ and (right axis) $B$.
(top right) The mean full length of the 68.3\% CI relative to $\sqrt{B}$ as reported by the profile likelihood, where the dashed lines show the aic and aic-o results without accounting for model-selection uncertainty.
(bottom left) Coverage reported at 95\% CL by each method, where the solid black line shows the expected 95\% and the dashed black lines show the approximate uncertainty in all results due to ensemble size.
(bottom right) The fraction of data sets that report a $p$-value corresponding to a significance $>3\sigma$ for the NULL hypothesis (the expected value is 0.27\% when $S=0$ as a two-sided test is performed due to the limited number of data sets).
{\em N.b.}, in all examples in this study, the statistical uncertainty on $\langle \hat{S} - S \rangle/\sqrt{B}$ is small and well approximated by $\left[\langle || \Delta\Lambda < 1|| \rangle / \sqrt{B} \right] / (2\sqrt{1000}) \in [0.03,0.06]$.
}
  \label{fig:exp0}
\end{figure}

\begin{figure}[t]
  \centering
  \includegraphics[width=0.49\textwidth]{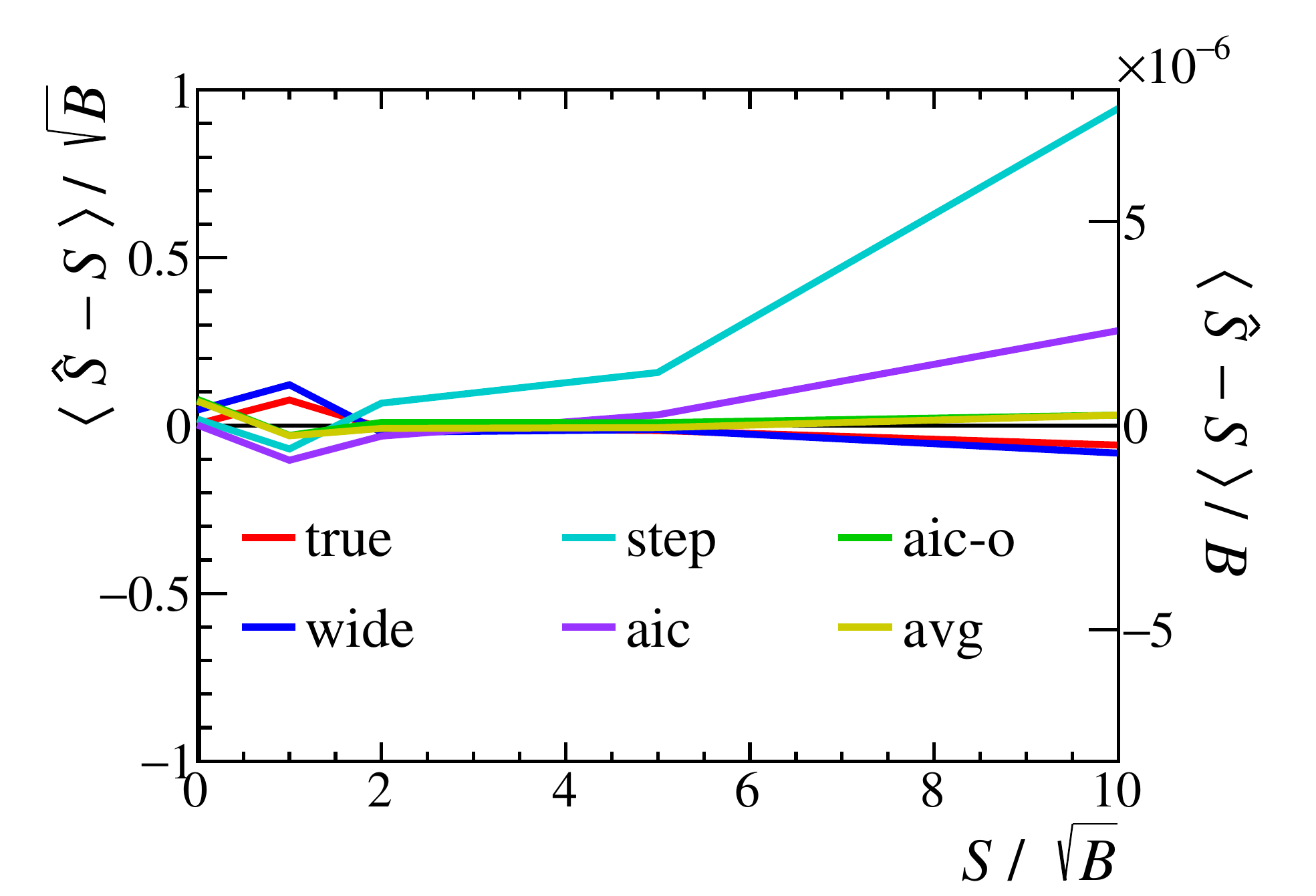}
  \includegraphics[width=0.49\textwidth]{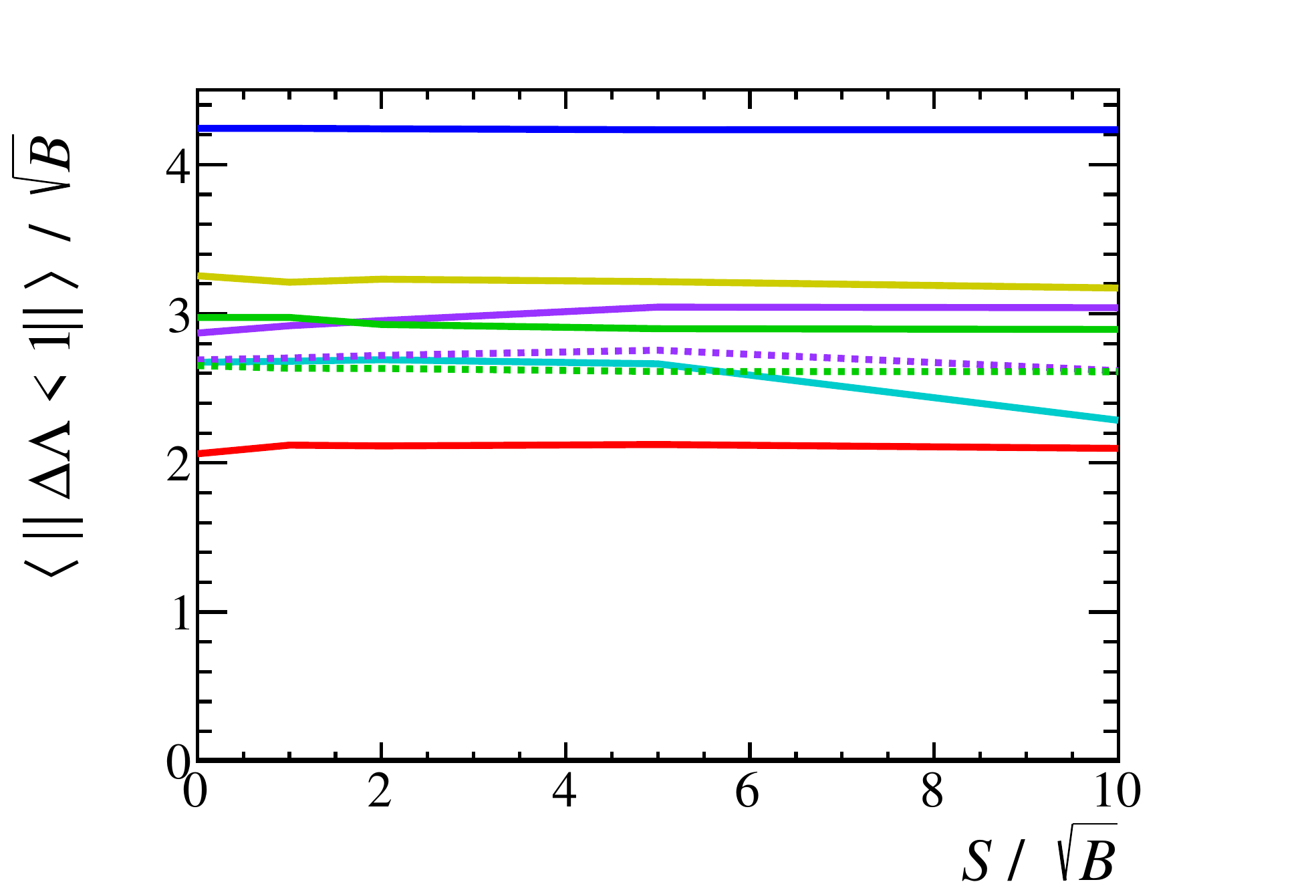}
  \includegraphics[width=0.49\textwidth]{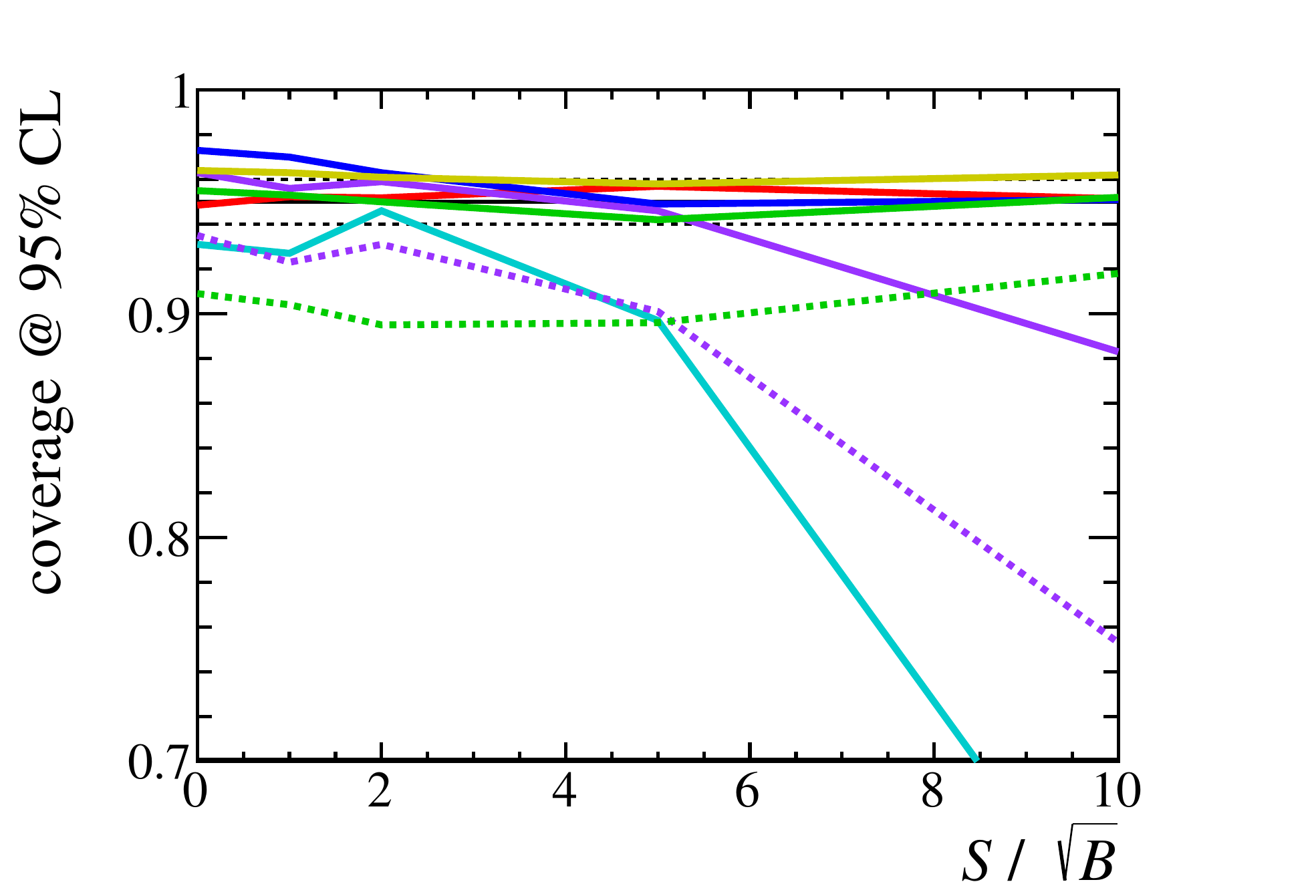}
  \includegraphics[width=0.49\textwidth]{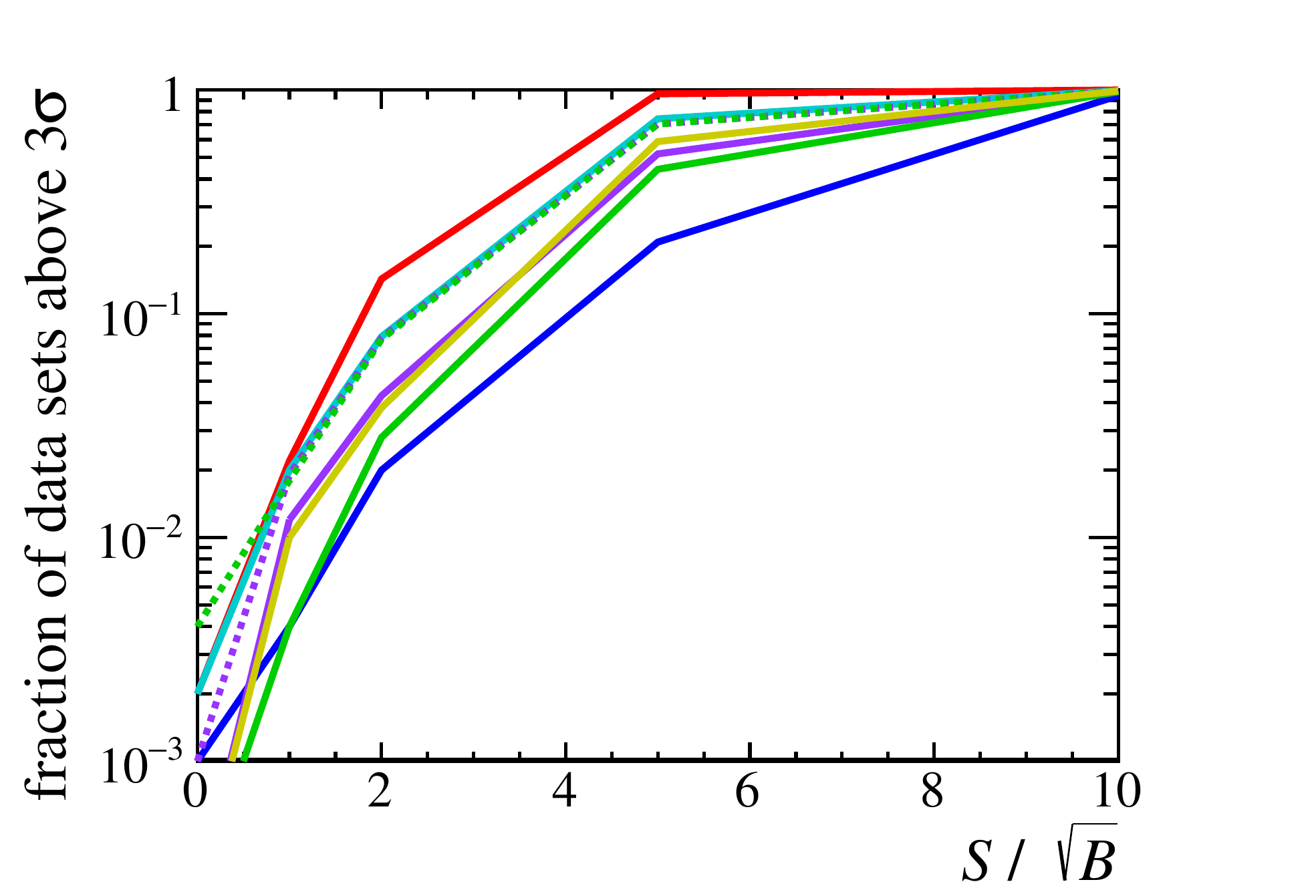}
\caption{Same as Fig.~\ref{fig:exp0} but for $\langle {\rm bkg} \rangle = 10^9$.}
  \label{fig:exp2}
\end{figure}

\begin{figure}
  \centering
  \includegraphics[width=0.329\textwidth]{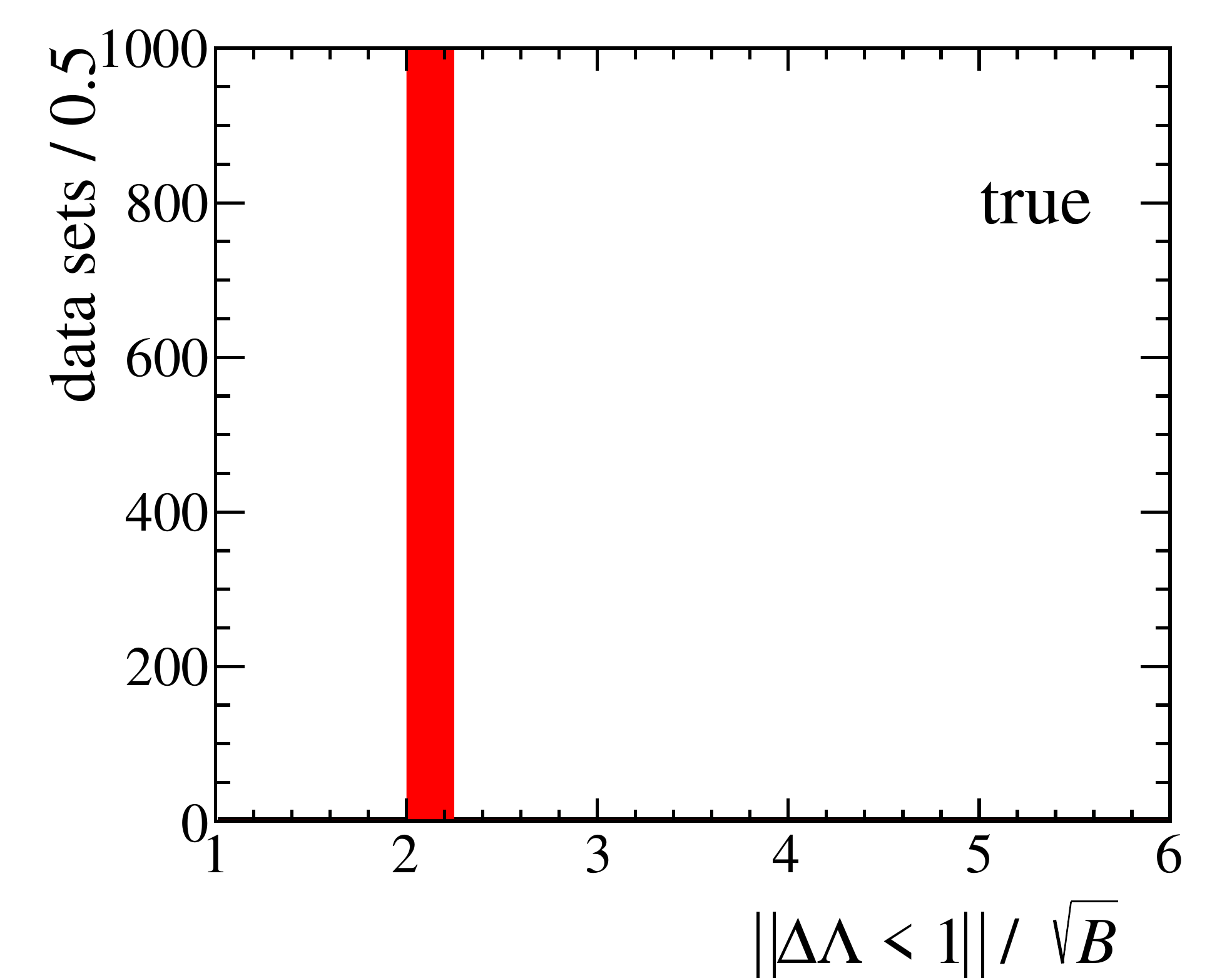}
  \includegraphics[width=0.329\textwidth]{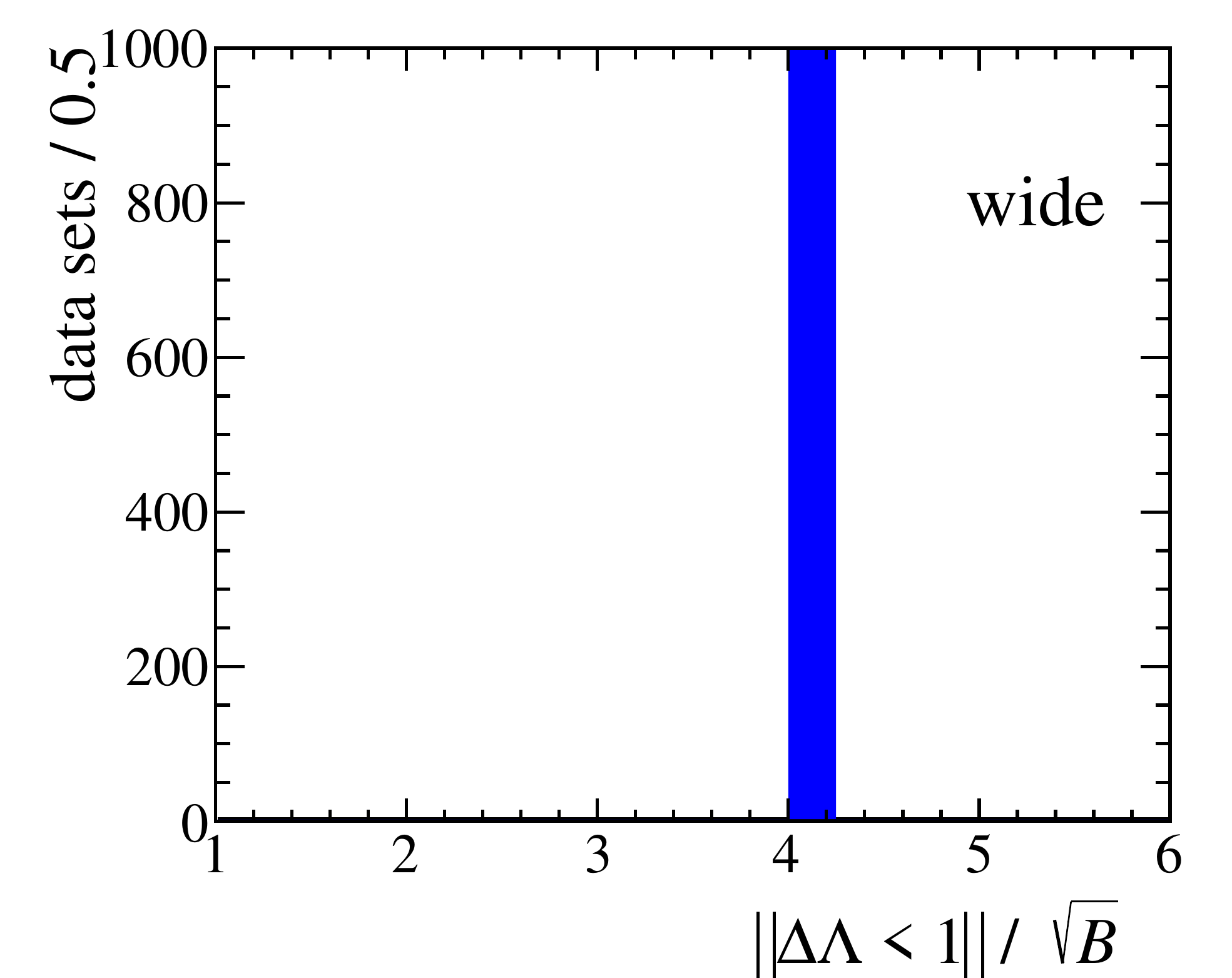}
  \includegraphics[width=0.329\textwidth]{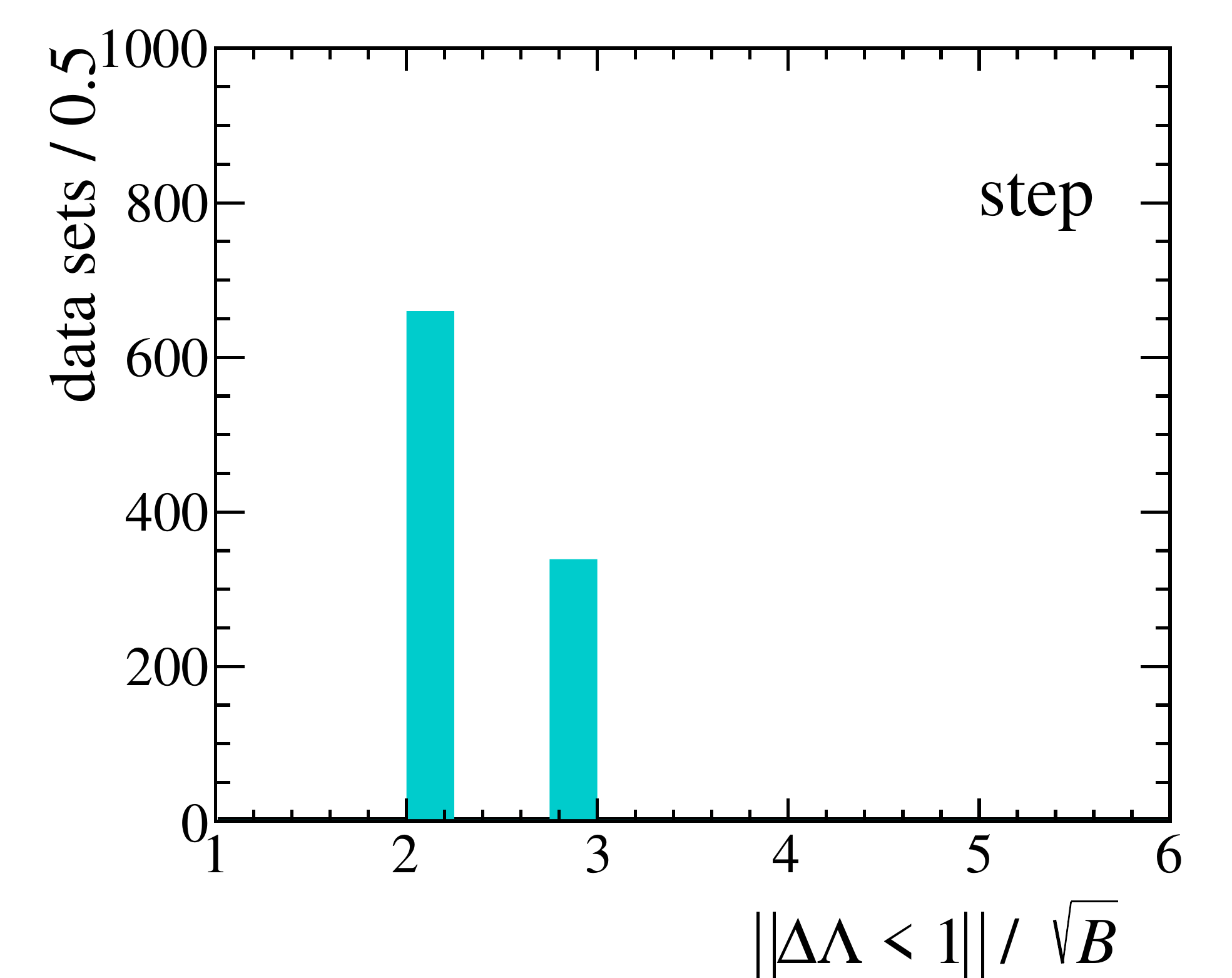}
  \includegraphics[width=0.329\textwidth]{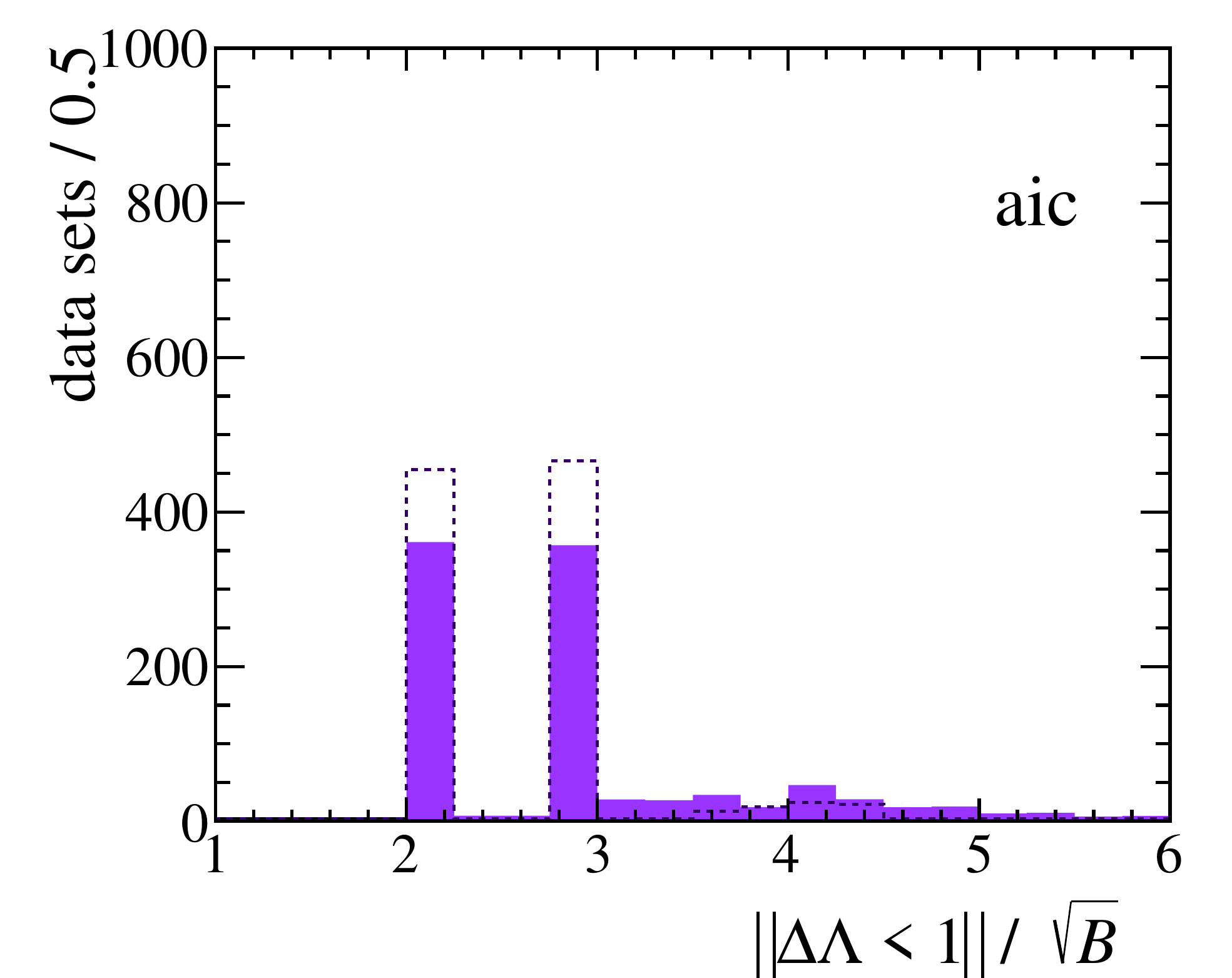}
  \includegraphics[width=0.329\textwidth]{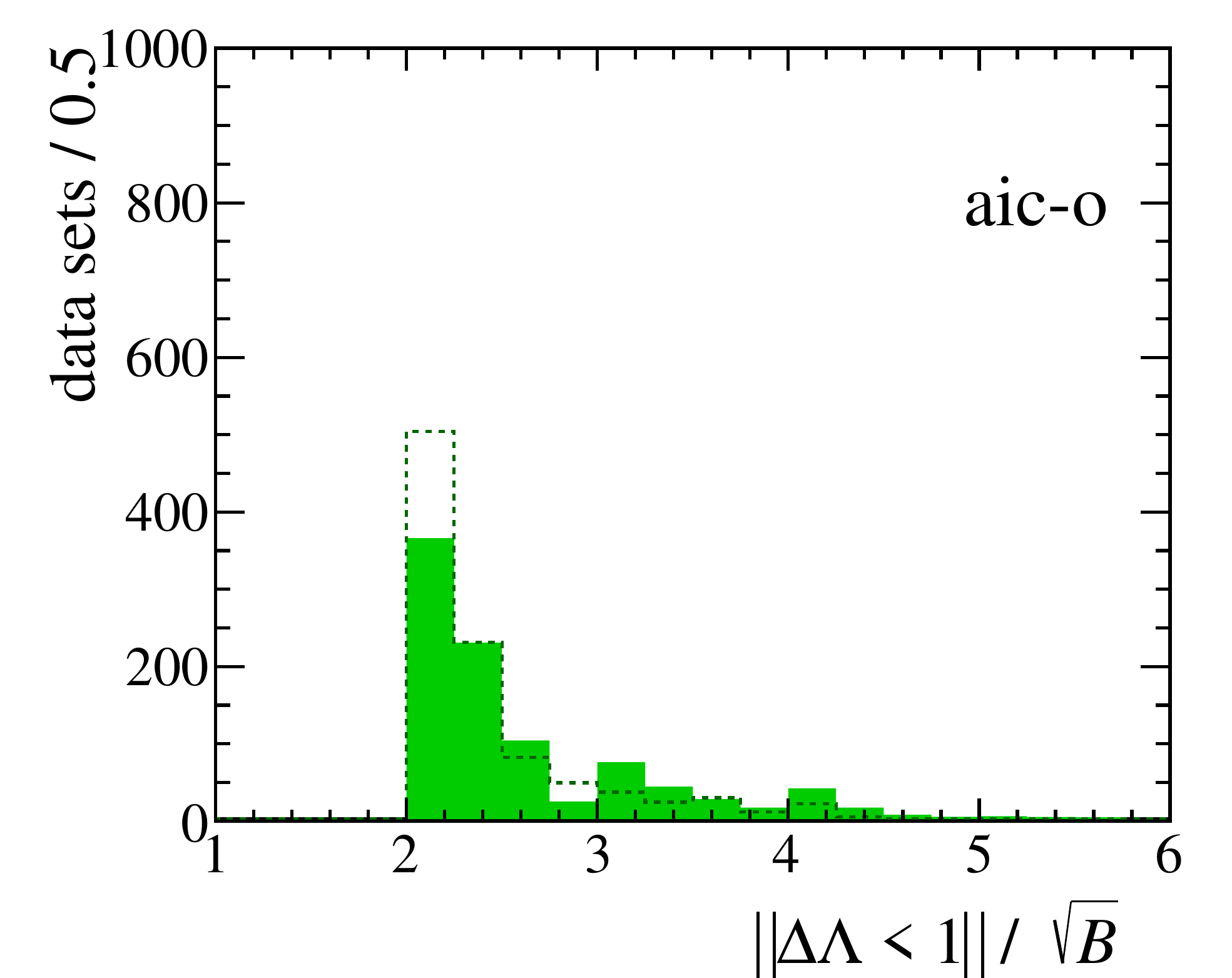}
  \includegraphics[width=0.329\textwidth]{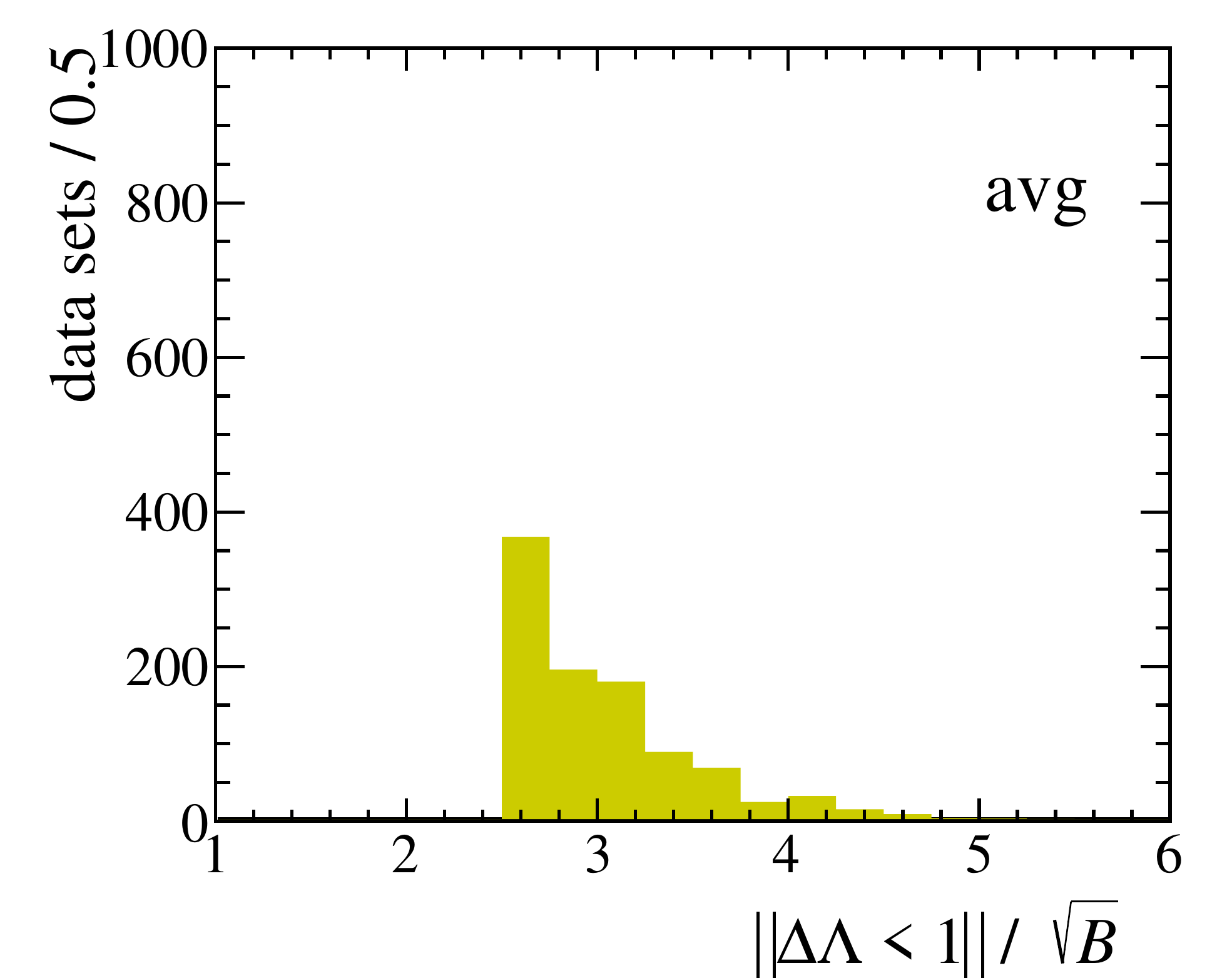}
  \caption{Distributions of the full length of the 68.3\% CIs relative to $\sqrt{B}$ as reported by the profile likelihood for the toy-model example for $\langle {\rm bkg} \rangle = 10^6$ and $S=10\sqrt{B}$.
  In the aic and aic-o plots, the dashed lines show the CI lengths without accounting for model-selection uncertainty.
  }
    \label{fig:sigma_dist}
  \end{figure}

Figures~\ref{fig:exp0} and \ref{fig:exp2} show the results of applying each method to the toy-model problem described in Sec.~\ref{sec:toy} for $\langle {\rm bkg} \rangle = 10^6$ and $10^9$, respectively.
The results are summarized as:
\begin{itemize}
  \item the stepwise-in and standard AIC approaches yield large biases in $\hat{S}$ at larger $S$,\footnote{The sign of the bias changes going from $10^6$ to $10^9$ background events, which occurs because, on average, the order at which the background model is truncated increases by one for this specific example. }
   while all other methods are found to be unbiased (the potential bias is 10 times smaller than the statistical variance, which is comparable to the uncertainty due to ensemble sample size);
  \item the 68.3\% CI for the wide model is about twice the ideal value, which sets the scale of what can be gained by performing model selection ({\em i.e.}\ since, by construction, the wide model is unbiased and produces valid CIs and $p$-values, if this loss of sensitivity is acceptable, the wide model can be used and no model selection is required);
  \item including all odd modes has minimal impact on the variance as expected ({\em c.f.}\ green to purple lines, dashed or solid, in the top-right panels);
  \item conversely, and also as expected, including all odd modes greatly reduces the potential bias on $\hat{S}$ ({\em c.f.} green to purple lines in the top-left panels);
  \item including the model-selection uncertainty in the AIC-based methods increases the lengths of the CIs by about 10--20\%, these CIs are about 10--20\% shorter than the ones reported by the model-averaging approach;
  \item the stepwise-in and standard AIC approaches do not yield proper coverage, and while including model-selection uncertainty greatly improves the standard-AIC coverage, the method still undercovers at large $S$;
  \item all other methods have good coverage properties, though this is only true for the AIC-based approach that includes all odd modes if model-selection uncertainty is accounted for;
  \item the significance when $S=0$ is overestimated for the stepwise-in method, and for both AIC-based approaches if model-selection uncertainty is not included, but shows good (possibly conservative) performance for all other approaches (including the AIC-based methods that account for model-selection uncertainty);
  \item for $S=10\sqrt{B}$ the significance reported by the wide model is about half that of the true model, which is expected given that the wide-model CI is about twice as long as that of the true model (quantitatively, this relationship will depend on the value chosen for $\ell_{\rm max}$);
  \item and finally, for $S=10\sqrt{B}$ the significance reported by the AIC-based method including all odd modes is only about 10\% larger than that reported by the wide model, while the model-averaging approach reports a significance about halfway between the true and wide models.\footnote{I have not found any direct discussion about $p$-values in the frequentist model-averaging literature; therefore, here the $p$-value is computed under a Gaussian assumption using $\hat{S}_{\rm avg}$ and $\sigma_{\rm avg}$. While this approach works for the problems studied, one could certainly question its validity---especially for small $p$-values.}
\end{itemize}
It is also interesting to examine how the 68.3\% CI lengths are distributed in the ensembles.
Figure~\ref{fig:sigma_dist} shows this for the toy-model example for $\langle {\rm bkg} \rangle = 10^6$ and $S=10\sqrt{B}$.
Since no model-selection criteria are applied for the true and wide models, they report approximately the same CI length for each data sample.
The stepwise-in approach is bimodal here, since it chooses to truncate the background PDF at either $\ell = 3$ or 4 in each data set, and the standard AIC approach is similar.
Both the AIC-based method including all odd modes and the model-averaging approach have much smoother distributions, especially when the model-selection uncertainty is included for the former.

\begin{table}
  \begin{center}
    \begin{tabular}{cl}
      legend label & background PDF used in the fit\\
      \hline
      true & the PDF used to generate the data with no free shape parameters \\
      wide & all even and odd modes up to the maximum considered ($\ell\leq10$ here)\\
      step & model selected by the stepwise-in approach \\
      aic & model selected by the standard AIC approach  \\
      aic-o & all odd modes in the wide model plus even modes selected by AIC \\
      avg & frequentist averaging over models (all models contain all odd modes) \\
      \hline
    \end{tabular}
  \end{center}
  \caption{Labels used to denote the various methods in figure legends. The uncertainty due to model selection is (by construction) not required for the true and wide models and not included for the stepwise-in approach (it is not clear how this would be done for the stepwise-in method). The AIC-based results are shown both with and without accounting for model-selection uncertainty, while this uncertainty is always included for frequentist model averaging. {\em N.b.}, the solid-purple-line aic results are the $c=2$ method of Ref.~\cite{ref:iic} (the same method but with $c=1$ was used by CMS in $H\to\gamma\gamma$~\cite{ref:hgg}).}
  \label{tab:labels}
\end{table}

Based on these results, it is clear that even in a simple problem like this---one that has no peaking-background structures---the stepwise-in approach is prone to bias, undercoverage, and overestimation of significance.
Conversely, the wide model produces unbiased $\hat{S}$ values, CIs, and $p$-values.
The only drawback to using the wide model is that the lengths of its CIs are about twice the optimal length.
The potential bias in the AIC-based approach is greatly reduced by only considering background models containing all odd modes included in the wide model when performing the model selection.
Furthermore, to obtain proper coverage and $p$-values from this AIC-based method requires accounting for model-selection uncertainty.
Frequentist model averaging also produces valid results, though its CIs are typically 10--20\% longer than those of AIC (after including model-selection uncertainty).

In this example, the best performance
%for determining $\hat{S}$ and its CI
is achieved by the AIC-based approach to arbitrating the inclusion of even modes in the background model, where model-selection uncertainty is accounted for using the discrete-nuisance-parameter profiling technique.
%This method can also be used to produce valid local $p$-values.
For highly significant signals, however, the model uncertainty drives the $p$-values to be close to those obtained by simply using the wide model.
This is expected since if one does not expect all components of the wide model to be relevant at the $5\sigma$ level then why include them?
An alternative approach for assigning local $p$-values is to take those obtained using the wide model with no additional model selection and, therefore, no additional model-selection uncertainty, which provides a mildly conservative significance estimate relative to the model-selection-based one.
Either way, following the procedures outlined above should yield valid results---provided that a proper wide model is chosen; {\em i.e.}\ provided that the wide model has sufficient complexity to accommodate any structure that could be manifest in the data in the absence of a signal, or equivalently, that the assumptions made about the background when constructing the wide model are valid.

\subsection{Alternative Methods}
\label{sec:other}

\begin{figure}[t]
  \centering
  \includegraphics[width=0.49\textwidth]{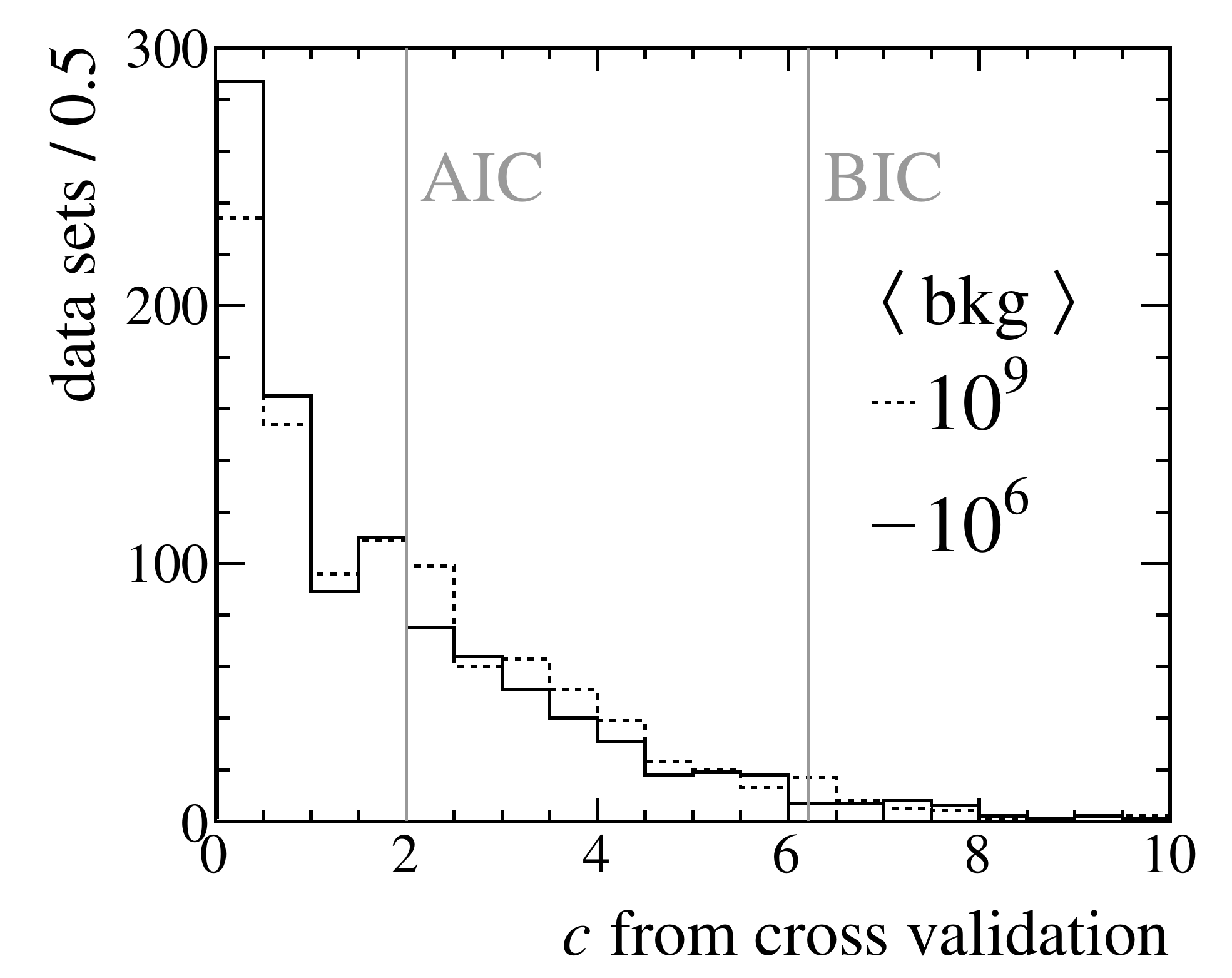}
\caption{Data-driven determination of the penalty parameter $c$ using 10-fold cross validation for $S=0$.
The mean values are 1.9 and 1.8 for the $10^6$ and $10^9$ samples, respectively.
As expected, the $c$ values determined using cross validation are, on average, close to the AIC value of $c=2$, though determining $c$ this way introduces sizable noise.
}
  \label{fig:cv}
\end{figure}

In the previous subsection, only the cases of $c=0$ and $c=2$ were considered, {\em i.e.}\ no likelihood penalty and the AIC penalty.
One alternative approach is to instead choose $c$ for each data set using cross validation.
Figure~\ref{fig:cv} shows the distribution of optimal $c$ values chosen using 10-fold cross validation for the toy-model problem.
In 10-fold cross validation, the data sample is divided into 10 subsamples, where each background model is determined by fitting to 9 subsamples, and then---with all parameters fixed---the mean square error (MSE) is computed for the remaining subsample, {\em i.e.}\ the one not used in the fit.
This process is repeated 10 times, and the total MSE for each model is computed as the mean of the 10 MSE values obtained.
The value of $c$ is chosen as the one that selects the model with the smallest MSE.\footnote{The model selection could be done without determining $c$ by choosing the model with the smallest MSE; however, $c$ must be known to incorporate the model-selection uncertainty using the method employed above.  }
The $c$ values determined using cross validation are, on average, close to the AIC value of $c=2$, though determining $c$ in this way introduces sizable noise.
Indeed, it is well known that asymptotically AIC is equivalent to leave-one-out cross validation~\cite{ref:aicloo}.
While this approach is feasible, determining $c$ using  cross validation substantially increases the CPU required with no obvious benefits over using the AIC-based method from the previous subsection.

Another alternative approach is to {\em regularize} the fitting process, which involves modifying the likelihood according to
\begin{equation}
-2\log{\mathcal{L}} + \lambda \sum |\alpha_i|^{\beta},
\end{equation}
where $\lambda$ is a constant that controls the degree of regularization, $\alpha_i$ are the model parameters, $\beta=2$ corresponds to Ridge regression, and $\beta=1$ is the LASSO~\cite{ref:lasso}.
Both methods reduce overfitting by penalizing large $|\alpha_i|$ values, since overfitting often involves some parameters being driven to large positive values and others to large negative ones (induced by the fact that the overall normalization must match the observed data counts).
The value of $\lambda$ is determined using 10-fold cross validation for each data sample.

\begin{figure}[t]
  \centering
  \includegraphics[width=0.49\textwidth]{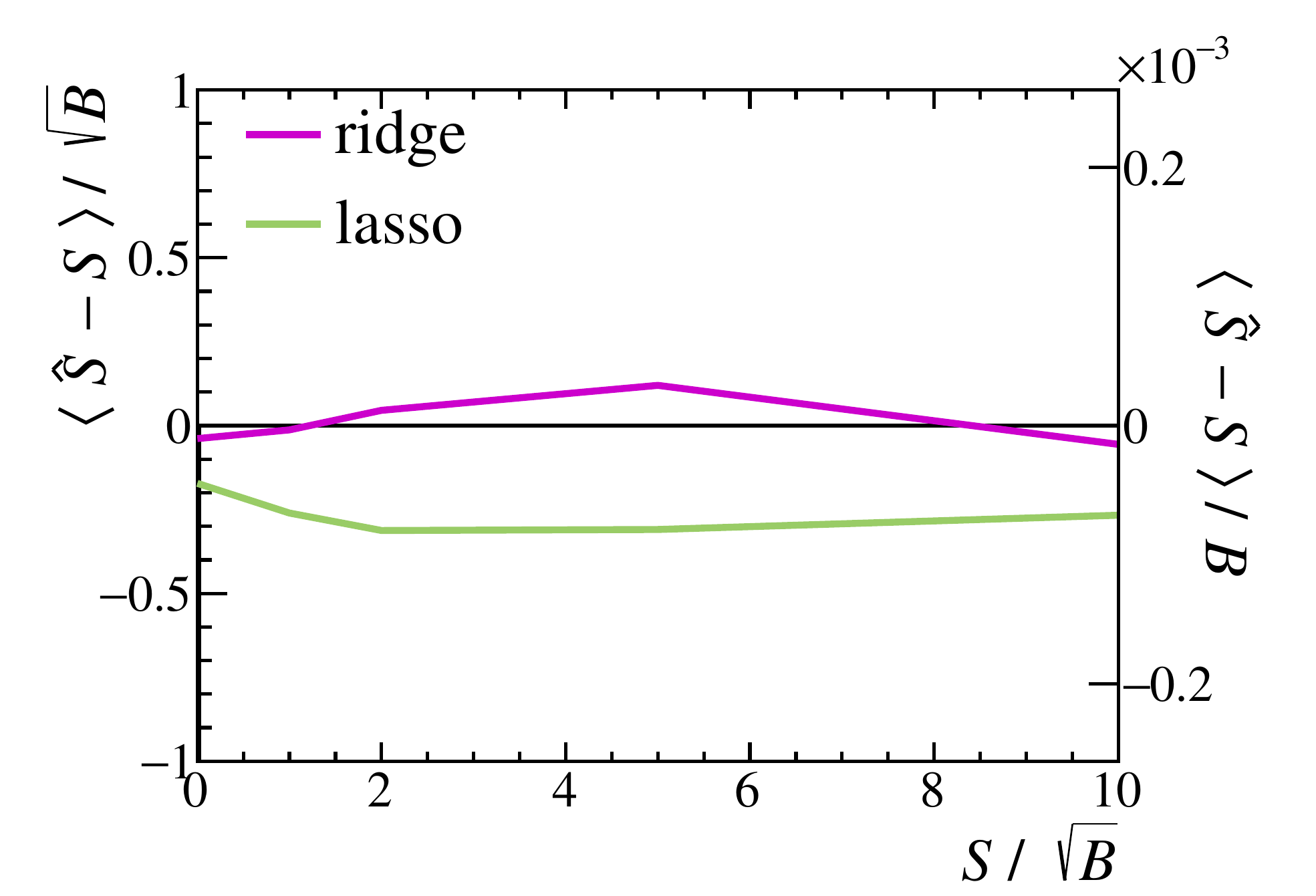}
  \includegraphics[width=0.49\textwidth]{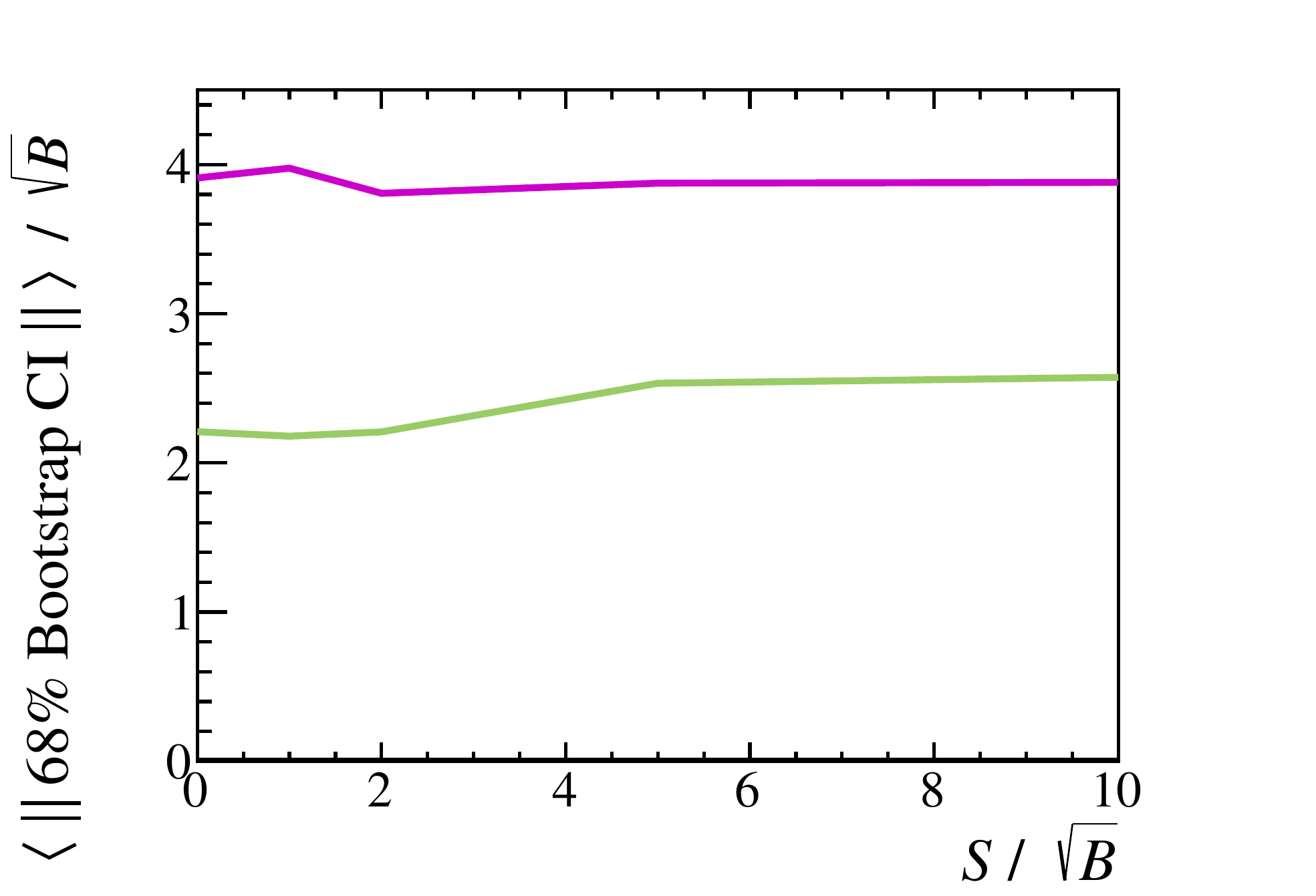}
  \includegraphics[width=0.49\textwidth]{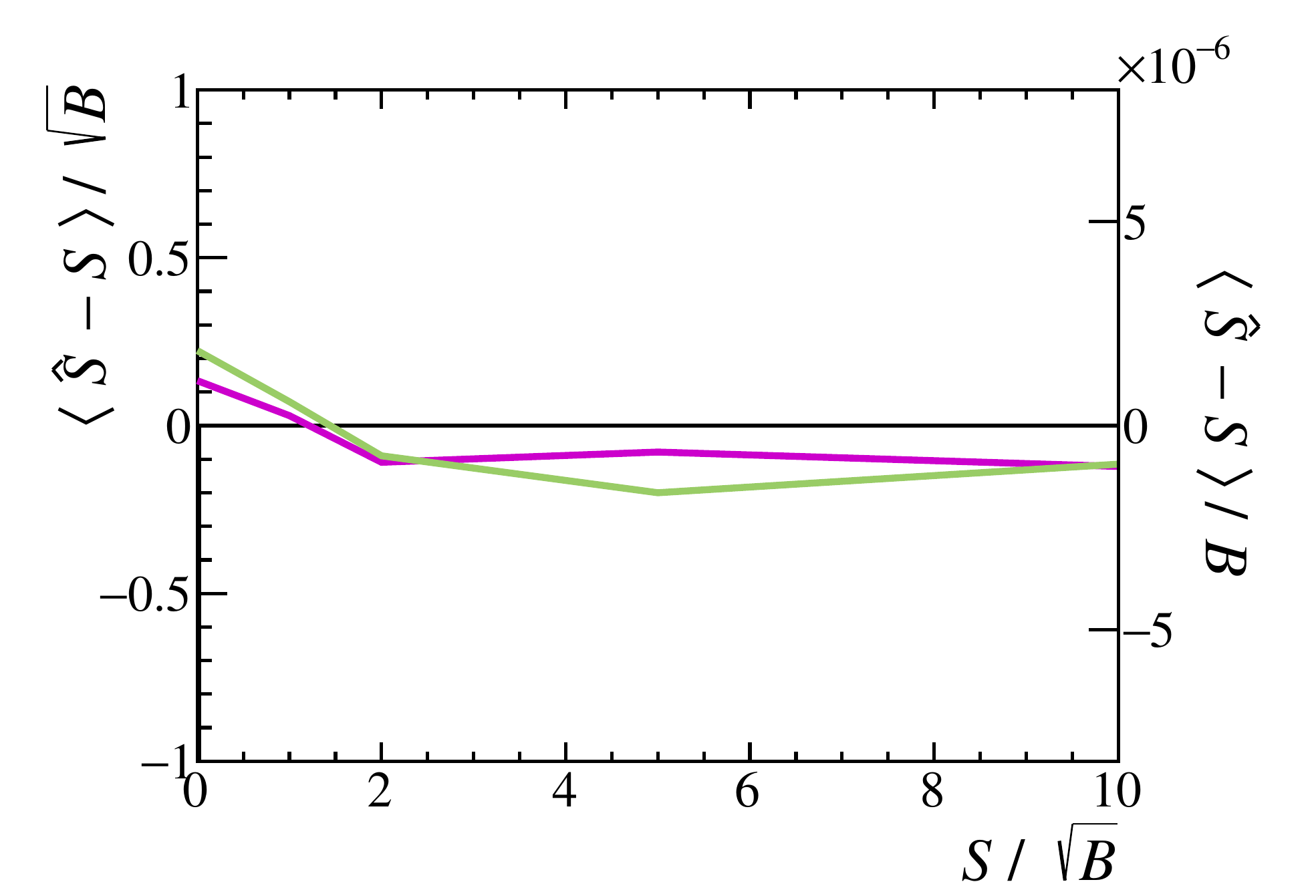}
  \includegraphics[width=0.49\textwidth]{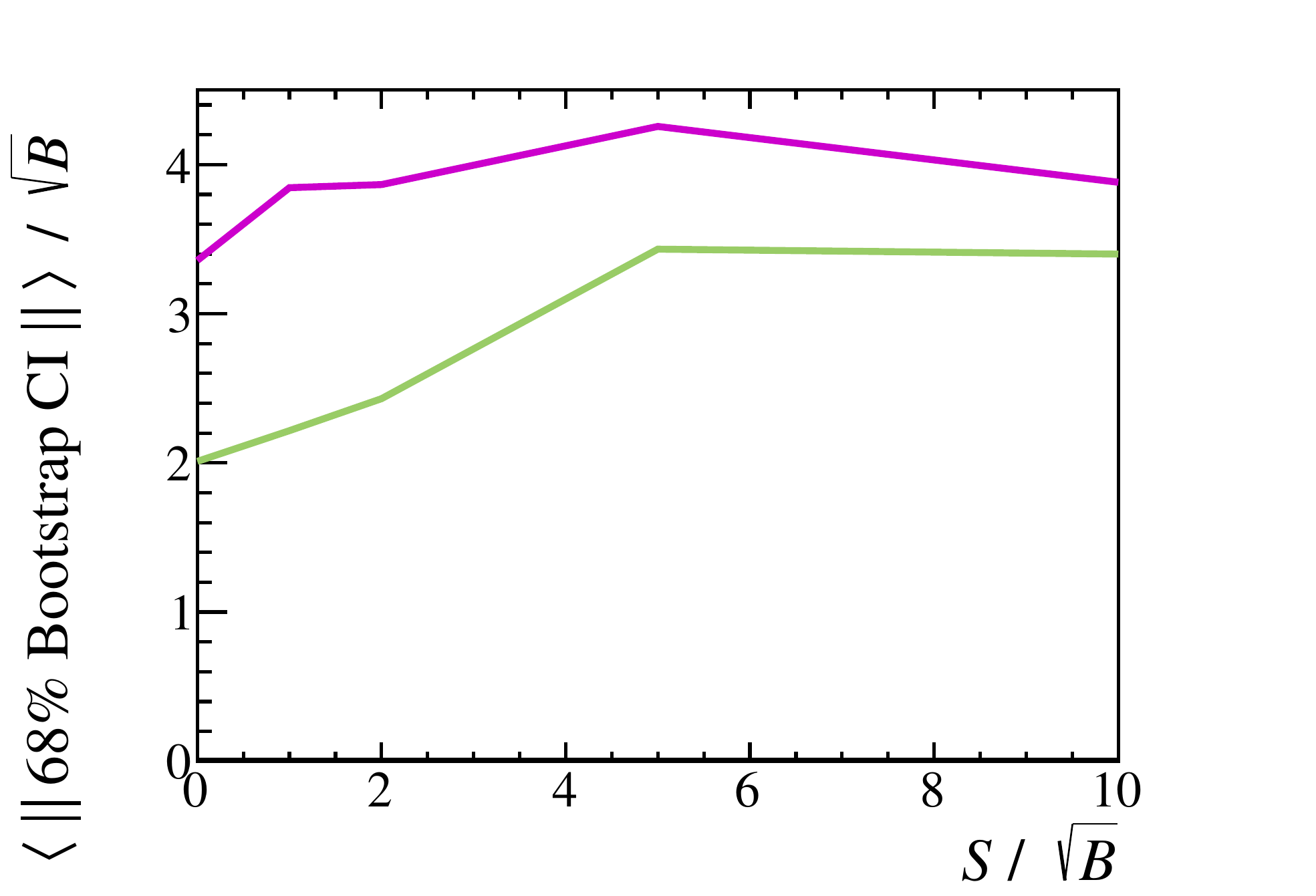}
\caption{Results using Ridge and LASSO regularization for the toy-model  example for (top) $\langle {\rm bkg} \rangle = 10^6$ and (bottom) $10^9$.
(left) The mean bias in signal estimator $\hat{S}$ relative to (left axis) $\sqrt{B}$ and (right axis) $B$.
(right) The mean full length of the 68.3\% CI relative to $\sqrt{B}$ as determined using the bootstrap resampling approach.
For both the Ridge and LASSO, the $\lambda$ parameter is determined using 10-fold cross validation.
}
  \label{fig:lr}
\end{figure}

Figure~\ref{fig:lr} shows the bias and 68.3\% CI lengths obtained by applying the Ridge and LASSO to the toy-model problem.
For both methods, the CIs are obtained using the bootstrap~\cite{ref:boot} due to the presence of the regularization term; however, in this example, the likelihood-based CIs are consistent with the bootstrap-based ones.
The Ridge penalty decreases quickly as $\alpha_i \to 0$; therefore, the Ridge does not perform model selection, {\em i.e.}\ it does not zero out terms in the model.
Because of this, one can see that the Ridge results are similar to those of the wide model, with the Ridge-based CIs $\approx 5\%$ shorter than those obtained using the wide model.
The LASSO does perform model selection (it zeroes out model components), and produces CIs with similar lengths to those  obtained using the AIC-based model-selection approach in the previous subsection; however, in this example, the LASSO results have a small bias in $\hat{S}$.
The advantage of the LASSO is that it does not require running a separate fit for all possible models under consideration (though it does require determining $\lambda$ from cross validation).
Since in this case only 64 models are considered, it is possible to determine the $\Lambda$ value (with $c=2$) for all models and perform the model selection.
%\footnote{Recall that the wide model contains all Legendre modes with $\ell \leq 10$. The largest model space considered during model selection in this study includes all odd modes up to 9 in all models; therefore, considering all combinations of even modes being on or off is $2^6 = 64$ models. Note that the full model space for $\ell \leq 10$ contains $2^{11}=2048$ models, which could not be fully explored using AIC making the LASSO an excellent candidate for such a problem.}
For applications where the model space is large (see, {\em e.g.}, Ref.~\cite{ref:amp}), the LASSO is likely the best feasible option.
For a bump hunt, however, regularization does not perform as well as the AIC-based approach, where all odd modes are included in all models and model-selection uncertainty is accounted for.

Finally, an alternative to the frequentist model averaging approach used above is Bayesian model averaging, where the AIC values in the model weights are replaced by the corresponding BIC values ({\em i.e.}\ $c=2$ is replaced by $c = \log{[n({\rm bins})]}$).
This type of model averaging is natural in the Bayesian framework, since the BIC values provide approximations of the Bayes factors from which the model weights discussed above are easily obtained by assuming a uniform prior over models.
Indeed, the form of the model weights in the frequentist model averaging method studied in this article were inspired by their Bayesian counterparts.
However, since our field is focused on frequentist CIs and $p$-values, I will not present any detailed study of Bayesian methods here.

\subsection{Peaking Background}

\begin{figure}[t]
  \centering
  \includegraphics[width=0.49\textwidth]{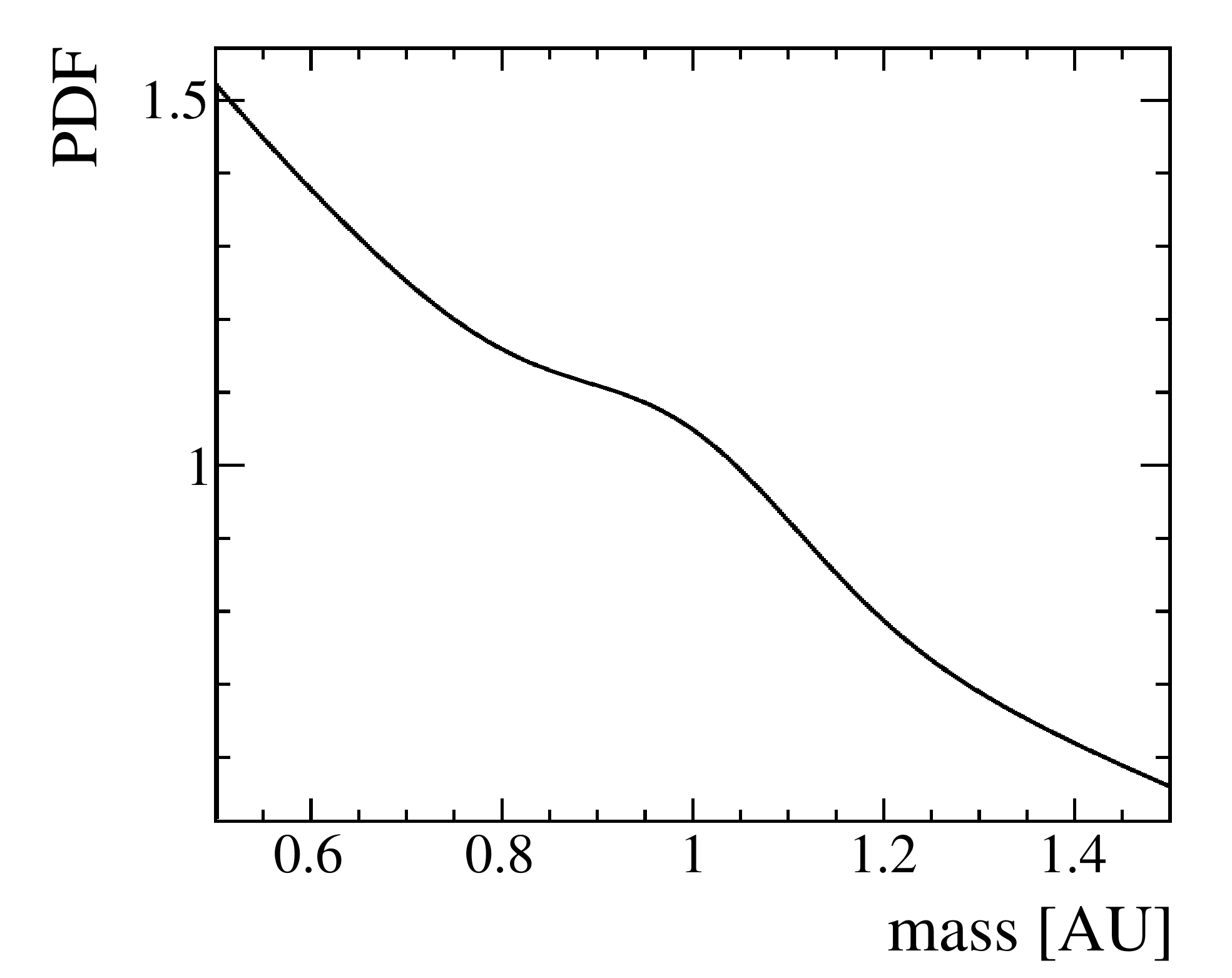}
  \caption{Peaking background-only PDF with $\langle {\rm bkg} \rangle = 10^6$, where the peaking background structure has a width of $3\sigma$ and a yield of $100\sqrt{B}$ ($S=0$ in this example).}
  \label{fig:sig}
\end{figure}

The most difficult background to confront in a bump hunt is an unexpected peaking structure.
Figure~\ref{fig:sig} shows the toy-model background modified to include a Gaussian structure with a width 3 times larger than that of the signal, and an expected yield of $100\sqrt{B}$.
Such a structure is potentially large enough to result in rejecting the NULL hypothesis  even when using the wide model chosen above (this situation violates the assumptions made when choosing $\ell_{\rm max}$ for the wide model).
That said, it is interesting to investigate what happens if no dedicated component is added to the background-only PDF for this peaking structure.
{\em N.b.}, the case considered here is where the peaking background is centered on the signal mass. For a Gaussian background, this is where the largest bias is expected. As the peaking structure is moved away from the signal mass, the bias decreases before flipping sign and then gradually becomes smaller as the peaking structure is moved farther from the signal mass.

Table~\ref{tab:adv} gives the results of applying each method to this challenging problem.
The stepwise-in approach totally fails in this case because the peaking-background structure is an even function; therefore, the stepwise-in procedure terminates when adding $\ell=5$ does not provide significant reduction in $\Lambda$, even though adding $\ell=6$, 8, and 10 would each significantly reduce $\Lambda$.
Interestingly, the AIC-based and model-averaging approaches effectively select the wide model, which is the best possible choice given the model space provided for consideration.
While $\hat{S}$ is biased and there is about 5\% undercoverage, the model-selection and model-uncertainty procedures have performed as well as possible given the wide model provided.
Indeed, in this situation, the analyst has failed to provide an adequate wide model (a peaking-background component is clearly required), though the consequences are rather mild.

\begin{table}[t]
  \begin{center}
    \begin{tabular}{c|ccccc}
      & wide & step & aic & aic-o & avg \\
      \hline
      $\langle \hat{S}-S \rangle / \sqrt{B}$ & 1.1 & 22.3 & 1.8 & 1.2 & 1.3 \\
      $\langle \| \Delta\Lambda < 1 \| \rangle$ & 4.2 & 2.3 & 4.1 & 4.1 & 4.2 \\
      coverage \@ 95\% CL (\%) & 92 & 0 & 85 & 90 & 91 \\
      $>3\sigma$ (\%) & 0.7 & 100 & 1.5 & 0.5 & 0.5 \\
      \hline
    \end{tabular}
  \end{center}
  \caption{Results obtained for the peaking background shown in Fig.~\ref{fig:sig}.}
  \label{tab:adv}
\end{table}

\section{Real-World Examples}
\label{sec:real}

Having fully explored the toy-model example of Sec.~\ref{sec:method}, the same study will now be repeated on two (simulated) real-world examples. The first is a search for dark photon decays to electron-positron pairs at the APEX experiment, while the second is a search for the decay of a Higgs boson into muon-antimuon.
The APEX study includes regions where the background is steeply increasing, steeply decreasing, and nearly uniform.
The \htomm study is complicated by a huge $Z\to\mu^+\mu^-$ contribution that extends into the fit region, and by a non-Gaussian signal PDF that arises due to bremsstrahlung.
In both studies, it is again assumed that small peaking-background structures with widths $\gtrsim 3\sigma$ could appear in the spectra and $\ell_{\rm max} = 10$ is chosen.
While the real-world backgrounds for these specific problems may not be so difficult, allowing for such narrow peaking backgrounds provides the opportunity for more stringent testing of the bump-hunting strategies.
{\em N.b.}\ the use of examples with parametric backgrounds facilitates performing these studies without using vast CPU resources.
The same approach can be applied to situations where the background must be obtained from simulation, provided that the resources exist to generate the required background ensembles.

\subsection{\atoee}
\label{sec:aee}

\begin{figure}[t]
  \centering
  \includegraphics[width=0.7\textwidth]{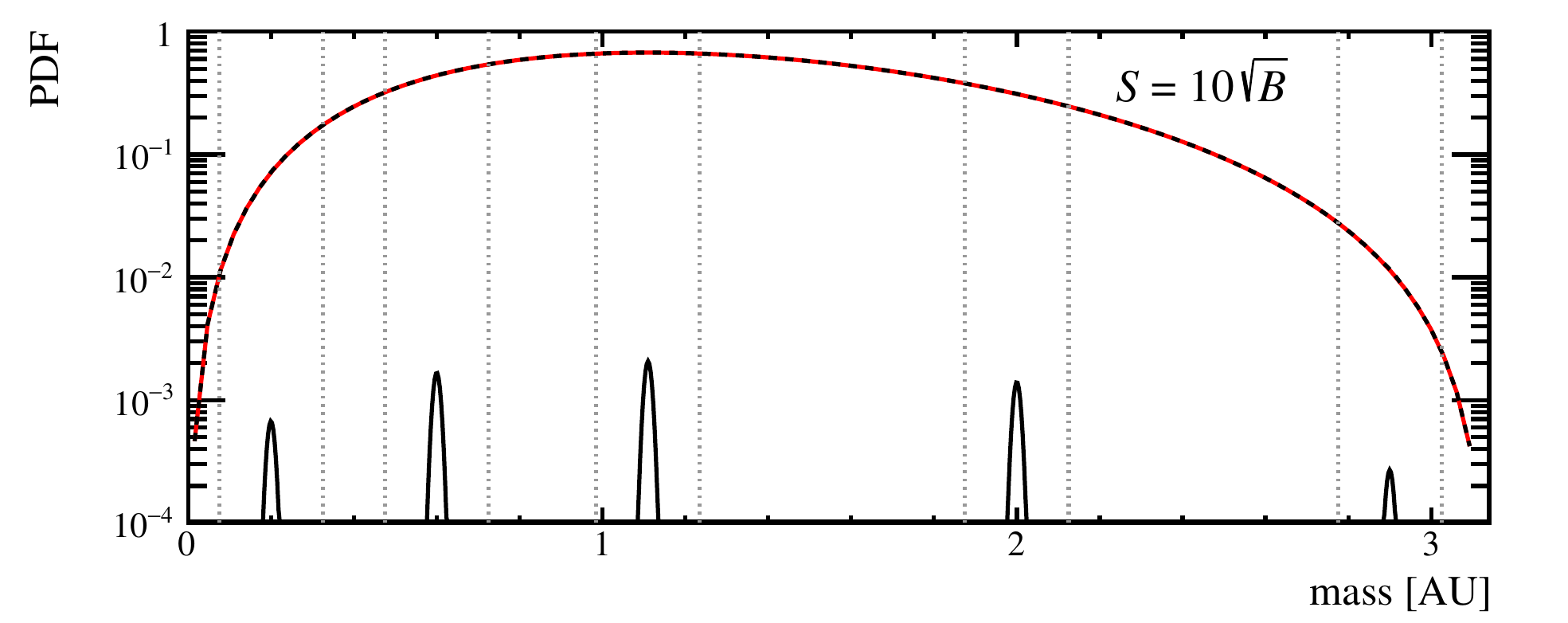}\\
  \includegraphics[width=0.7\textwidth]{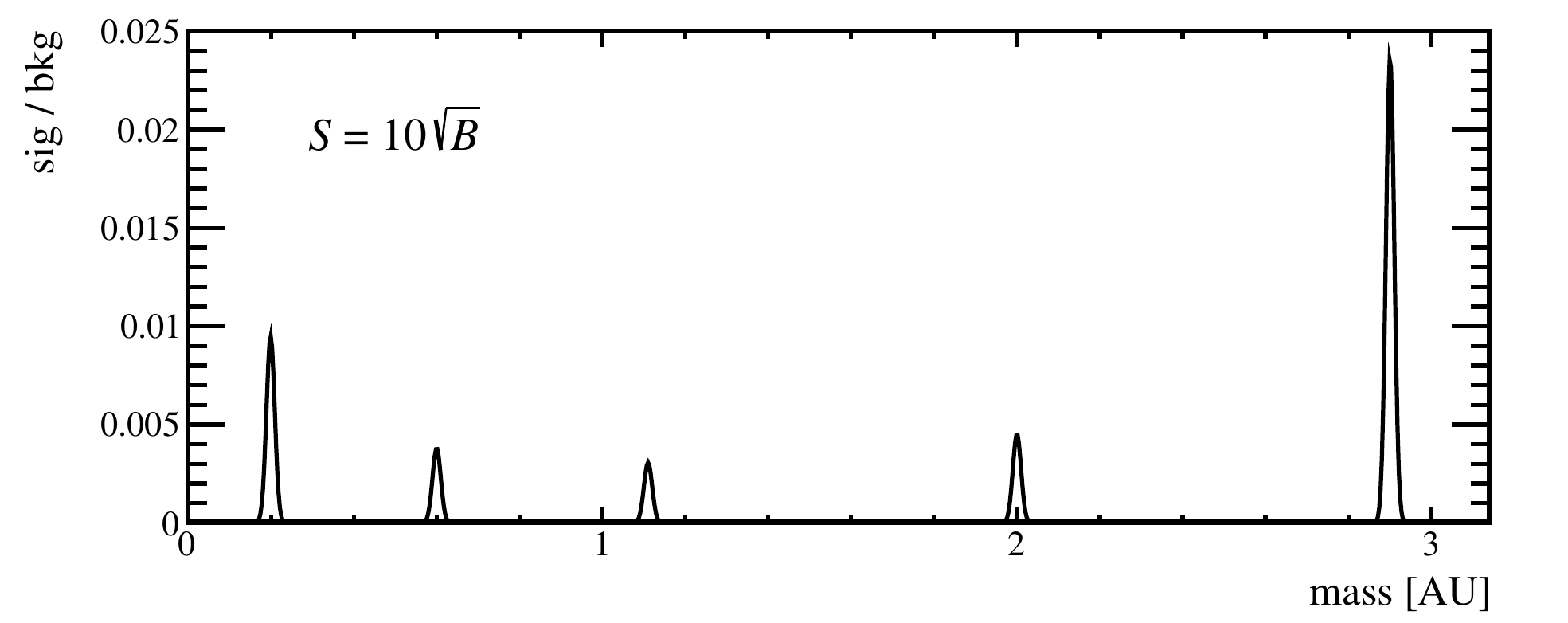}\\
  \includegraphics[width=0.7\textwidth]{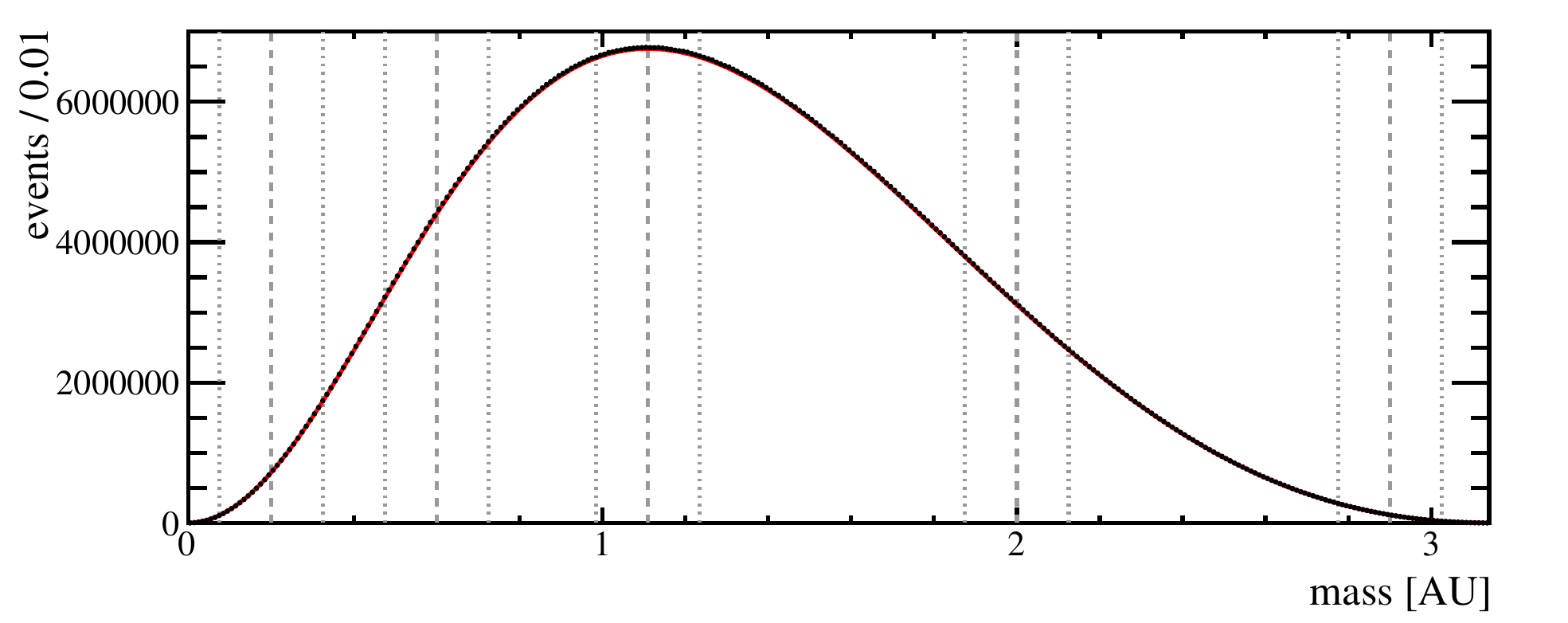}
\caption{APEX model for $S=10\sqrt{B}$ and $\langle {\rm bkg} \rangle = 10^9$:
(top) the PDF, showing the (red) background, (solid black) Gaussian signals, and (dashed black) total (the edges of each $25\sigma$ mass window are denoted by the vertical dotted lines);
(middle) the signal-to-background ratio at each signal mass for $S=10\sqrt{B}$ (the largest $S$ values considered in this study);
and
(bottom)
an example data set sampled from the PDF shown in the top panel (the signals are not visible, their peak locations are denoted by the vertical dashed lines).
}
  \label{fig:apex_pdf}
\end{figure}

\begin{figure}[t]
  \centering
  \includegraphics[width=0.49\textwidth]{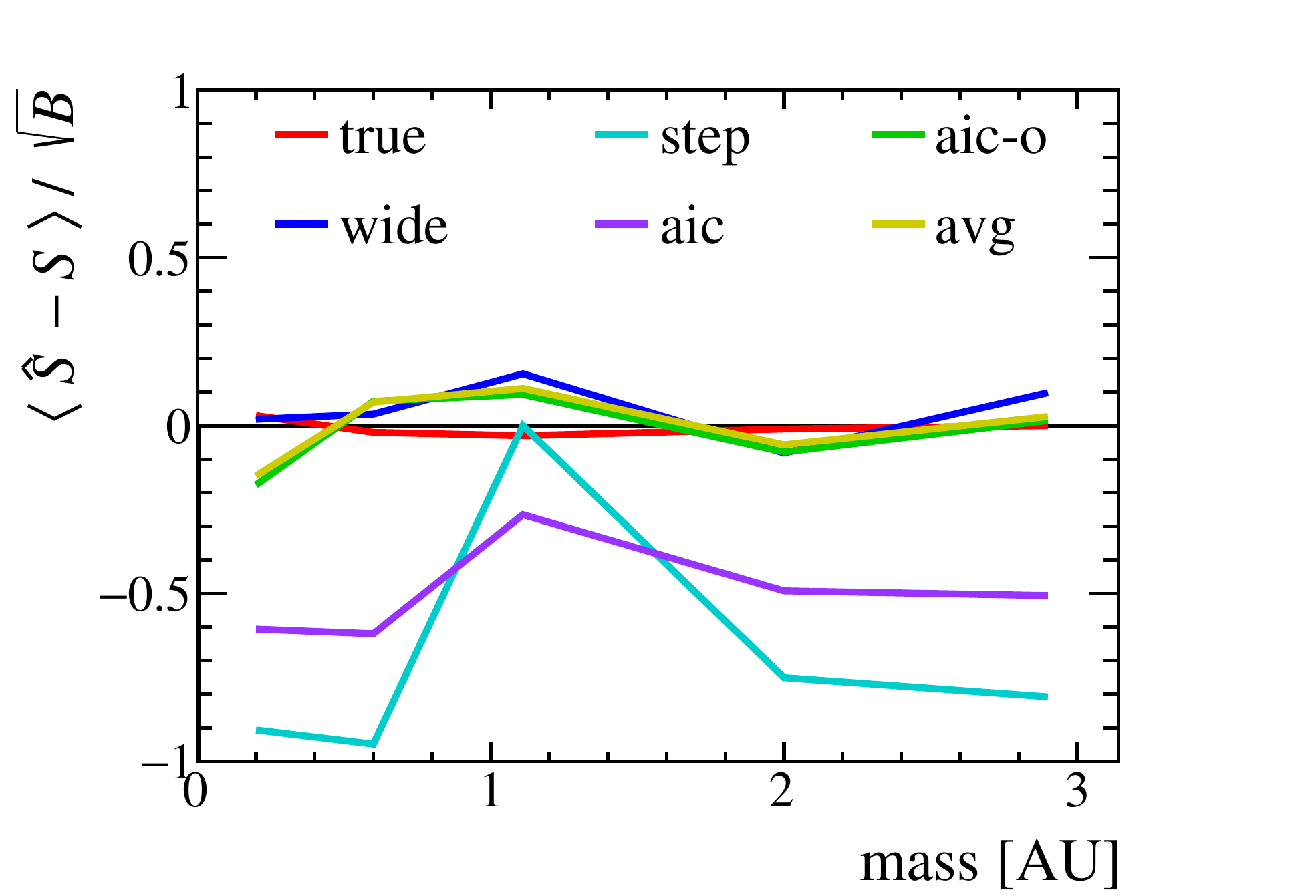}
  \includegraphics[width=0.49\textwidth]{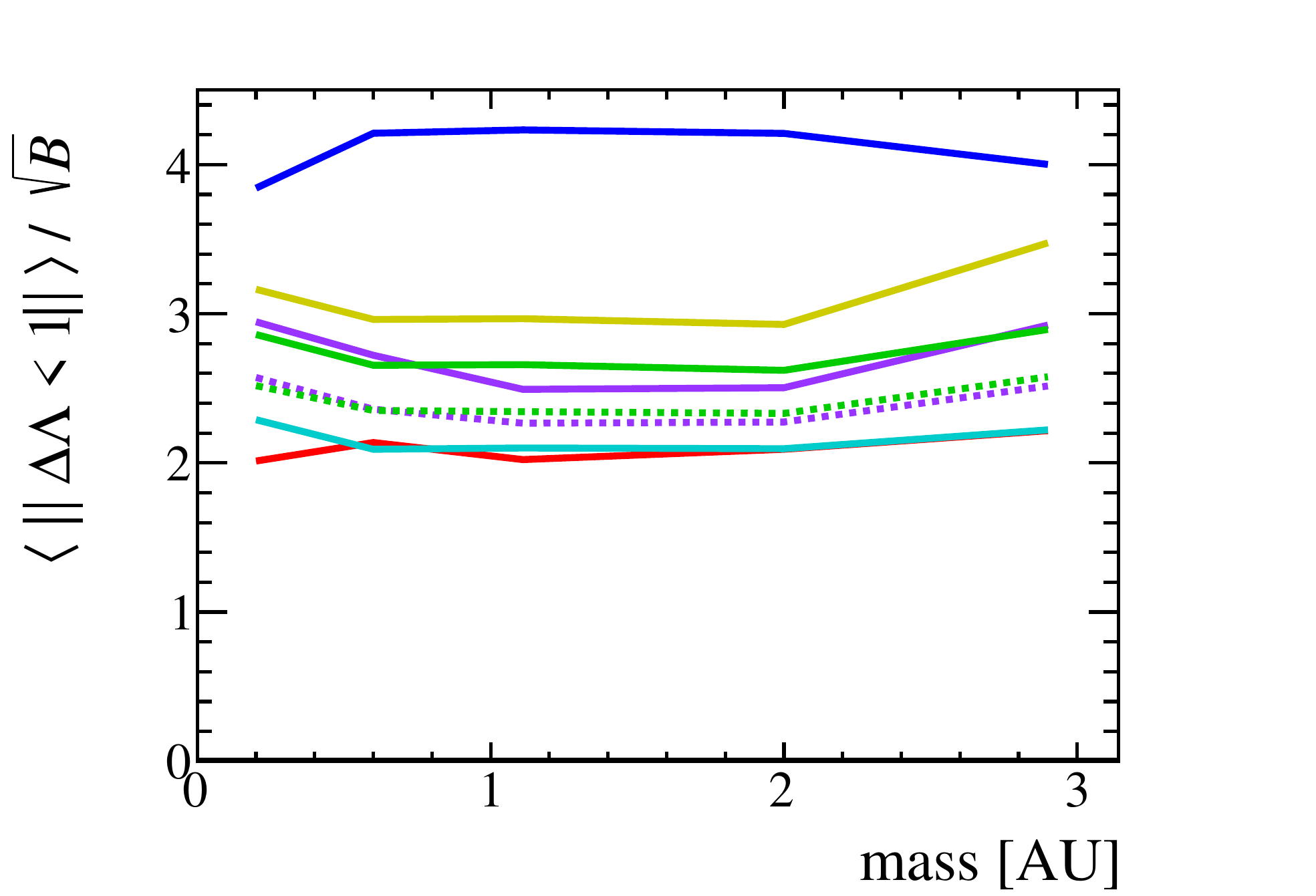}
  \includegraphics[width=0.49\textwidth]{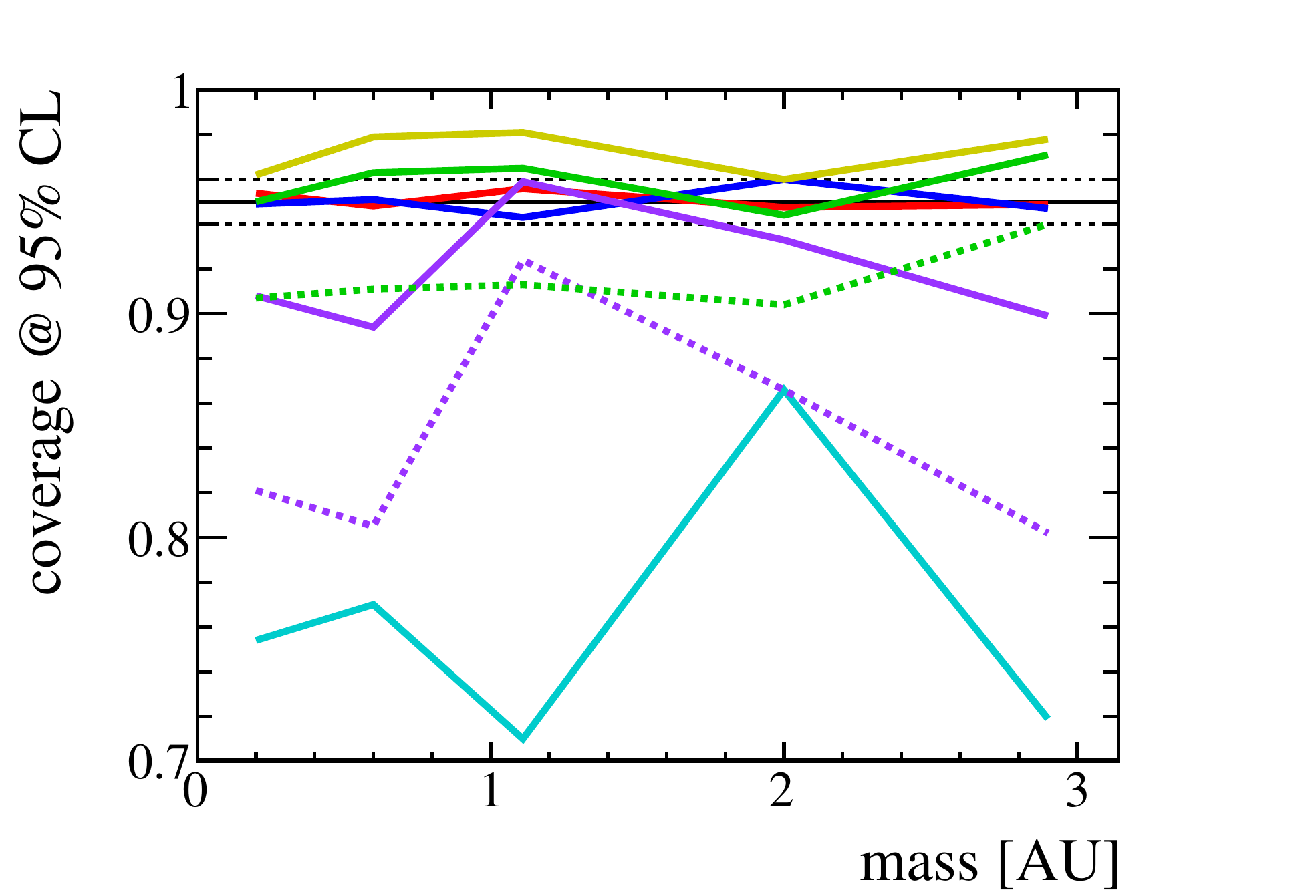}
  \caption{Results for the APEX example problem with $S=10\sqrt{B}$ versus test mass value.
(top left) The mean bias in signal estimator $\hat{S}$ relative to $\sqrt{B}$.
(top right) The mean full length of the 68.3\% CI relative to $\sqrt{B}$ as reported by the profile likelihood, where the dashed lines show the aic and aic-o results without accounting for model-selection uncertainty.
(bottom) Coverage reported at 95\% CL by each method, where the solid black line shows the expected 95\% and the dashed black lines show the approximate uncertainty in all results due to ensemble size.}
  \label{fig:apex_results}
\end{figure}

Figure~\ref{fig:apex_pdf} shows the PDF used for the APEX \atoee search, where the background function is $\sin^2(x)/(1+x^2)$ which gives a similar background shape as that of APEX,
%\footnote{I thank Natalia Toro for suggesting this parametrization of the APEX background.}
and the expected total background yield is chosen to be $10^9$ (this is about 10 times larger than the expected APEX data sample at each beam energy, which is chosen to provide a more stringent test)~\cite{ref:apex}.
Five test masses are selected, and signal strengths of $0 \leq S \leq 10\sqrt{B}$ are considered, where for each mass $B$ is again defined as the expected background yield in a $\pm2\sigma$ region centered on the test mass value (see Fig.~\ref{fig:apex_pdf}); the signal PDFs are each Gaussian.
The same studies performed in the previous section on the toy-model problem are repeated for each test-mass value considered in the APEX problem.
The full results are presented in the appendix, while the mass-dependence for a single $S$ value is shown in Fig.~\ref{fig:apex_results}.

Overall, the results obtained in the APEX example are consistent with those from the toy-model example, except for the ${m=1.1}$ test mass value.
At this mass, the APEX background is close to uniform, which results in less bias in the stepwise-in and standard-AIC approaches (though the stepwise-in coverage is still not valid).
An interesting feature is that the size of the CIs increases at large and small masses where the background PDF undergoes its most rapid variations.
In this example, no dedicated mass-dependent long-range components are added to the wide model; therefore, this rapid background variation must be accounted for entirely by the Legendre modes, which requires using most of the wide model.
This works---$\hat{S}$ is not biased, the coverage is valid, {\em etc.}---but is not optimal.
The sensitivity would be improved by adding some problem-specific PDF components that roughly capture the long-distance background behavior.
The Legendre modes would then be left free to accommodate any local deficiencies in these terms, including any unexpected peaking-background-like structures.
The example presented next provides an excellent opportunity to examine this topic in greater detail.

\subsection{$H\to\mu^+\mu^-$}
\label{sec:hmm}

\begin{figure}[t]
  \centering
  \includegraphics[width=0.329\textwidth]{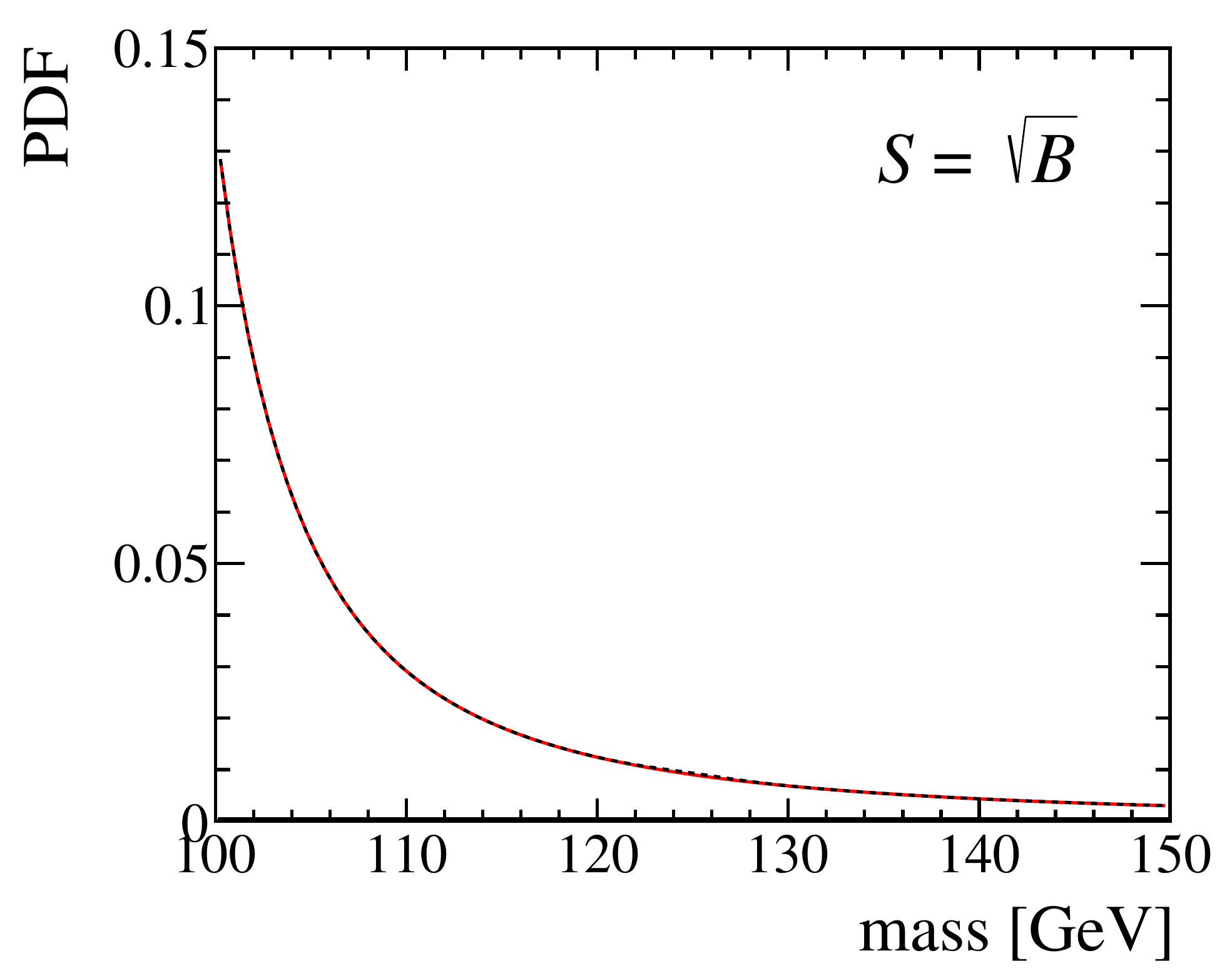}
  \includegraphics[width=0.329\textwidth]{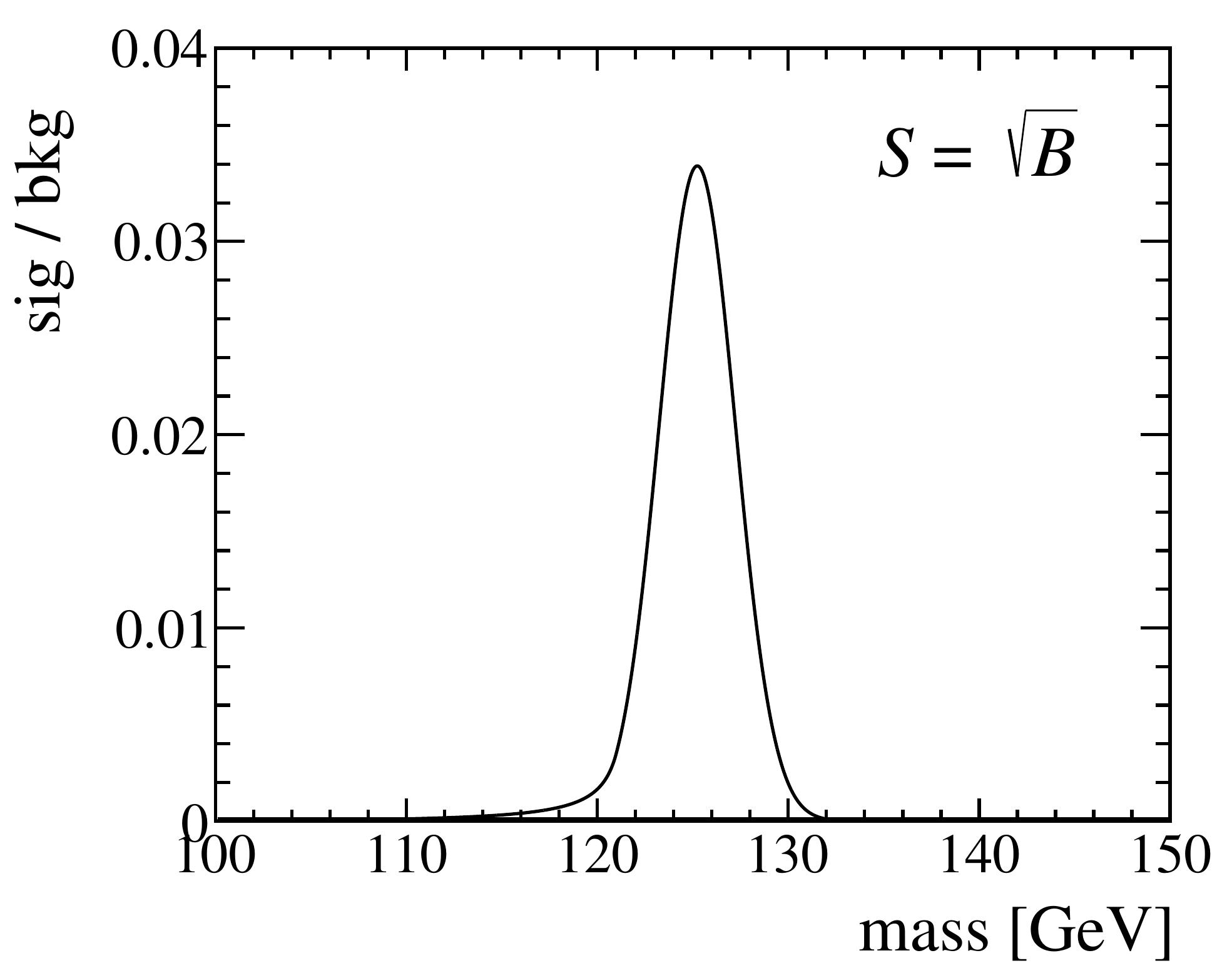}
  \includegraphics[width=0.329\textwidth]{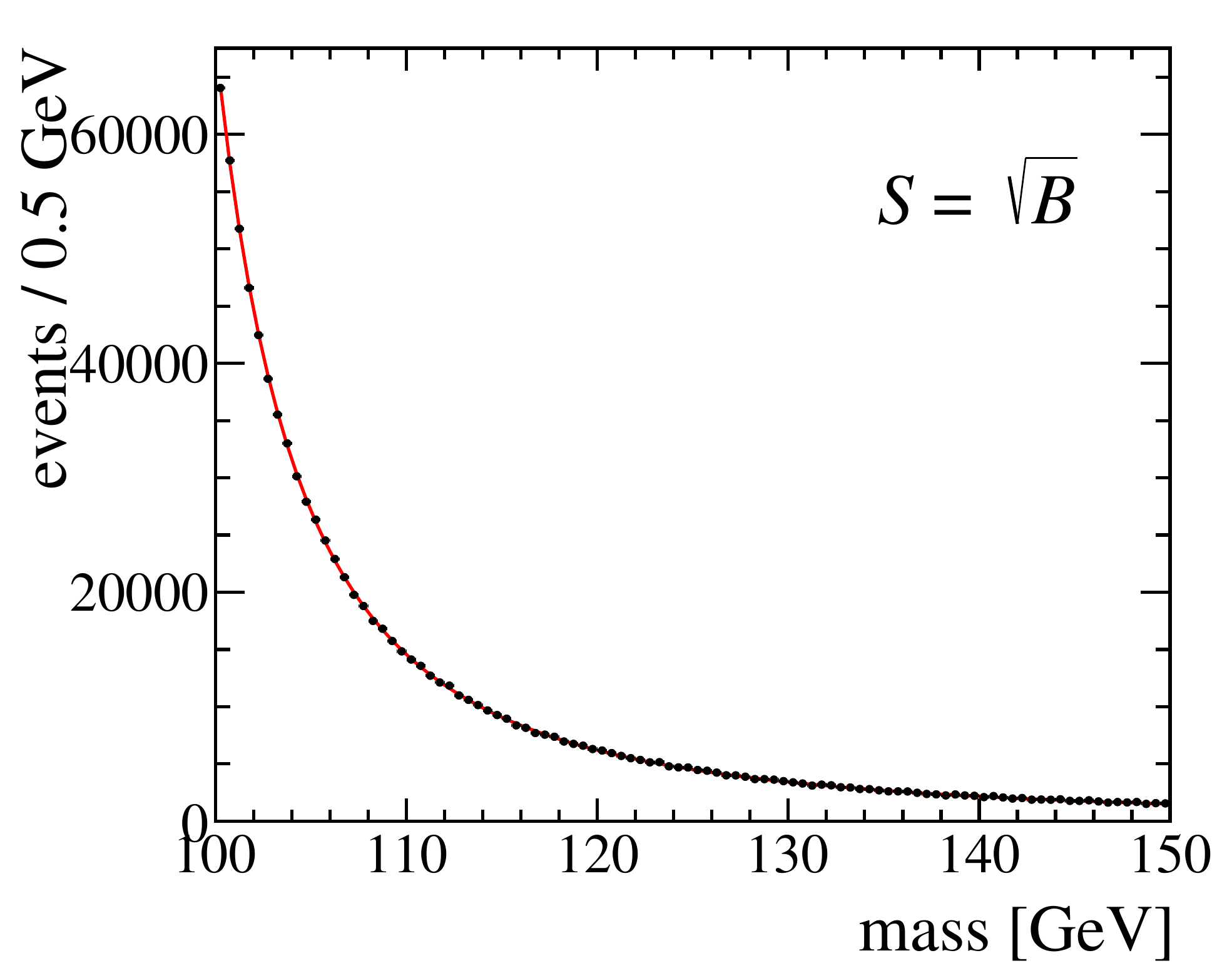}
\caption{Higgs-decay model for $S=\sqrt{B}$, where $B$ is the background yield in a $\pm2\sigma$ window around 125~GeV (this is approximately the Standard Model value of $S$ for 50/fb at 13\,TeV).
The left panel shows the PDFs, including the (red) $Z$ boson background, (solid black) Crystal Ball signal, and (dashed black) total.
The middle panel presents the signal-to-background ratio.
The right panel displays an example data set sampled from the PDF, with the background PDF shown (the signal contribution is too small to be visible).
The signal has a mass of 125~GeV and a Gaussian-core-width of 2~GeV, making the full mass range considered a $25\sigma$ window.
}
  \label{fig:hmm_pdf}
\end{figure}

Figure~\ref{fig:hmm_pdf} shows the $H\to\mu^+\mu^-$ PDF, where the background is taken to be a Breit-Wigner distribution for the $Z$ boson convolved with a Gaussian resolution function and the signal PDF is a Crystal Ball function~\cite{ref:cb}.
Signal strengths are again studied in the range $0 \leq S \leq 10\sqrt{B}$, where $S \approx \sqrt{B}$ is the SM value using about 50/fb of LHC data collected at 13~TeV.
This example is similar to the toy-model one, with two important differences: the signal PDF is not a pure even function;
and the tail of the $Z$ boson provides a much more difficult background, one that varies by a factor of $\approx 50$ over the $25\sigma$ mass window.
Fitting the $Z$ background using only the Legendre-based wide model with $\ell_{\rm max} = 10$ results in the full wide model being selected in every data sample for all model-selection methods---but still provides a poor description of the data as determined using the wide-model $\chi^2$.
A fit that uses the full wide model and includes a signal component should not be rejected at a high confidence level if the wide model is adequate.
Situations where the wide model is inadequate can often be identified using this $\chi^2$, which does not spoil the blinding provided that the value of $\hat{S}$ is not inspected.
A potential strategy is to first perform a fit at every test-mass value using the full wide model including a signal component whose peak value is free to vary to ensure that all such models are capable of adequately describing the data.

\begin{figure}[t]
  \centering
  \includegraphics[width=0.49\textwidth]{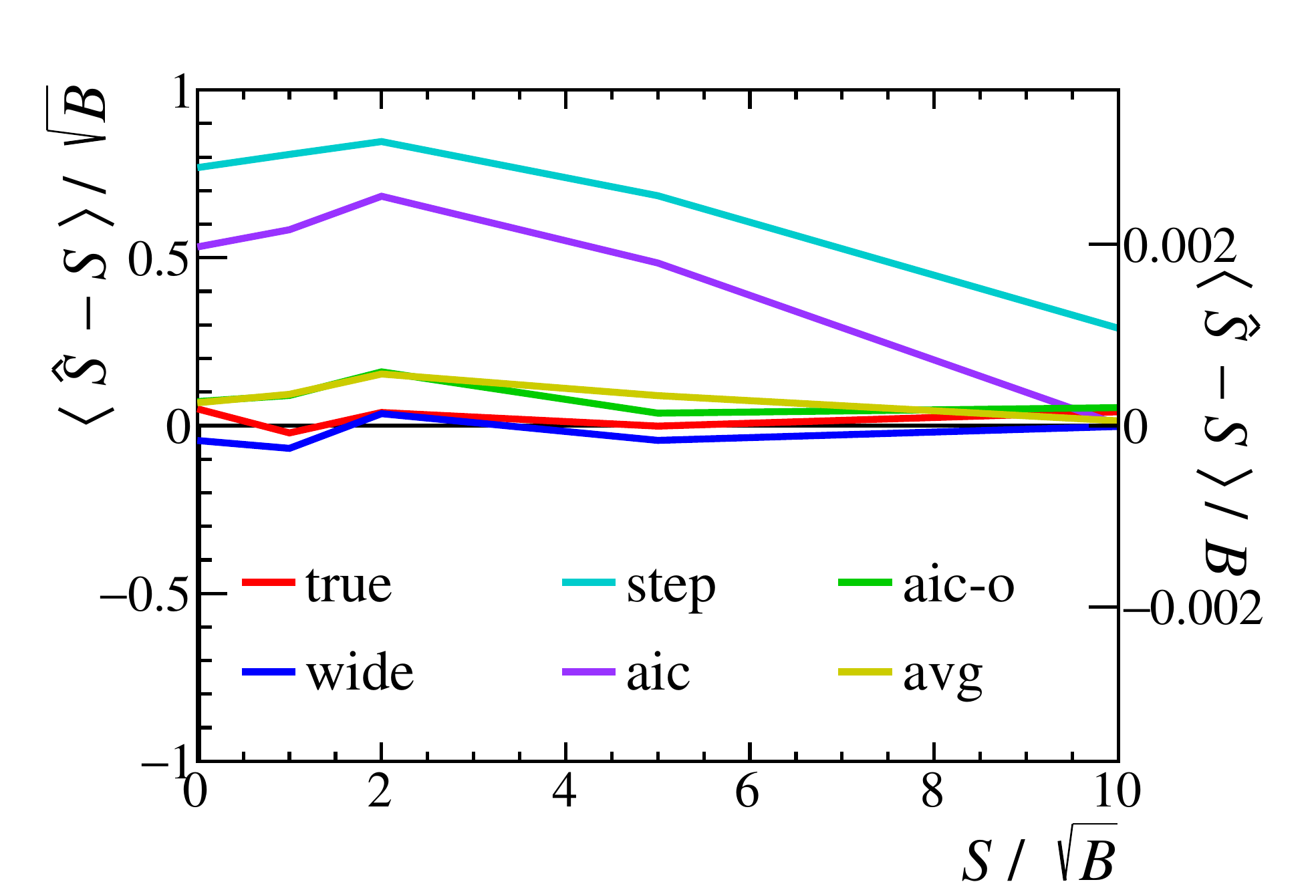}
  \includegraphics[width=0.49\textwidth]{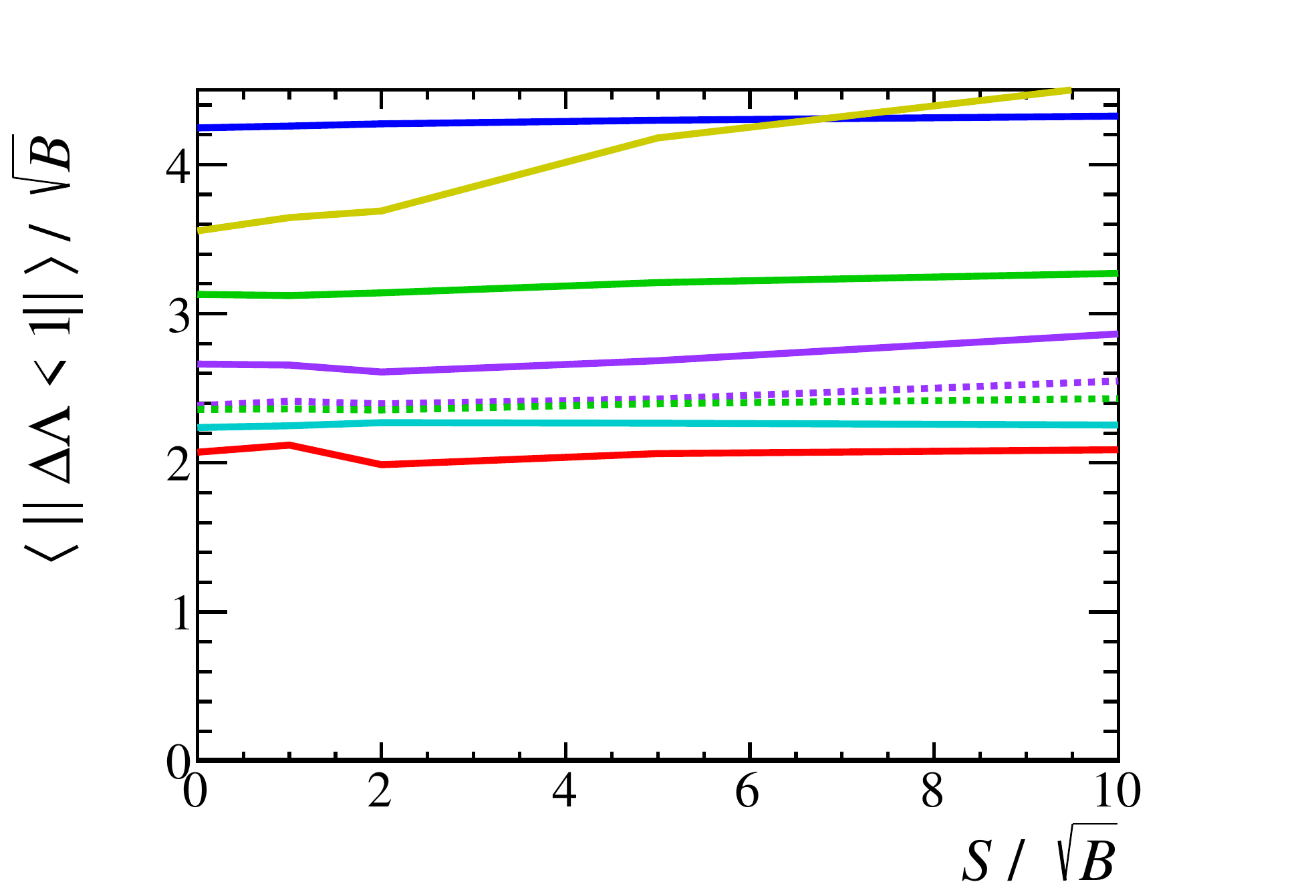}
  \includegraphics[width=0.49\textwidth]{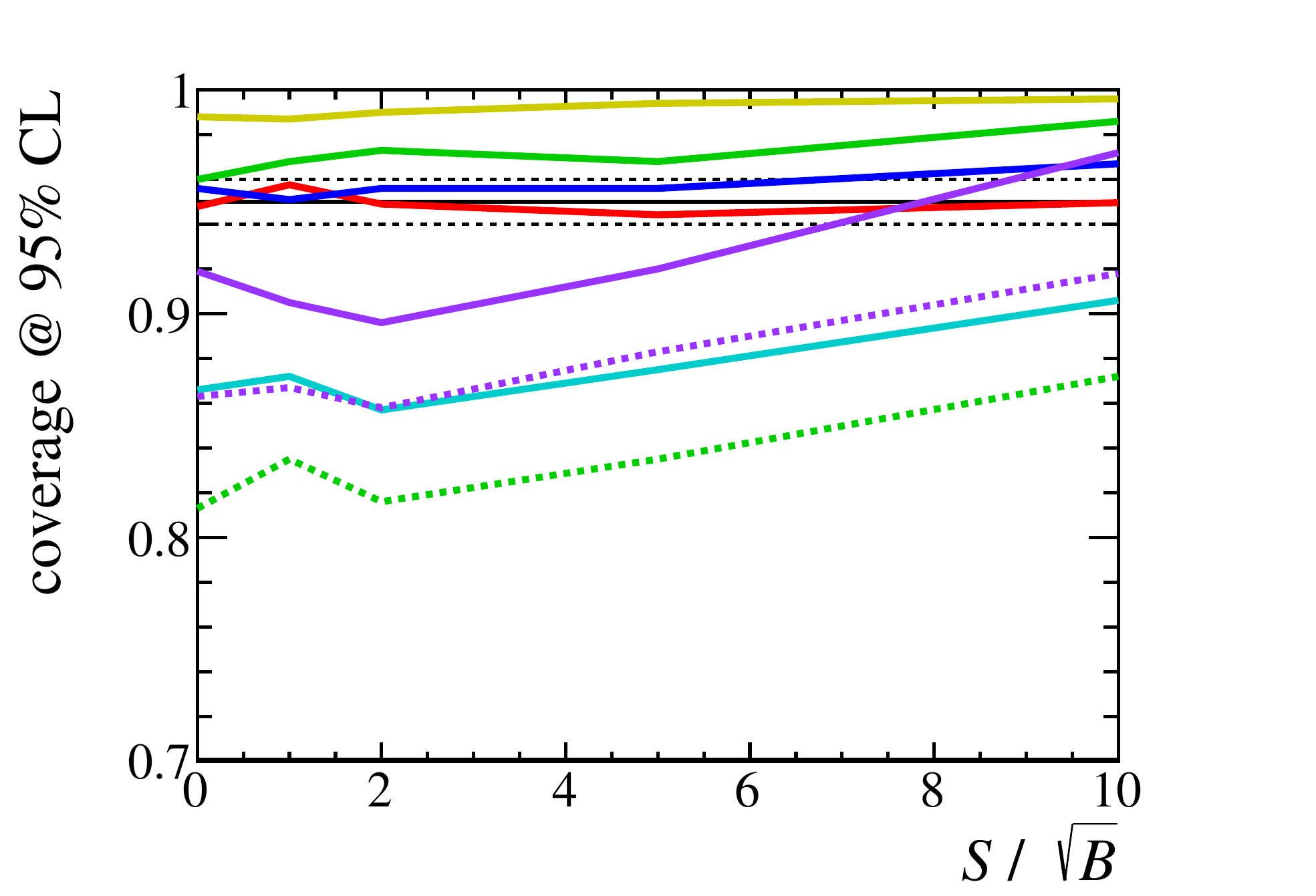}
  \includegraphics[width=0.49\textwidth]{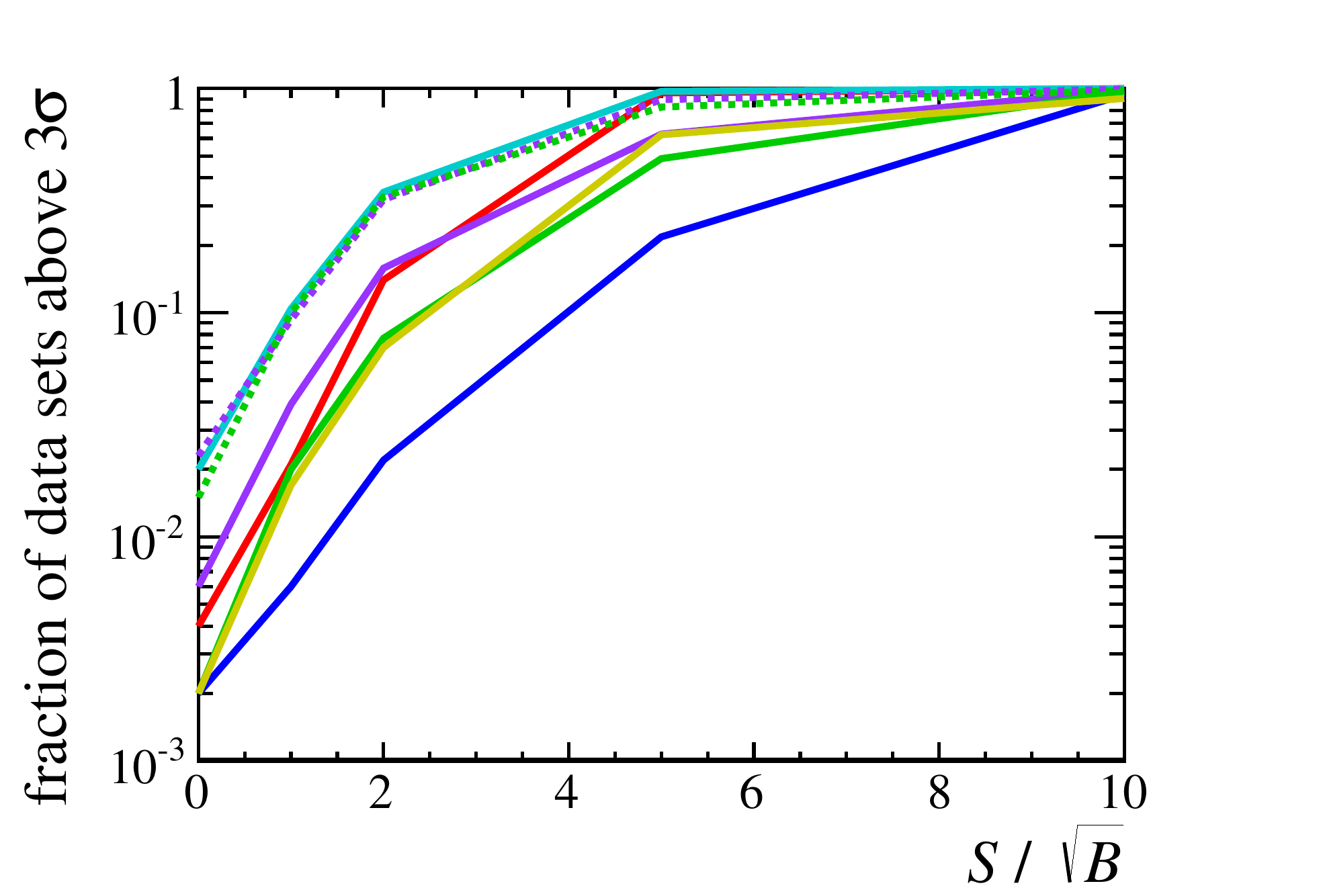}
\caption{Results for the Higgs-decay model versus the generated signal $S$ relative to $\sqrt{B}$, where $B$ is the background yield in a $\pm2\sigma$ window around the Higgs mass.
(top left) The mean bias in signal estimator $\hat{S}$ relative to (left axis) $\sqrt{B}$ and (right axis) $B$.
(top right) The mean full length of the 68.3\% CI relative to $\sqrt{B}$ as reported by the profile likelihood, where the dashed lines show the aic and aic-o results without accounting for model-selection uncertainty.
(bottom left) Coverage reported at 95\% CL by each method, where the solid black line shows the expected 95\% and the dashed black lines show the approximate uncertainty in all results due to ensemble size.
(bottom right) The fraction of data sets that report a $p$-value corresponding to a significance $>3\sigma$ for the NULL hypothesis (the expected value is 0.27\% when $S=0$ as a two-sided test is performed due to the limited number of data sets in the ensemble).
}
  \label{fig:hmm_results}
\end{figure}

It is clear that some dedicated background PDF components are required to properly handle the tail of the $Z$ boson.
While the optimal approach certainly utilizes the known properties of the $Z$, for illustrative purposes this information is not used here.
Instead, three exponential terms are added to the wide model, which enables obtaining a decent description of the data as evaluated using the wide-model $\chi^2$.
Figure~\ref{fig:hmm_results} shows the results, which are similar to those obtained for the toy-model example.
%\footnote{Just to reiterate, this is not the optimal approach for describing background from the $Z$ boson. The results are encouraging and would likely be even better if a $Z$-boson-specific component was added instead of the series of exponentials.}
I stress here that, unlike in a traditional bump-hunt approach, the quality of the problem-specific PDF components does not need to be exceedingly high.
The Legendre modes in the wide model are available to the model-selection procedure and can account for discrepancies in the problem-specific PDF, including any small unexpected peaking structures.
Indeed, in this example using only three exponentials to describe the $Z$-boson line shape results in some structure in the residuals, nevertheless valid results are obtained.
Despite the extreme challenges presented in this example, the same AIC-based strategy---including all odd modes and accounting for model-selection uncertainty---produces unbiased $\hat{S}$ values, CIs, and $p$-values after dedicated $Z$-boson background components are added (despite the fact that the $Z$-boson terms used here are not optimal).
{\em N.b.}, quantitatively the lengths of the AIC-based CIs depend on the choice $\ell_{\rm max}=10$. %, which was chosen here to allow for potential peaking-background structure at 125~GeV.
The use of a smaller $\ell_{\rm max}$, corresponding to a less conservative wide model, will produce AIC-based CI lengths closer to the optimal values.

\section{Summary \& Discussion}
\label{sec:sum}

A critical aspect of searching for unknown particles is properly assessing the uncertainty in the background model, including the model-selection contribution which can be large when the signal-to-background ratio $S/B$ is small.
Background-model selection involves a tradeoff between bias and variance, since increasing the background-model complexity decreases the potential bias on $\hat{S}$ while increasing its variance.
The key principles to take away from the studies presented in this article are:
\begin{itemize}
\item The NULL hypothesis is generally implicitly taken to be the absence of any unknown particles in the mass spectrum; however, in reality, the NULL is the lack of any features in the data that cannot be explained by the background-only PDF.
\item It is vital that the wide model, {\em i.e.}\ the model containing all background PDF components under consideration, is able to describe the data in the absence of a signal contribution well enough to provide unbiased $\hat{S}$ estimates and valid CIs and $p$-values.
\item I proposed transforming each fit region onto the interval $[-1,1]$ centered on the test mass. Provided that the length of each fit region (prior to this transformation) is defined in terms of the mass resolution, it is straightforward to relate the highest-order even mode included in the background PDF, $\ell_{\rm max}$, to the width of the narrowest peaking-background structure that can be accommodated by the NULL hypothesis. This approach has the desirable feature that the choice of $\ell_{\rm max}$ for the wide model is largely driven by the potential peaking backgrounds that may occur, which must be considered in detail anyway, rather than on some large-scale simulation study at each mass. %A caveat to this is presented in the $H\to\mu^+\mu^-$ study, where a
\item Furthermore, on the $[-1,1]$ interval signals are predominantly even functions; therefore, adding complexity to the background PDF in the form of odd modes has minimal impact on the variance of $\hat{S}$ (due to orthogonality), while greatly reducing the potential bias. Therefore, all odd modes up to (possibly even beyond) $\ell_{\rm max}$ should be included in every background model considered in the model-selection process.
\item Even modes, however, do affect the variance on $\hat{S}$; therefore, adding all even modes up to $\ell_{\rm max}$ will reduce the sensitivity of the search, which means that some form of data-driven model selection is required to arbitrate between background models that include different subsets of the even modes contained in the wide model.
The common principle invoked by data-driven model-selection methods is to reward goodness of fit while punishing model complexity, though different methods employ different approaches in practice.
\end{itemize}
This study explored various data-driven methods for performing background-model selection, and for assigning uncertainty on $\hat{S}$ that arises due to the choice of background model.
Various realistic example problems were considered, and the results are summarized as follows:
\begin{itemize}
\item It was demonstrated that the stepwise-in approach often used in physics is prone to bias, undercoverage, and overestimation of significance; clearly this approach should not be used.
\item Conversely, the wide model produced unbiased $\hat{S}$ values, CIs, and $p$-values.
The only drawback to using the wide model is that the lengths of its CIs are about twice the optimal length (specifically, this value is for $\ell_{\rm max}=10$ and will vary depending on how much complexity is required in the wide model for each fit region in each search).
\item While the AIC-based approach is well known and widely used outside of physics, it is prone to producing biased $\hat{S}$ values and invalid CIs and $p$-values.
The potential bias in a bump hunt is greatly reduced by including all odd modes from the wide model in all submodels considered in the background-model selection process;
however, obtaining proper coverage and $p$-values still requires accounting for model-selection uncertainty.
In this study, the discrete-nuisance-parameter approach of Ref.~\cite{ref:iic} was employed due to its simplicity, and was found to produce valid CIs and $p$-values in all problems studied.
\item Frequentist model averaging was also found to produce valid results, though its CIs were typically 10--20\% longer than those of AIC (including model-selection uncertainty).
\end{itemize}
In all examples studied in this work, the best results for $\hat{S}$ and its CI were obtained using the novel approach proposed here.
This involves transforming the fit interval onto $[-1,1]$ centered on the test mass and choosing an appropriate wide model, {\em i.e.}\ choosing a wide model that is capable of describing the data well enough in the absence of a signal contribution to provide an unbiased $\hat{S}$ and a valid CI and $p$-value.
Model selection is then performed using the AIC-based approach, where all odd modes are included in every model considered and all even modes are arbitrated on.
The model-selection uncertainty is determined by treating the model index as a discrete nuisance parameter in the profile likelihood.
This same approach can also be used to obtain the local $p$-values; however, for highly significant signals, model uncertainty drives the $p$-values to be close to those obtained using the wide model as expected.
An alternative approach for assigning local $p$-values is to take those obtained using the wide model with no additional model selection and, therefore, no additional model-selection uncertainty, which provides a mildly conservative significance estimate relative to the model-selection-based one.

The key role of the analyst then is to input an adequate wide model for each test-mass value in their bump hunt.
This is accomplished largely using knowledge about any potential peaking-background structures and any long-distance rapid variation in the background PDF, {\em e.g.}\ due to a large resonance contribution nearby, and can be aided by studying simulated backgrounds.
It is not possible to guarantee that any method will provide unbiased $\hat{S}$ estimates and valid CIs and $p$-values in all possible situations; however, given the range of backgrounds studied here, and the consistently good performance observed from the method proposed in this article, valid results are expected provided that an adequate wide model is chosen---and a background model with $\ell_{\rm max}=10$ and any dedicated resonance terms is likely to be appropriate in the absence of a large and narrow peaking background.
Finally, I note that this method was successfully employed by LHCb in a recent search for dark photons~\cite{ref:lhcbatomm}.

\acknowledgments

I thank N.~Toro for suggesting the APEX problem, M.~Klute for suggesting the Higgs-decay one, and N.~Wardle for helpful feedback.
This work was supported by DOE grant DE-SC0010497 and NSF grant PHY-1607225.

\clearpage

\appendix

\section*{Appendix: Full APEX Results}
\label{app:apex}

\begin{figure}[h]
  \centering
  \includegraphics[width=0.49\textwidth]{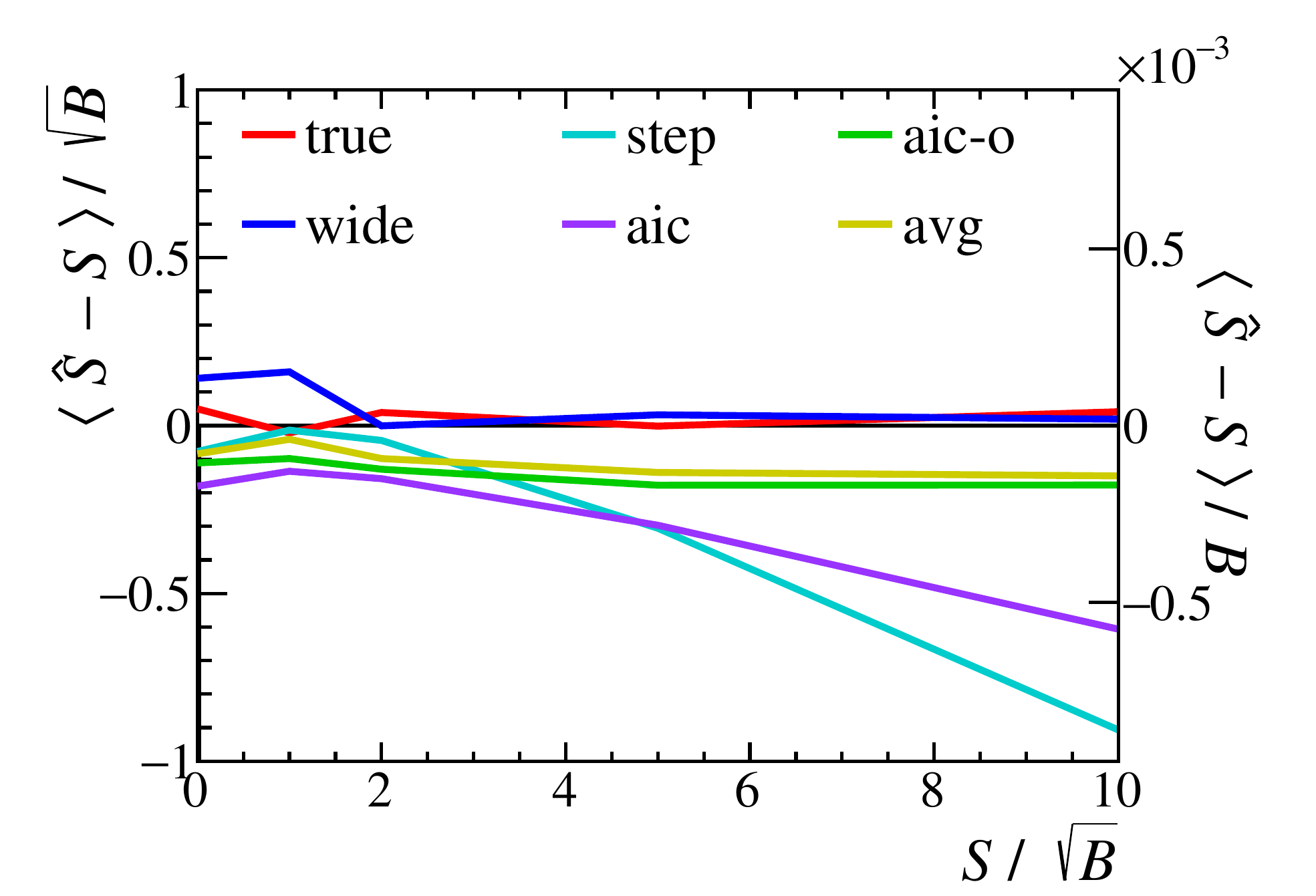}
  \includegraphics[width=0.49\textwidth]{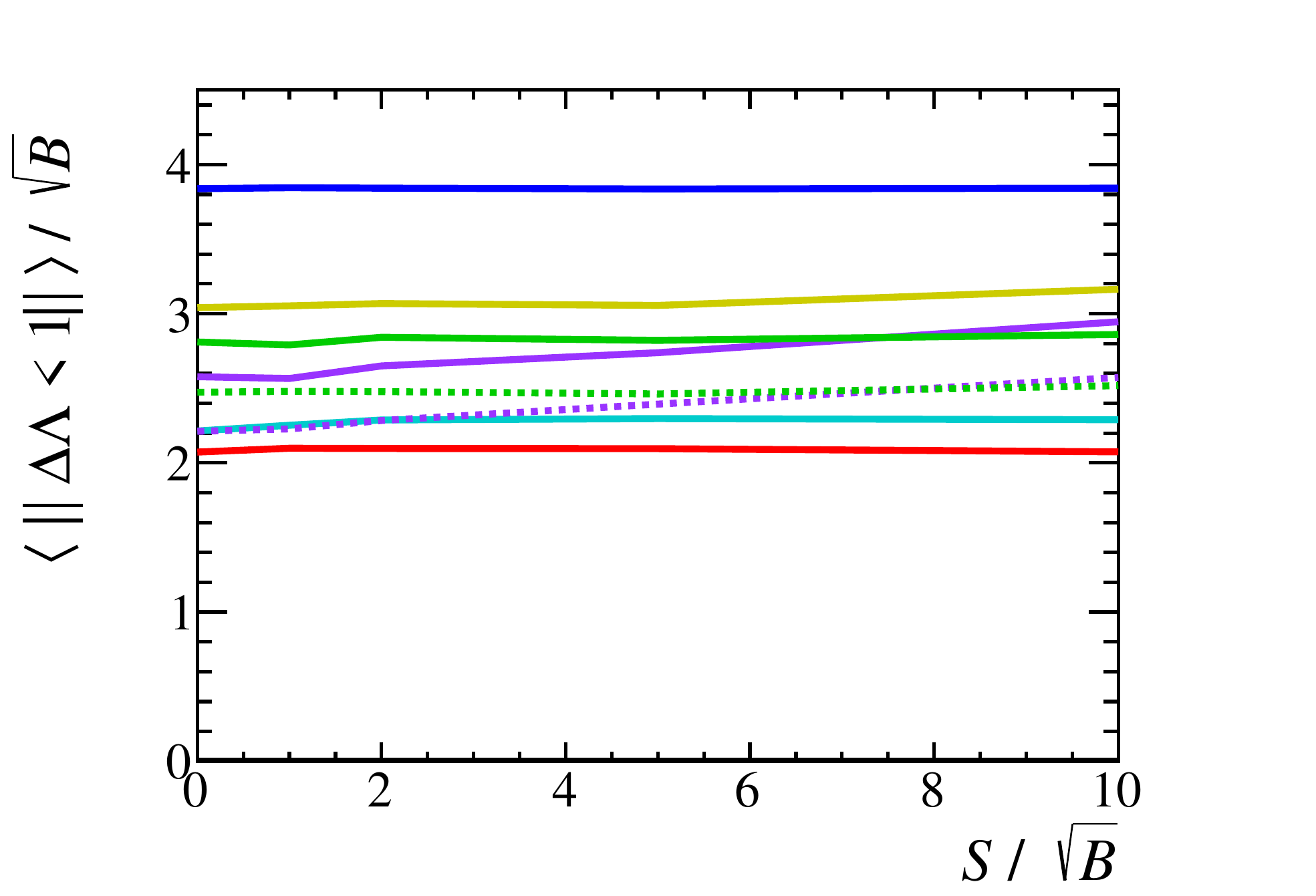}
  \includegraphics[width=0.49\textwidth]{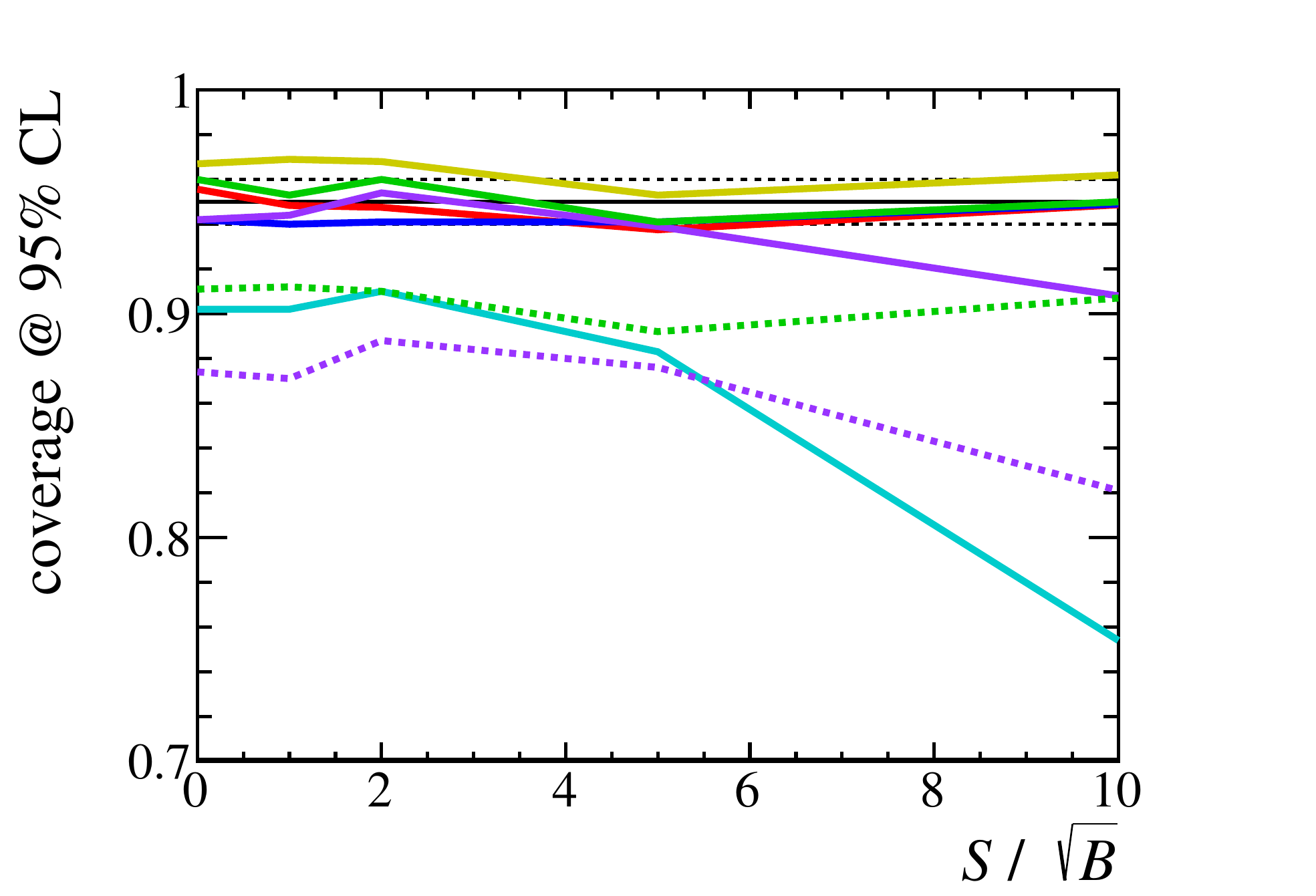}
  \includegraphics[width=0.49\textwidth]{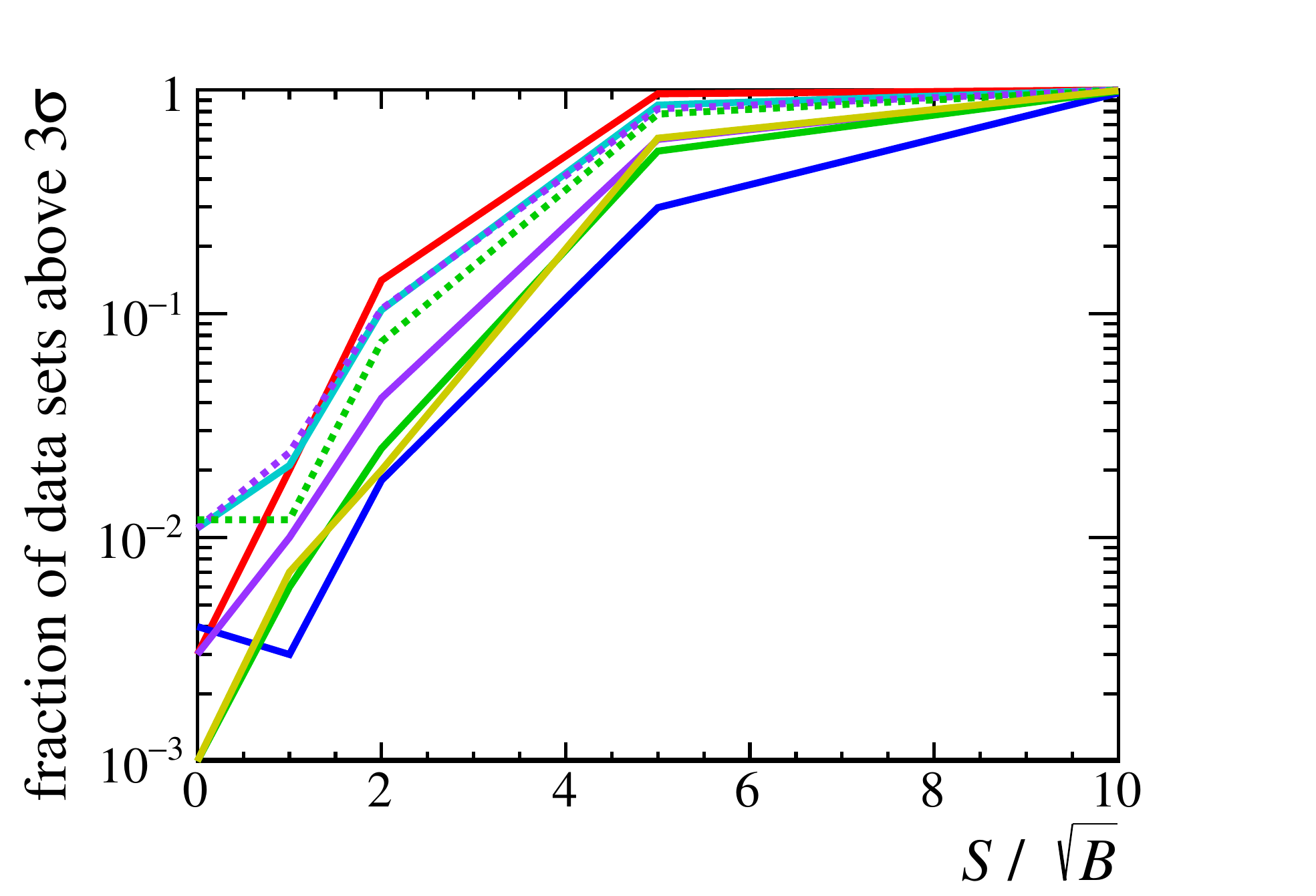}
\caption{Results for the APEX example for $m=0.2$ versus the true signal $S$ relative to $\sqrt{B}$, where $B$ is the background yield in a $\pm2\sigma$ window around the signal mass
(see Table~\ref{tab:labels} for legend label definitions).
(top left) The mean bias in signal estimator $\hat{S}$ relative to (left axis) $\sqrt{B}$ and (right axis) $B$.
(top right) The mean full length of the 68.3\% CI relative to $\sqrt{B}$ as reported by the profile likelihood, where the dashed lines show the aic and aic-o results without accounting for model-selection uncertainty.
(bottom left) Coverage reported at 95\% CL by each method, where the solid black line shows the expected 95\% and the dashed black lines show the approximate uncertainty in all results due to ensemble size.
(bottom right) The fraction of data sets that report a $p$-value corresponding to a significance $>3\sigma$ for the NULL hypothesis (the expected value is 0.27\% when $S=0$ as a two-sided test is performed due to the limited number of data sets in the ensemble).}
  \label{fig:apex_results_0}
\end{figure}

\begin{figure}
  \centering
  \includegraphics[width=0.49\textwidth]{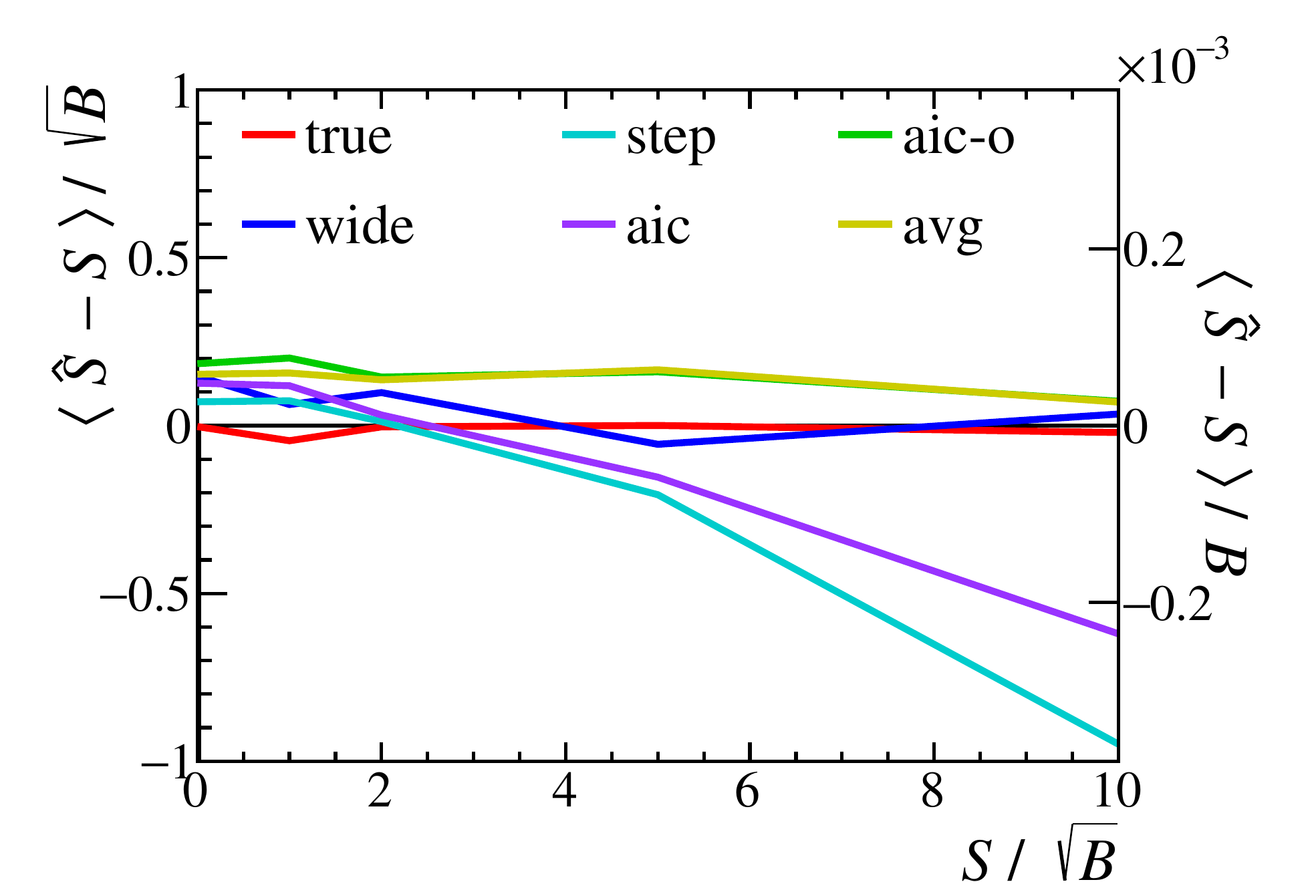}
  \includegraphics[width=0.49\textwidth]{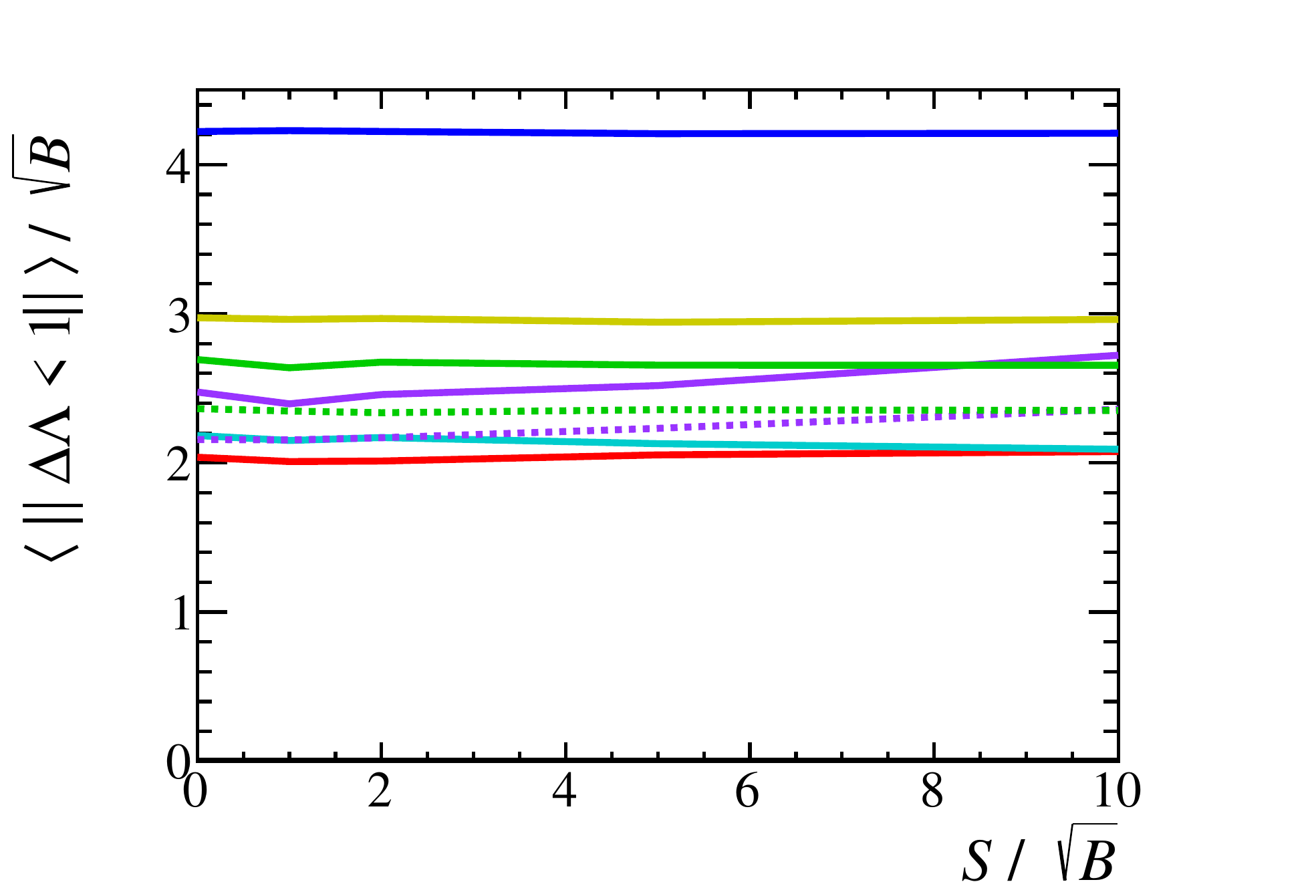}
  \includegraphics[width=0.49\textwidth]{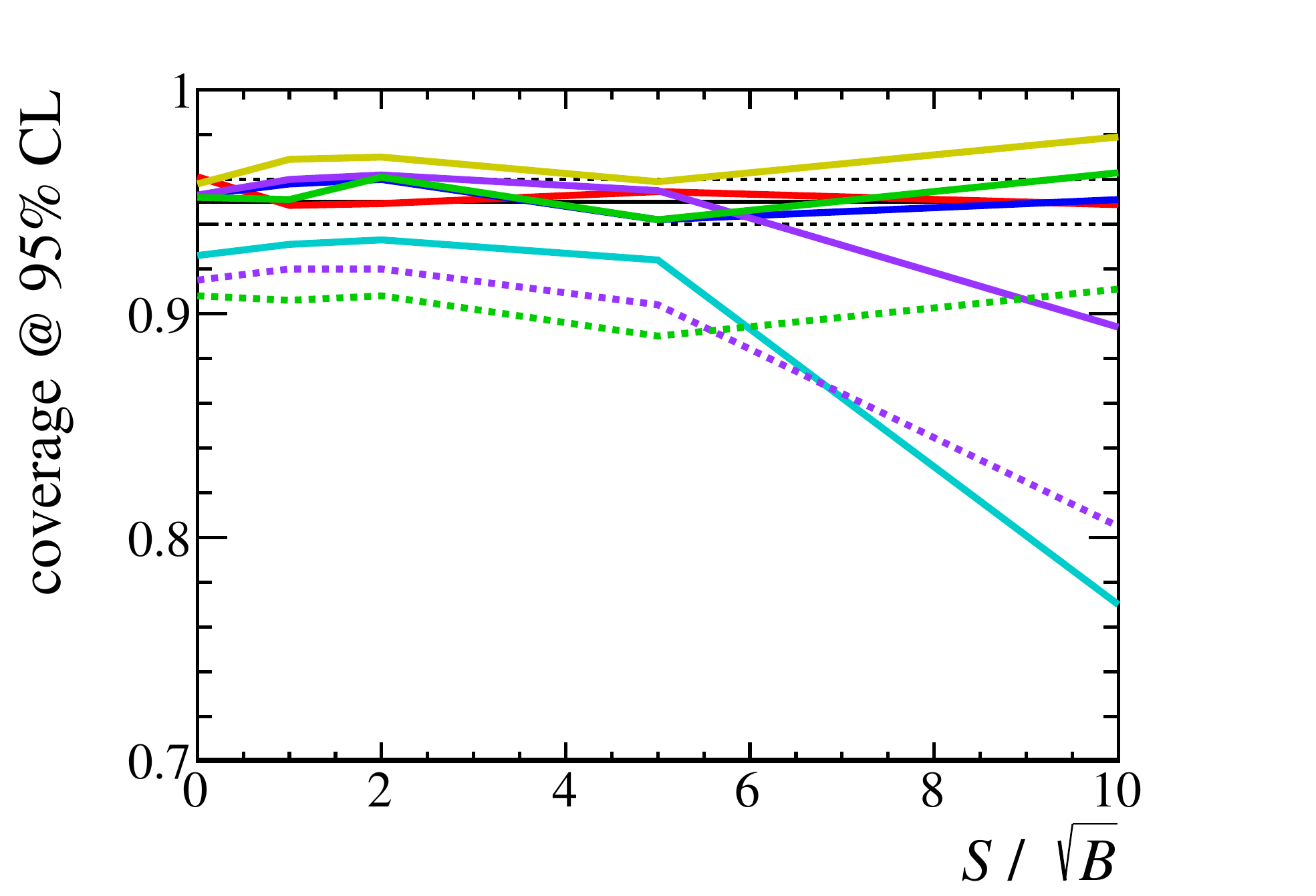}
  \includegraphics[width=0.49\textwidth]{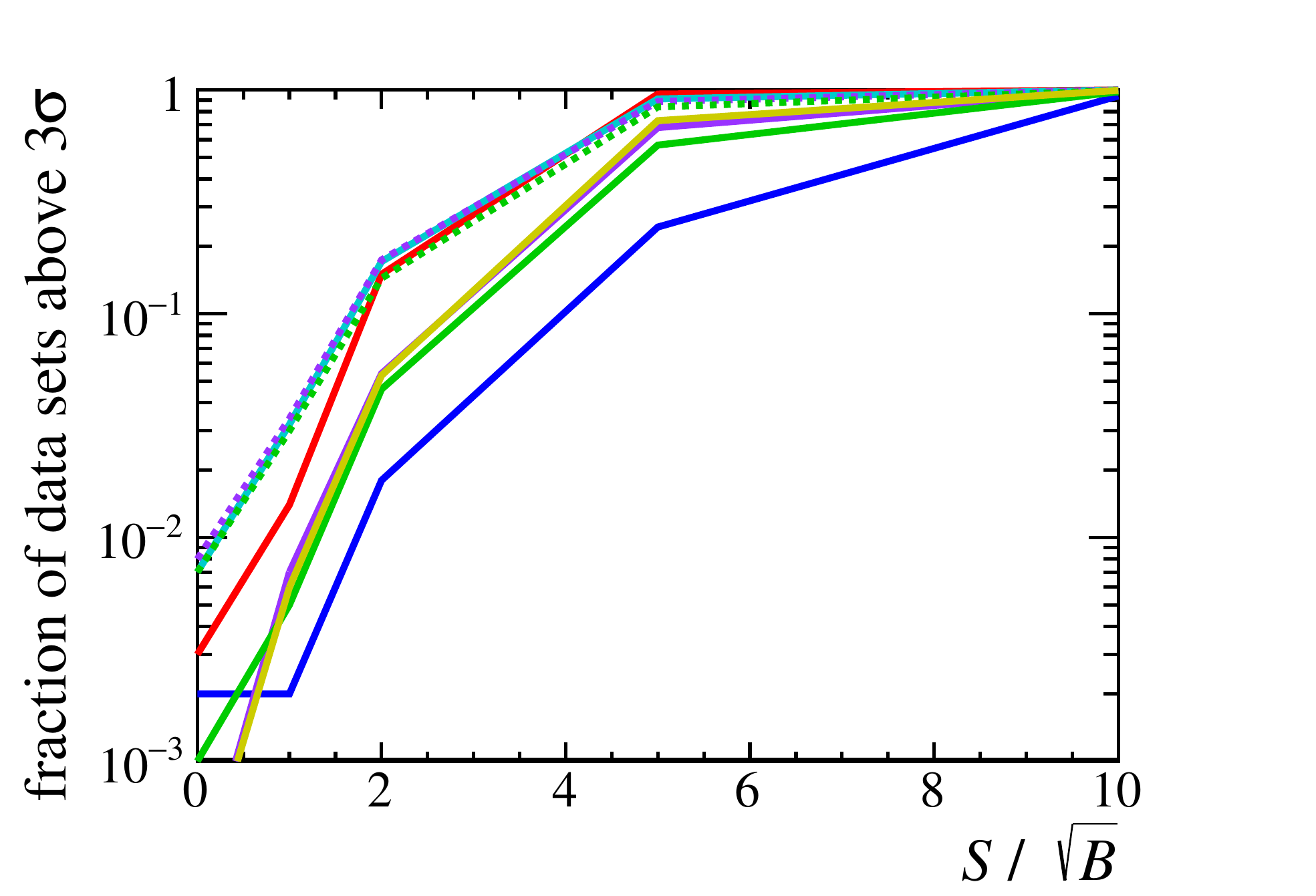}
\caption{Same as Fig.~\ref{fig:apex_results_0} but for $m=0.6$.}
  \label{fig:apex_results_1}
\end{figure}

\begin{figure}
  \centering
  \includegraphics[width=0.49\textwidth]{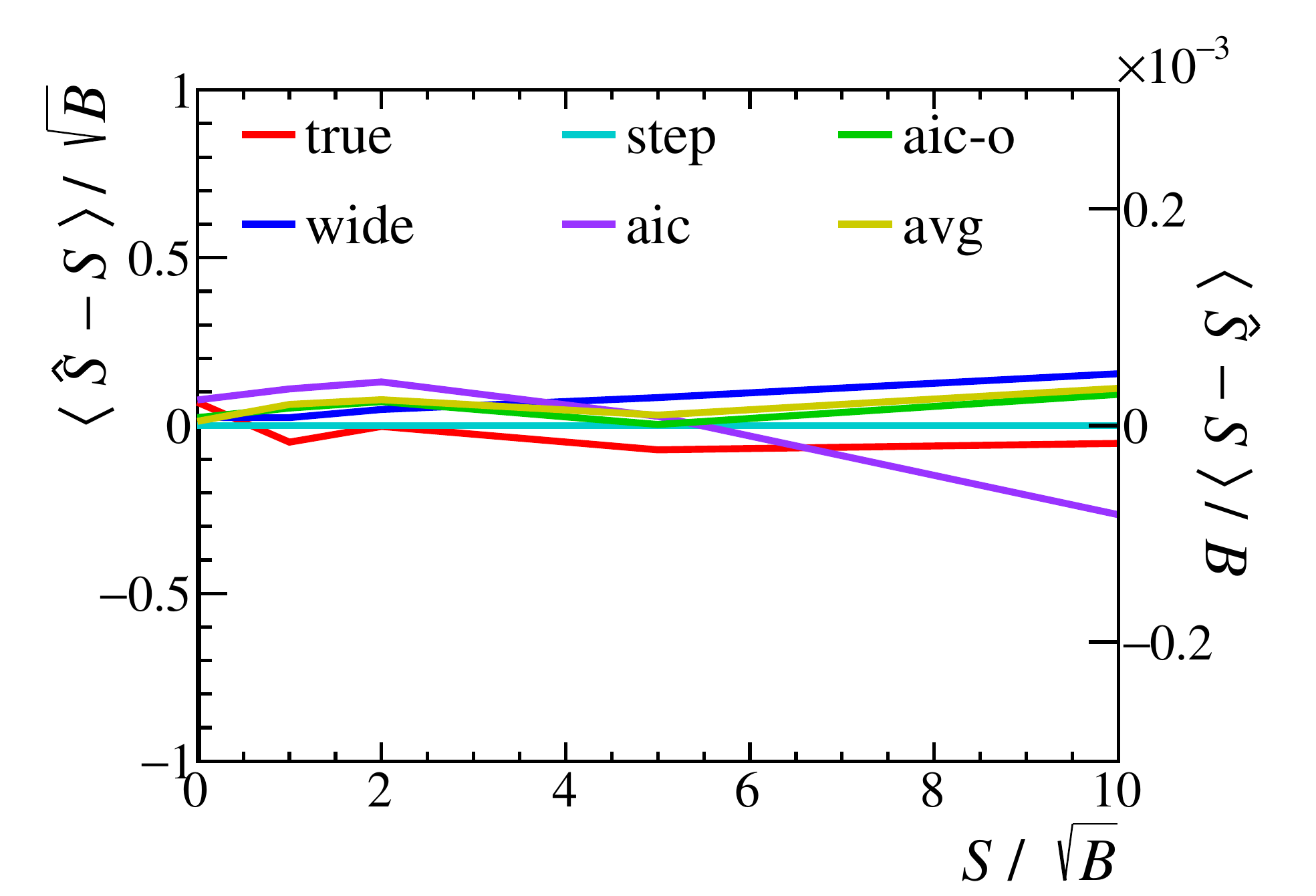}
  \includegraphics[width=0.49\textwidth]{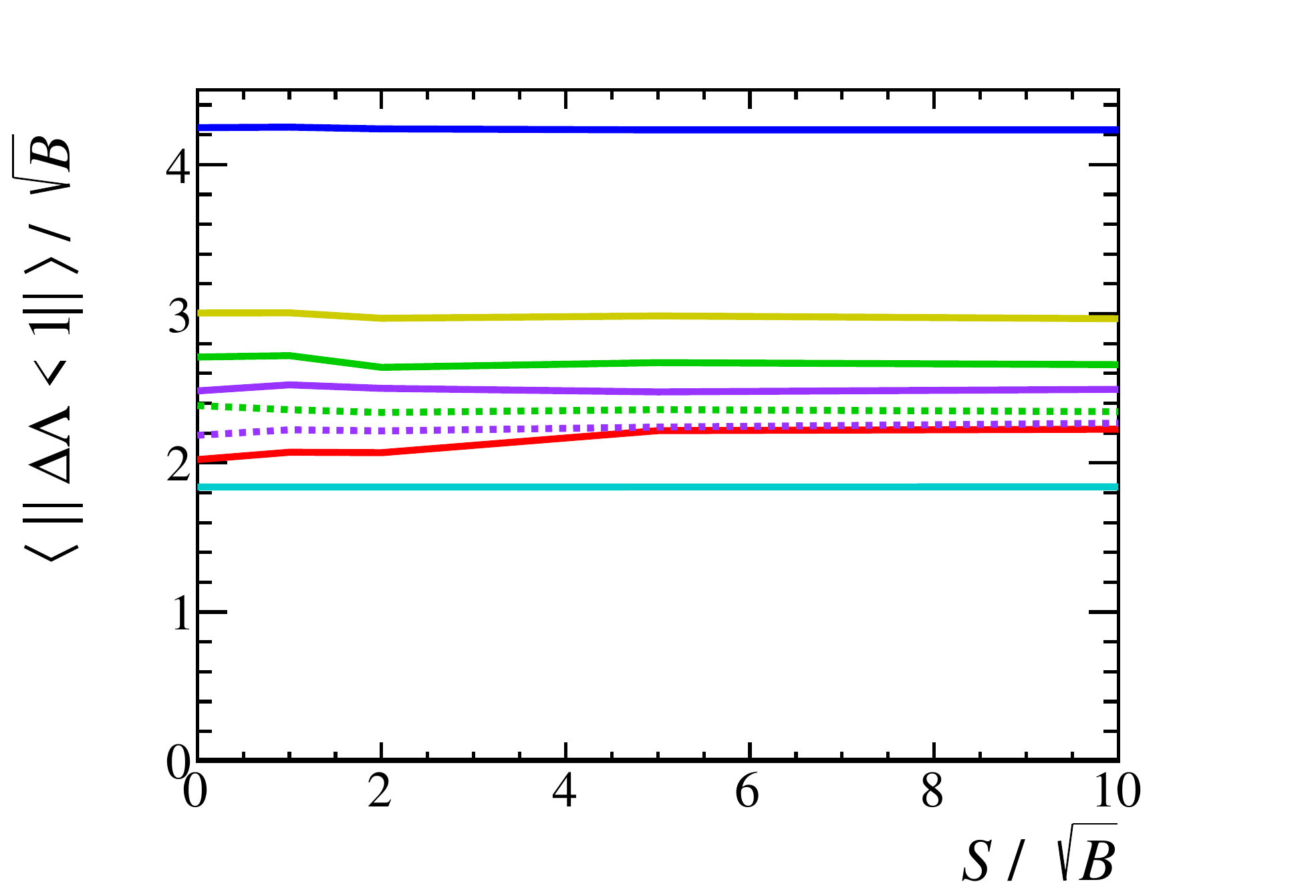}
  \includegraphics[width=0.49\textwidth]{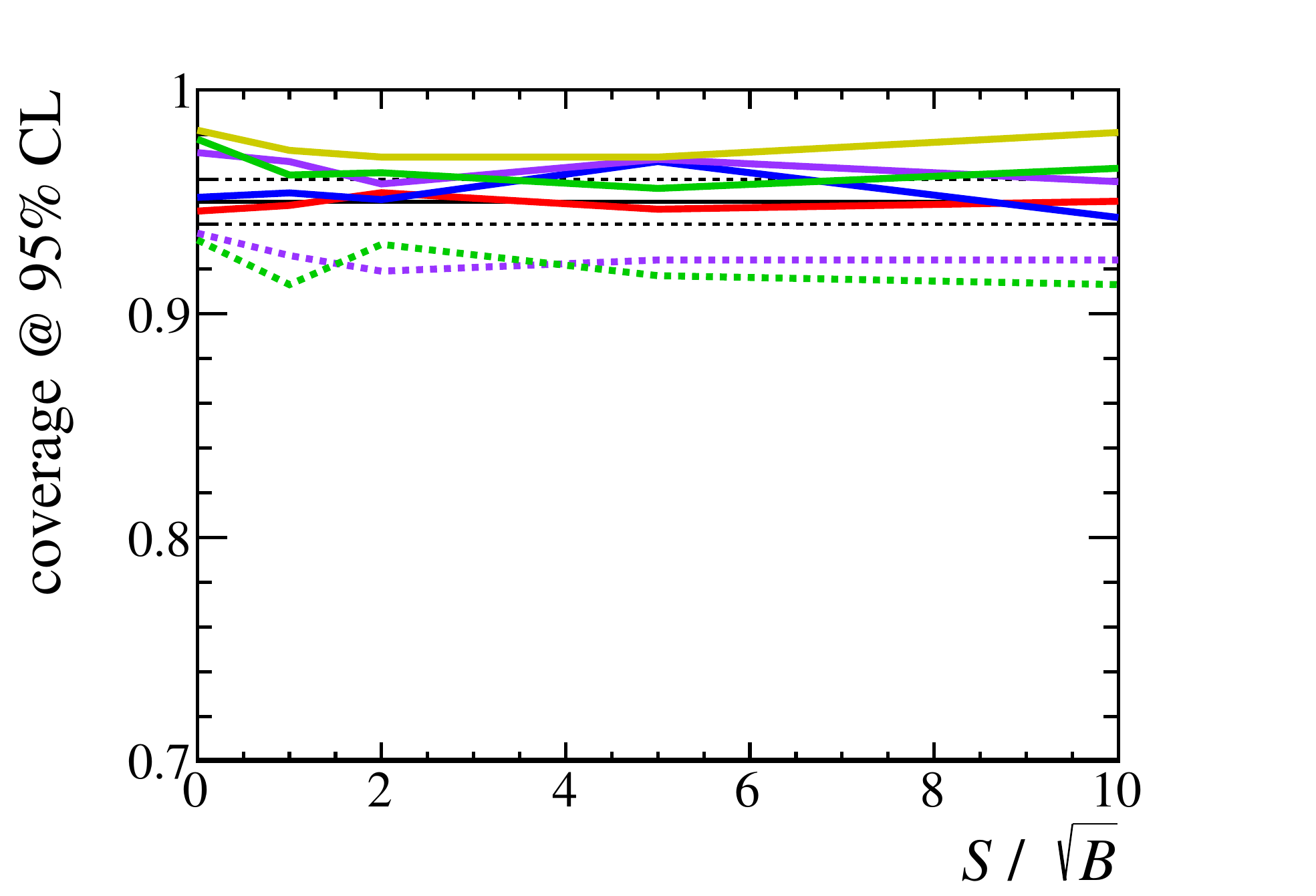}
  \includegraphics[width=0.49\textwidth]{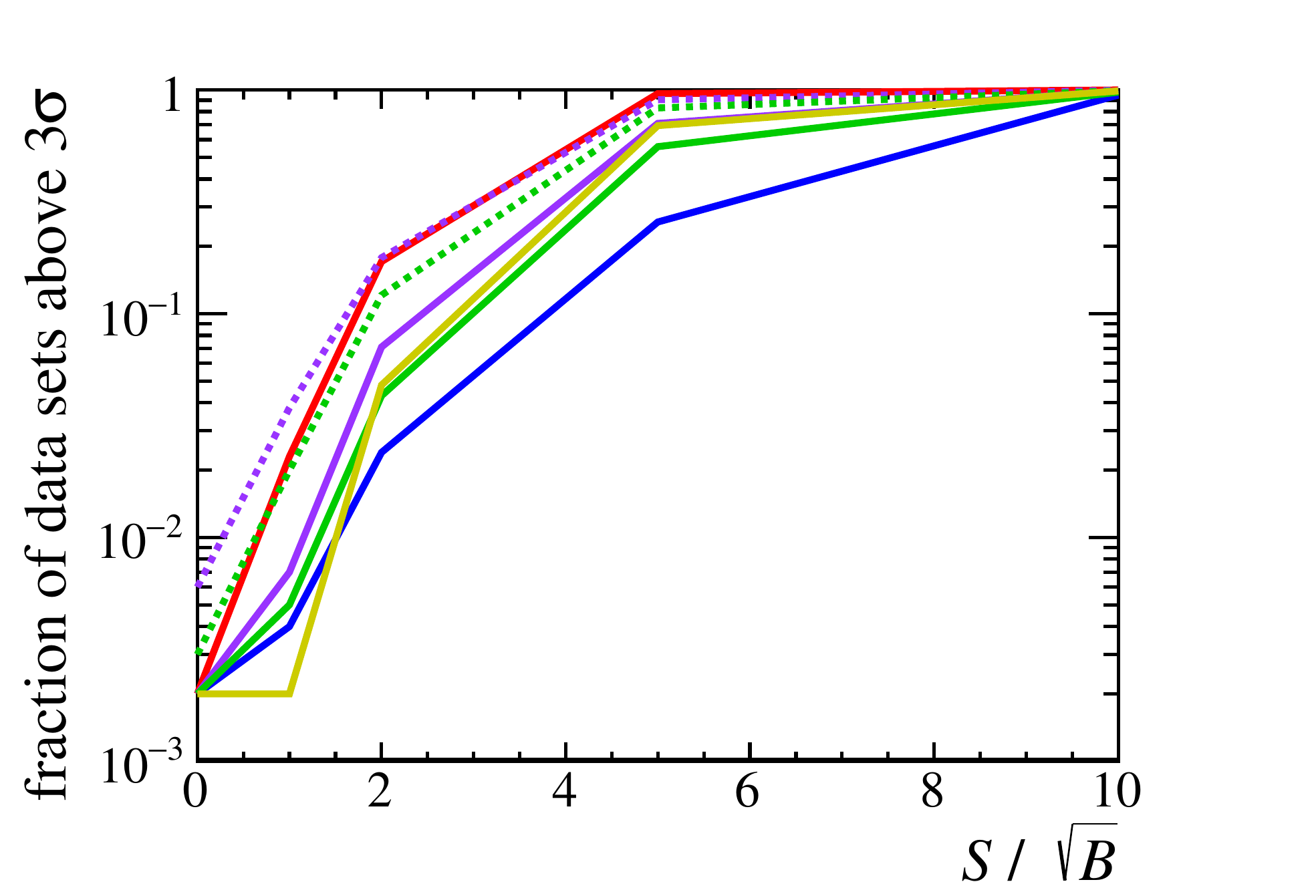}
\caption{Same as Fig.~\ref{fig:apex_results_0} but for $m=1.1$.}
  \label{fig:apex_results_2}
\end{figure}

\begin{figure}
  \centering
  \includegraphics[width=0.49\textwidth]{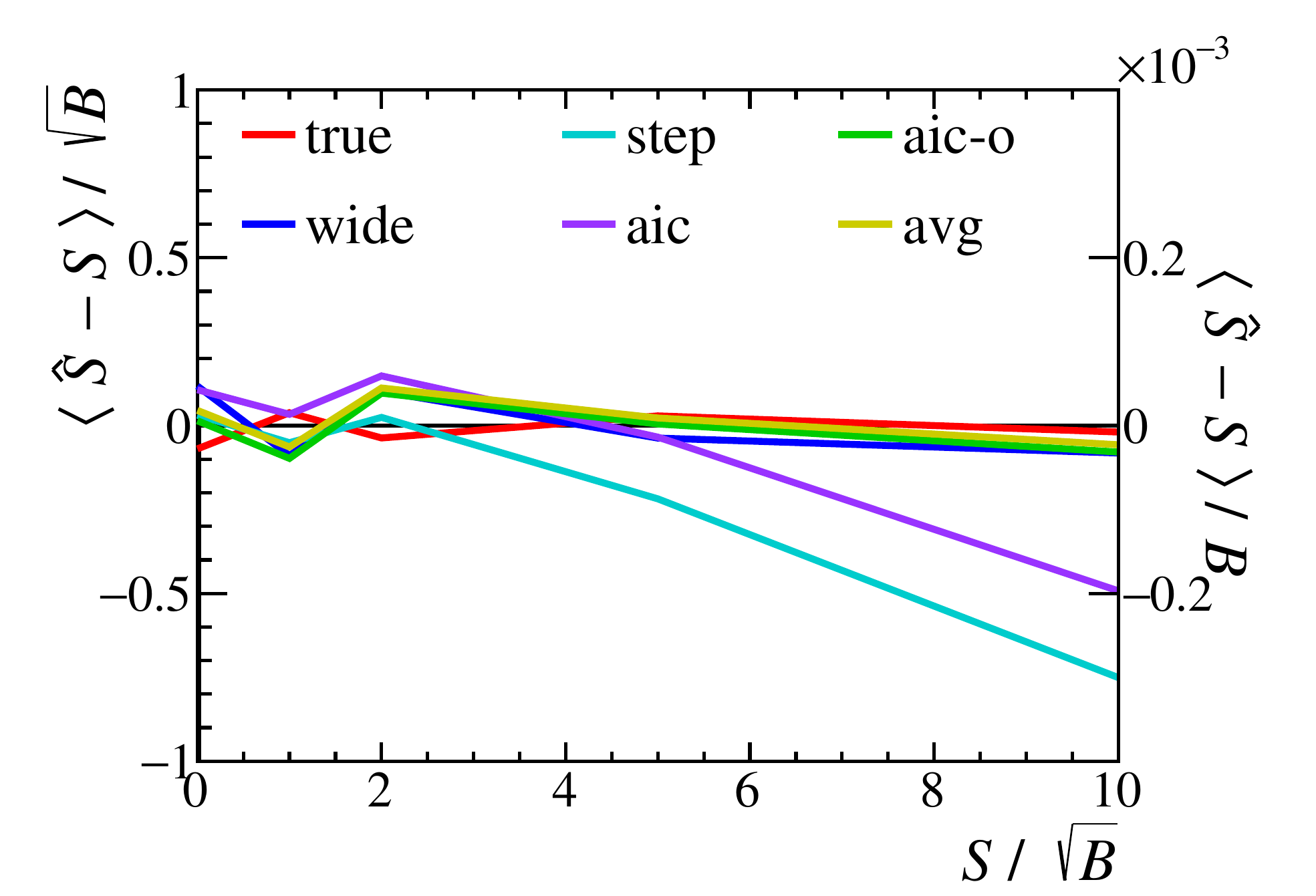}
  \includegraphics[width=0.49\textwidth]{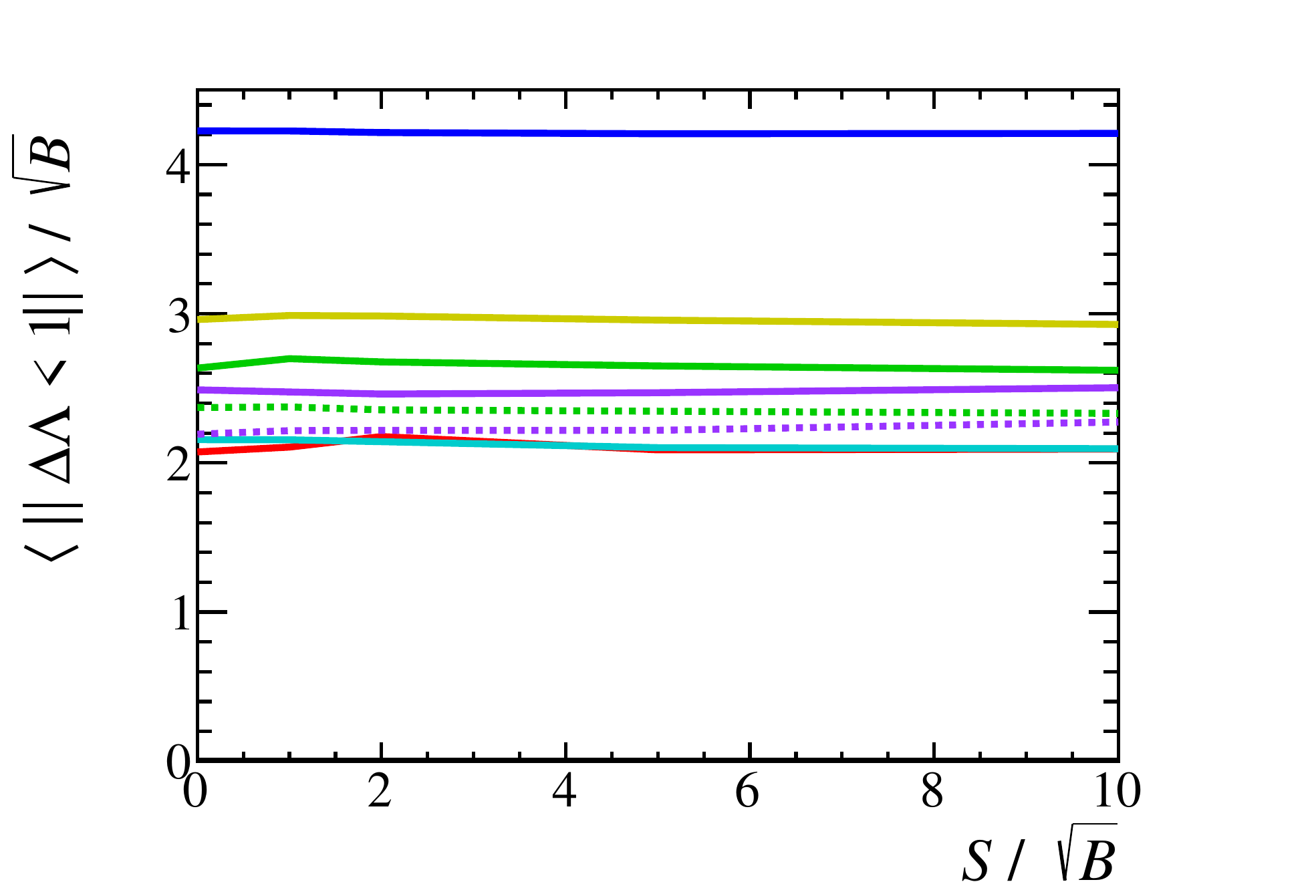}
  \includegraphics[width=0.49\textwidth]{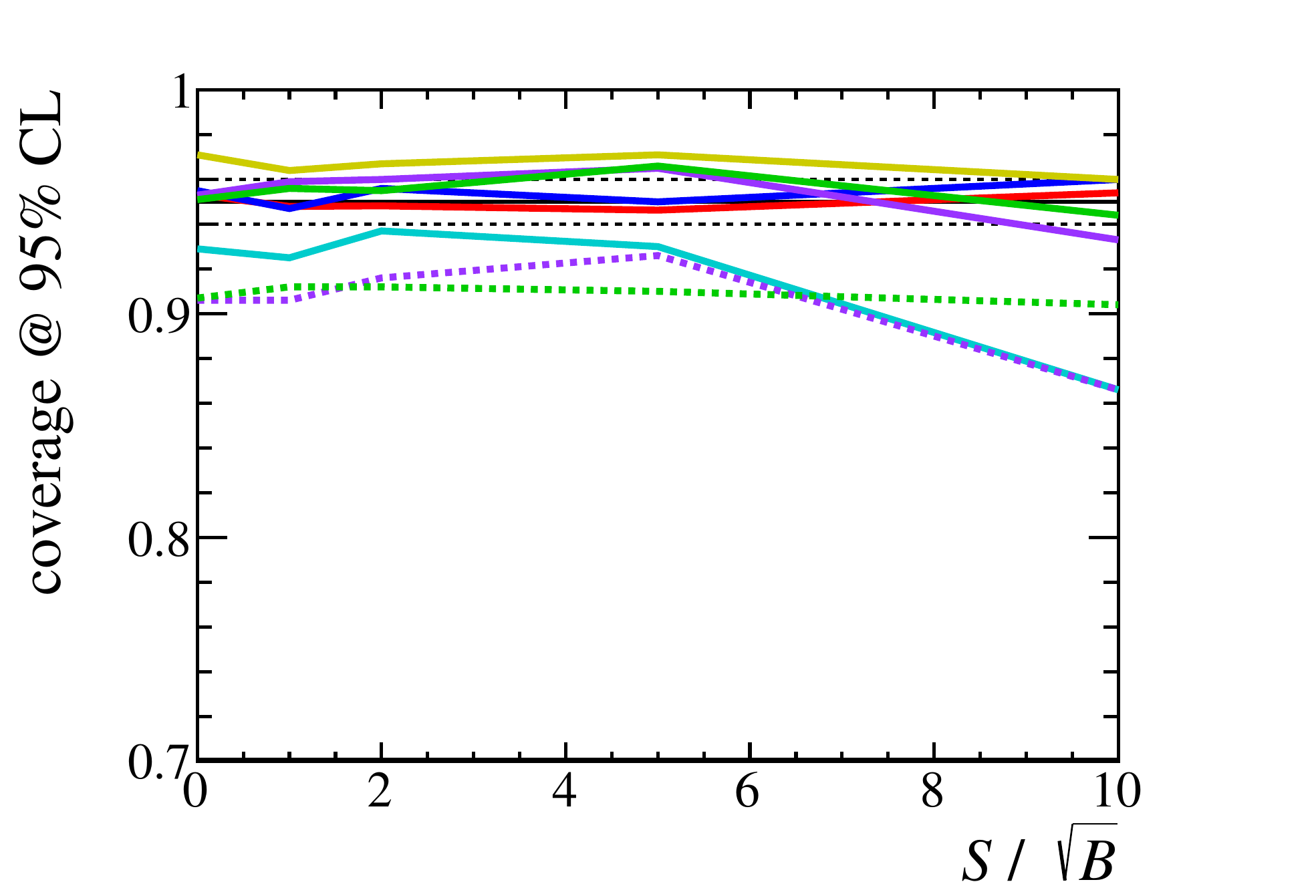}
  \includegraphics[width=0.49\textwidth]{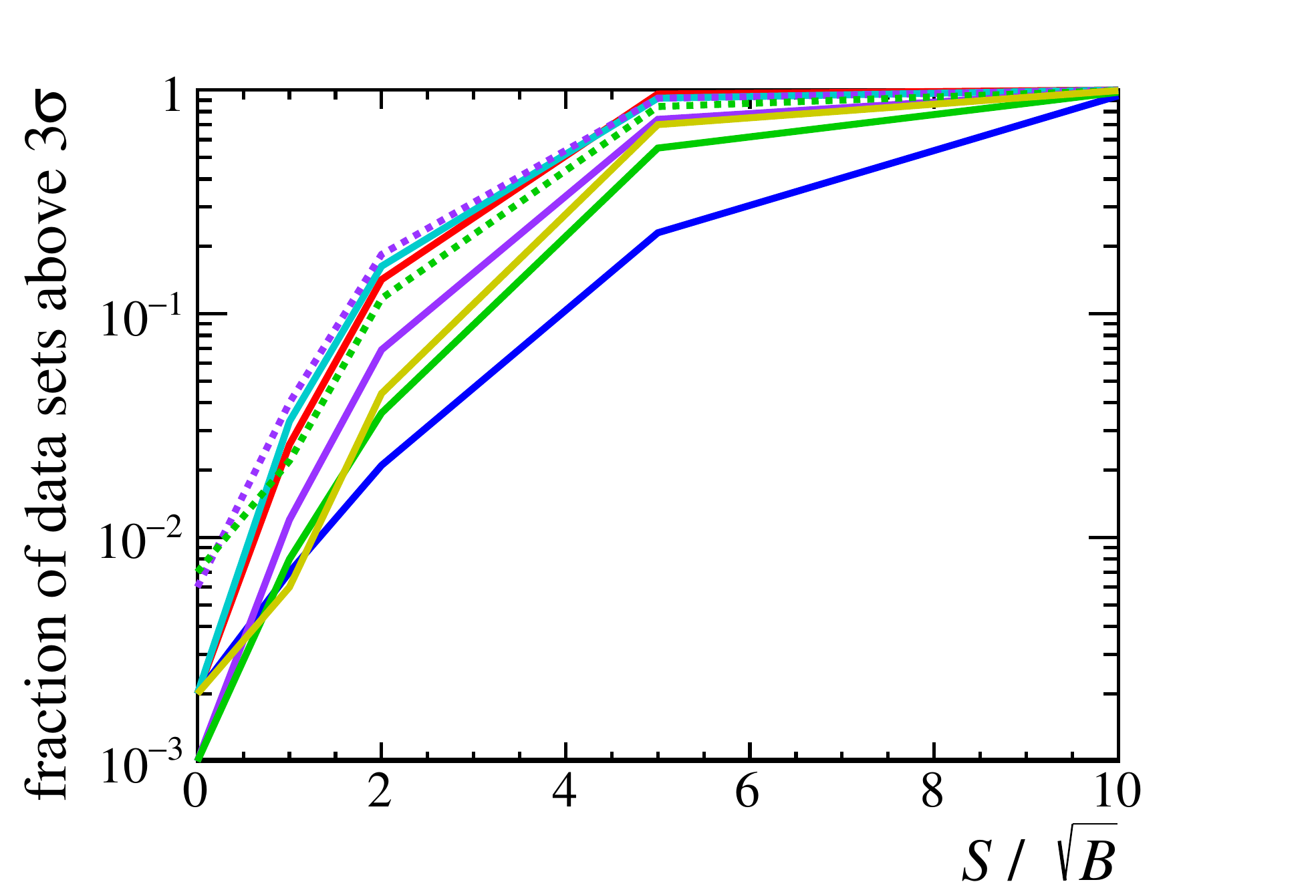}
\caption{Same as Fig.~\ref{fig:apex_results_0} but for $m=2.0$.}
  \label{fig:apex_results_3}
\end{figure}

\begin{figure}
  \centering
  \includegraphics[width=0.49\textwidth]{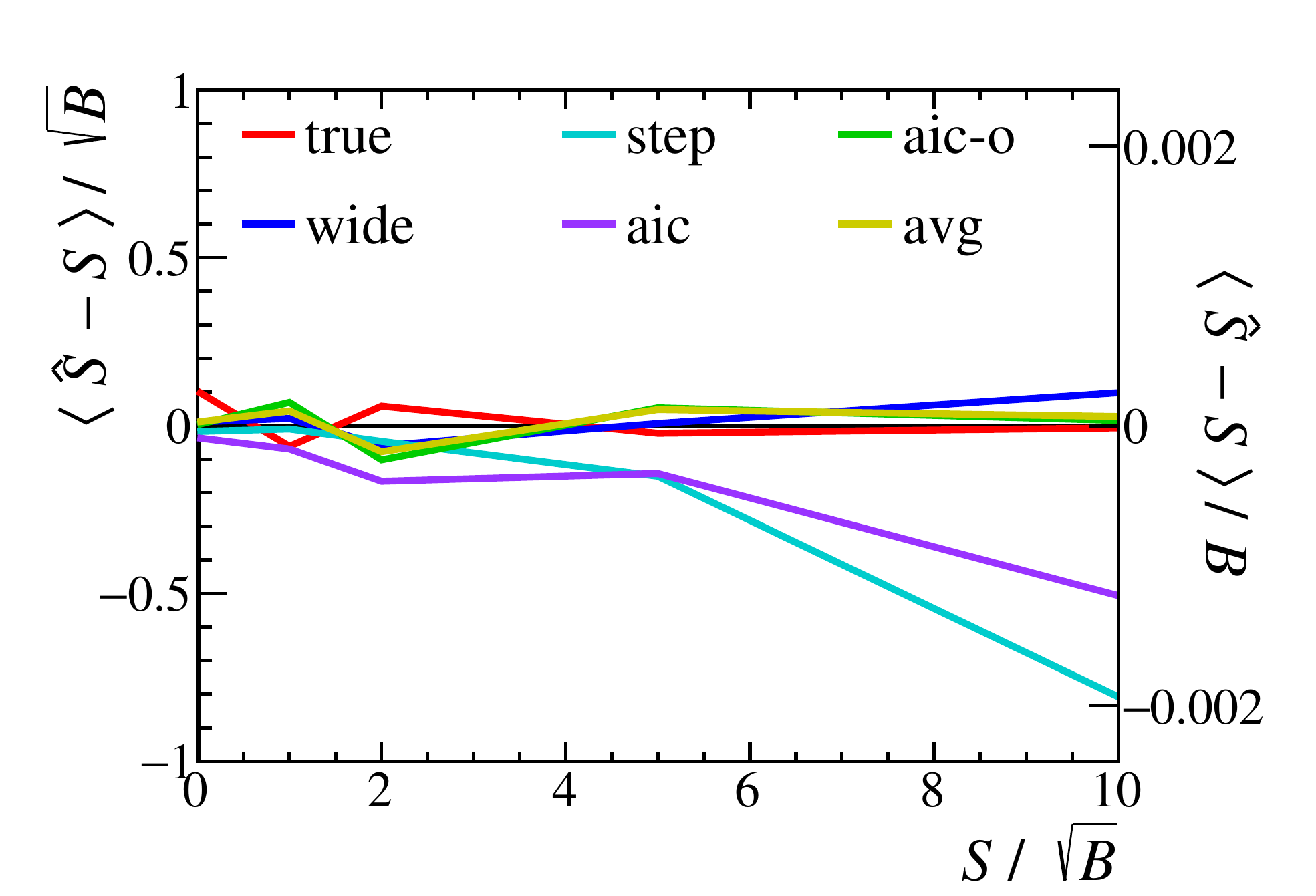}
  \includegraphics[width=0.49\textwidth]{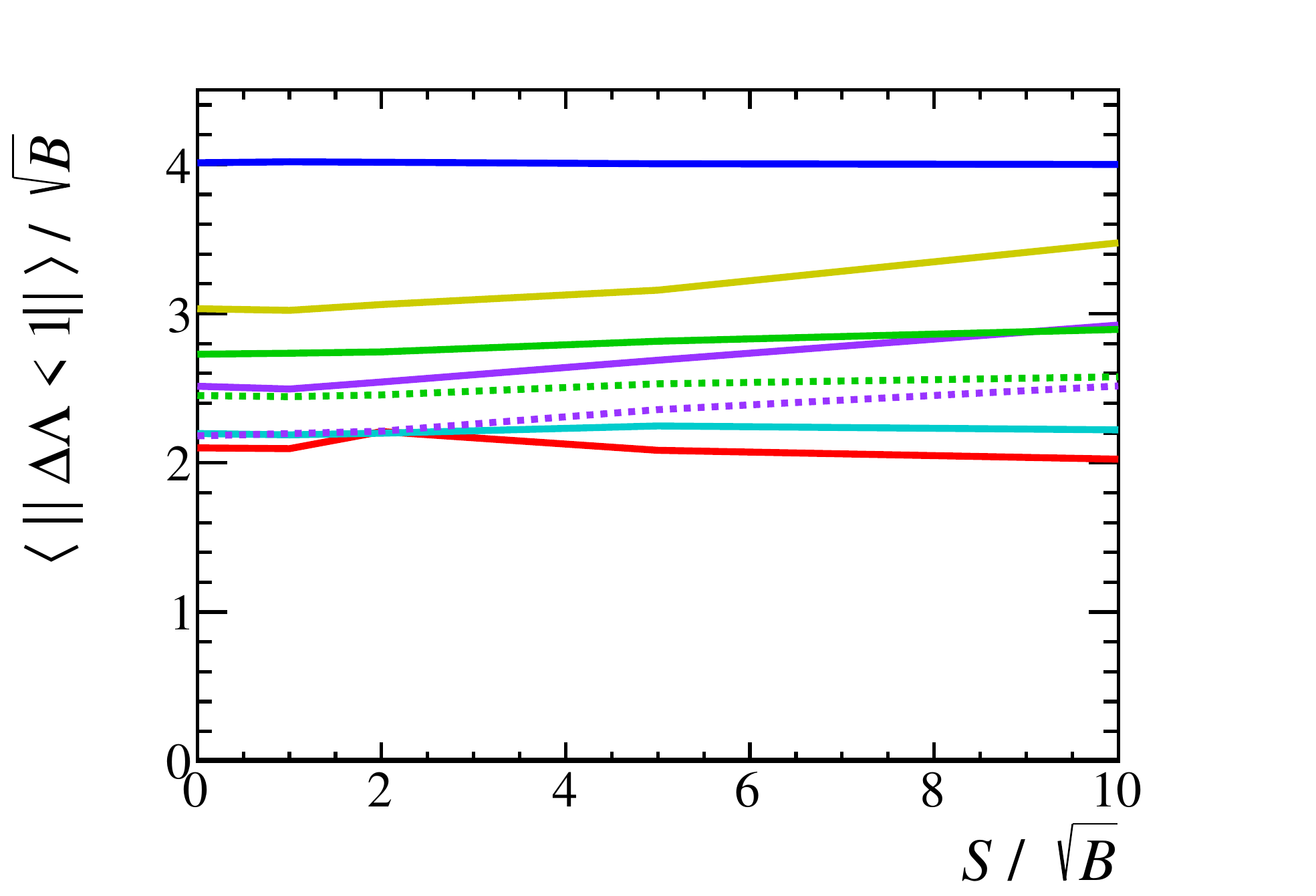}
  \includegraphics[width=0.49\textwidth]{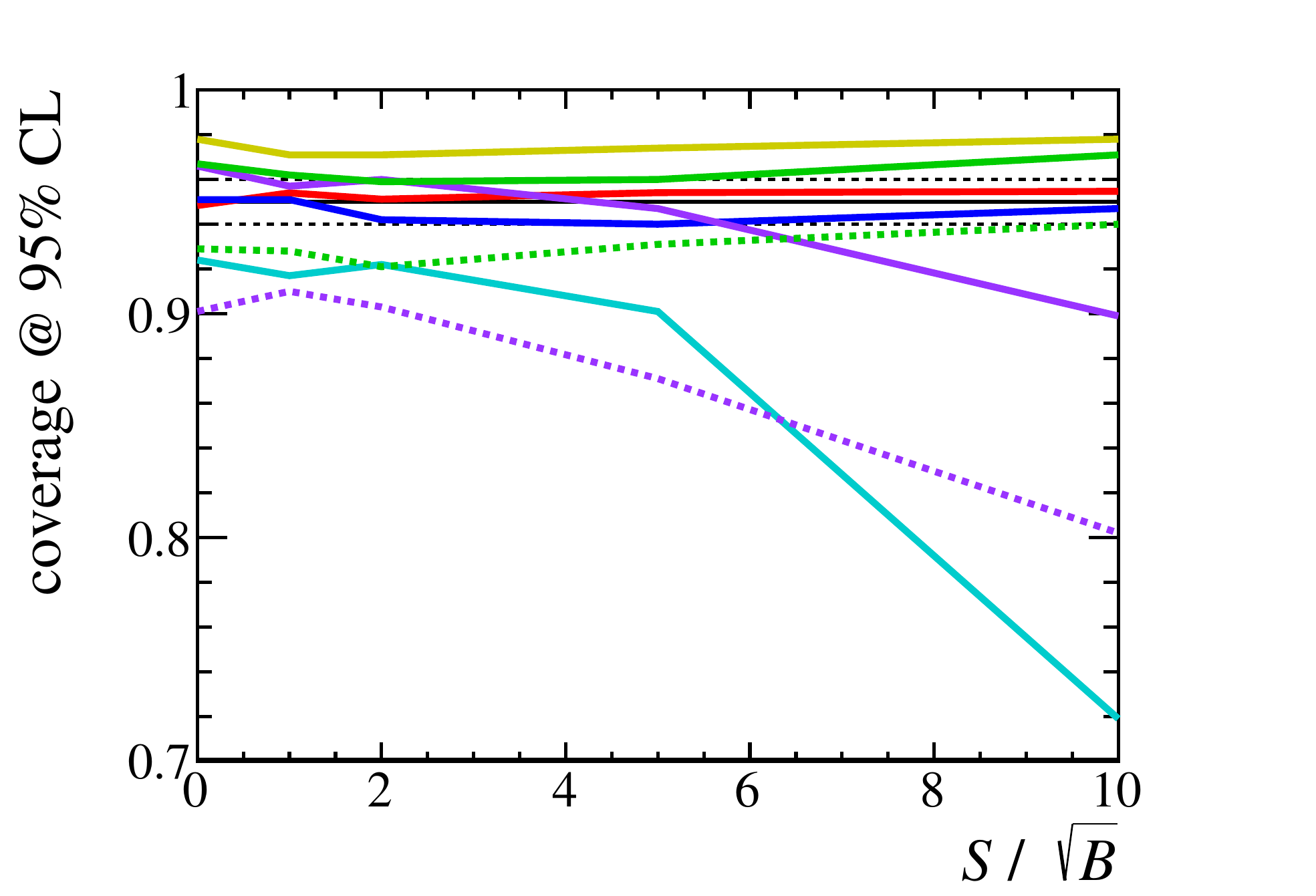}
  \includegraphics[width=0.49\textwidth]{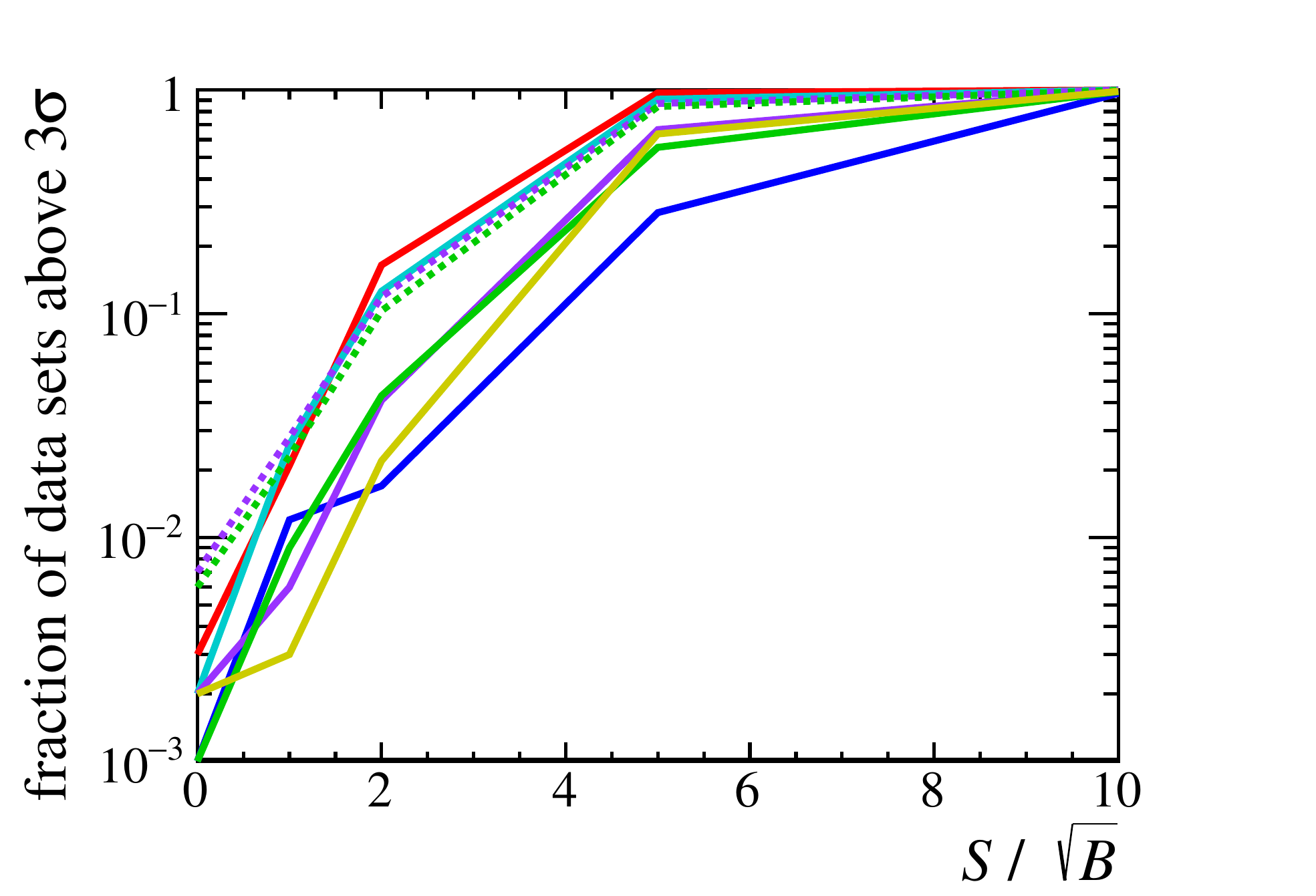}
\caption{Same as Fig.~\ref{fig:apex_results_0} but for $m=2.9$.}
  \label{fig:apex_results_4}
\end{figure}

\end{document}